\documentclass[12pt,a4paper]{book}%
\usepackage{graphicx}
\usepackage{epsfig}
\usepackage{psfrag}
\usepackage{bbm}
\usepackage{amsmath}
\usepackage{exscale}
\usepackage{amsfonts}
\usepackage{amssymb}%
\begin{document}

\author{Dissertation zur Erlangung des Doktorgrades\\vorgelegt von\\Konstantin Petrov}
\title{Dimensional Reduction of Lattice Gauge Theory in (2+1) dimensions}
\frontmatter
\maketitle
\tableofcontents

\mainmatter 

{\Large \textbf{Introduction}}\newline\vspace{1cm}

During last decades quantum chromodynamics (QCD), as a part of the Standard
Model, established itself as a working theory of the strong interactions. We
believe that quarks and gluons are the basic objects that make up the hadronic
matter. This theory has been very successful in predicting phenomena involving
large momentum transfer. Unfortunately, its direct applications are limited to
this particular case, when the coupling constant is small and perturbation
theory is reliable. However, in the hadronic world, the coupling constant is
of the order of one and the perturbation series do not converge. In this
domain the discrete space-time formulation of quantum chromodynamics provides
a non-perturbative tool for the calculation of the hadronic spectrum and
basically every other observable from first principles. Lattice QCD was used
to study the questions of confinement and chiral symmetry breaking.

Lattice QCD is usually formulated on a square Euclidean space-time grid. No
new parameters or field variables are involved in this formulation, hence, it
retains the fundamental character of QCD. In fact, it can serve two purposes.
First, it acts as a non-perturbative regularization scheme. At finite values
of the lattice spacing $a,$ which provides an ultra-violet cutoff there are no
infinities (as long as lattice is finite itself). On the other hand it gives
us an ability to do direct simulations of lattice QCD on the computer using
methods analogous to those used in statistical mechanics. One can in principle
calculate all correlation functions of hadronic operators and matrix elements
of any operator between hadronic states. These simulations are based on a
Monte Carlo integration of the Euclidean path integral and have therefore both
discretization and statistical errors.

The starting point for lattice QCD is the partition function in Euclidean
space-time%
\begin{equation}
Z=\int\mathcal{D}A_{\mu}\mathcal{D\psi D\bar{\psi}}e^{-S}%
\end{equation}
where $S$ is the QCD action%
\begin{equation}
S=\int d^{4}x\left(  \frac{1}{4}F_{\mu\nu}F^{\mu\nu}-\mathcal{\psi M\bar{\psi
}}\right)  \label{action}%
\end{equation}
Here $\mathcal{M}$ is Dirac operator, fermions are represented through
Grassmann variables $\mathcal{\psi}$ and $\mathcal{\bar{\psi}}$; $F_{\mu\nu}$
is the Yang-Mills field strength tensor for gauge fields $A_{\mu},$ which are
elements of the $SU(3)$ algebra. Fermion fields can be then integrated out
exactly, leaving us with
\begin{equation}
Z=\int\mathcal{D}A_{\mu}\det\mathcal{M}e^{-S_{G}}%
\end{equation}
The partition function is now an integral over background gauge
configurations. The fermionic contribution is contained only in the very
non-local term $\det\mathcal{M}.$ The model defined so describes zero
temperature physics. To calculate the thermodynamical quantities at finite
temperature one has to make the time direction periodic and set its length to
be the inverse temperature. We write then%
\begin{equation}
S=\int_{0}^{1/T}dx_{0}\int_{V}d^{3}x\left(  \frac{1}{4}F_{\mu\nu}F^{\mu\nu
}-\mathcal{\psi M\bar{\psi}}\right)  \label{ymaction}%
\end{equation}
where $x_{0}$ refers to the time coordinate and $V$ is the physical volume of
the system. The gauge fields have then periodic boundary conditions and
fermion fields - anti-periodic ones.

The fermionic determinant is the major complication in all simulations on the
lattice, due to its non-locality and other technical details. The model
without this term, so-called pure gauge theory, has been in the focus of the
interest for many years. It was successfully simulated by several groups
\cite{eos} and the equation of state and the phase diagram were successfully
studied and extrapolated to continuum

The full model, however, suffers from severe problems of both calculational
and theoretical type. Despite significant progress during last years
\cite{karschreview}, the time estimated for the continuum extrapolation of the
results from the direct simulations of any of the existing fermionic actions
(aiming, e.g. at the determination of the quark masses) varies from several to
hundred teraflops-years. Thus one would want to have an indirect method of
calculating the quantities of interest, which would require significantly less
computer time. One of such methods, applicable at high temperatures, is the
so-called dimensional reduction (\cite{ginsparg}-\cite{thomas}). In general it
is a method of turning (D+1)-dimensional system at high temperature into the
D-dimensional one. Using the fact that at high temperature the domain of
integration in (\ref{ymaction}) shrinks one can perturbatively get rid of the
non-static time-like degrees of freedom. Such an effective model is much
cheaper to simulate and is supposed to reproduce the long-distance behavior of
the model. This method was successfully applied to QCD at high temperature and
to the study of the electro-weak transition. At the same time the domain of
its validity is not well-studied, and this is the primary aim of the current work.

We focus on the (2+1)-dimensional pure gauge field theory, which is then
reduced to a two-dimensional one. The source model does not include the
dynamic fermions and can be simulated with a high precision using the present
computer power. While being different from the standard (3+1)-dimensional
system it shares with it many important properties, like static quark
confinement; the infrared singularities are even more severe. The reduced
model can be simulated even on personal computers.

Current work is organized as follows. In the first chapter we define our model
and operators of interest and review simple models of the dimensional
reduction. In the second one we present the dimensionally reduced model and
compare it to the original one. The third chapter is dedicated to the reduced
model \textit{per se: }its spectrum, phase structure and other properties. The
fourth chapter will address the restoration of the $Z(3)$ symmetry and is
followed by conclusions and outlook.

\chapter{Reducing the degrees of freedom}

\section{The Model}

Pure gauge chromodynamics on the lattice may be formulated in terms of closed
loops on the square lattice $\Lambda$ with link size $a$. We will denote its
extensions in space and in time directions by $L_{s}$ and $L_{0}$
correspondingly. The lattice is periodic in all directions and its length in
the time direction equals the inverse temperature%
\begin{equation}
L_{0}a=1/T\label{invT}%
\end{equation}
On each link of the lattice we put an $SU(3)$ matrix $U(x;\mu)$; it is linked
to the gauge fields $A_{\mu}(x)$ by%
\begin{equation}
U(x;\mu)=\exp iA_{\mu}(x)
\end{equation}
where $g_{0}$ is the bare strong coupling constant. The main requirement for
the action is the invariance under local gauge transformation generated by
$SU(3)$ matrices $G(x)$%
\begin{equation}
U(x;\mu)\rightarrow G^{\dagger}(x)U(x;\mu)G(x+\mu).
\end{equation}
The simplest such action is the Wilson one-plaquette action
\begin{equation}
S(U)=\beta_{3}\sum_{x\in\Lambda,\mu<\nu}\left(  1-\frac{1}{3}\operatorname{Re}%
TrU_{P}^{\mu\nu}(x)\right)
\end{equation}
Here%
\begin{equation}
\beta_{3}=\frac{6}{ag_{0}^{2}}%
\end{equation}
is the lattice coupling constant. In (2+1) dimensions $g_{0}^{2}$ has a
dimension and sets the scale of the model. The coupling $\beta_{3}$ is
inversly proportional to the lattice spacing and thus proportional to the
temperature (for constant $L_{0}$). The plaquette variable
\begin{equation}
U_{P}^{\mu\nu}(x)=U(x;\mu)U(x+\mu;\nu)U(x+\nu;\mu)^{-1}U(x;\nu)^{-1}%
\end{equation}
represents a loop around one plaquette. It is easy to show that in so-called
"naive" continuum limit when we send the link length to zero and expand the
$U$ fields around unity one recovers the Yang-Mills action%
\begin{equation}
S=\frac{1}{2}Tr\int d^{4}xF_{\mu\nu}F^{\mu\nu}%
\end{equation}

The partition function is%
\begin{equation}
Z=\int dU\exp\left(  -S\right)  ,
\end{equation}
where $dU$ is the invariant Haar measure. The expectation value of any
operator is defined thus as%
\begin{equation}
<O>=\frac{\int dUO(U)\exp\left(  -S\right)  }{\int dU\exp\left(  -S\right)  }.
\end{equation}
The operators which one frequently is interested in, and which will be the
major focus of our consideration are the Polyakov loop, which is a trace of
the product of the link variables along the time axis%
\begin{equation}
L\left(  \vec{x}\right)  =\frac{1}{3}Tr%
%TCIMACRO{\dprod \limits_{i=0}^{L_{0}-1}}%
%BeginExpansion
{\displaystyle\prod\limits_{i=0}^{L_{0}-1}}
%EndExpansion
U(\vec{x},ia;\hat{0}), \label{ploop3d}%
\end{equation}
and the spatial Wilson loop, which is a trace of the product of the spatial
link variables along the closed path. We will consider only rectangular Wilson
loops, so for a rectangle $R=[R_{1},R_{2}]$%
\begin{equation}
W(R_{1},R_{2})=\frac{1}{3}%
%TCIMACRO{\dprod \limits_{\vec{x},l\in R}}%
%BeginExpansion
{\displaystyle\prod\limits_{\vec{x},l\in R}}
%EndExpansion
U(\vec{x},l) \label{swl}%
\end{equation}
In (2+1) dimension this model has a deconfinement phase transition of the
second order. The order parameter (on infinite size lattices) is the Polyakov
loop. The critical coupling $\beta_{c}$ has been determined precisely in
\cite{karsch2p1} and is 14.74(5) for $L_{0}=4.$ We will later use this value
as the basic unit and express temperature in units of the critical
temperature, using for fixed $L_{0}$
\begin{equation}
\frac{T}{T_{c}}=\frac{\beta}{\beta_{c}}=\frac{\beta}{14.74}. \label{tcrit}%
\end{equation}
Above the critical point there exists a finite string tension and the Debye
screening mass is not zero; screening mass vanishes at the critical temperature.

\section{Naive reduction}

In this section we will briefly review so-called \textquotedblleft
naive\textquotedblright\ dimensional reduction. It is a rough approximation of
the full reduced model, which we will discuss in Chapter 2. In fact for very
high temperatures one may integrate out time-related degrees of freedom. Then
we are left with (in our case) a two-dimensional lattice $\Lambda_{2}$ with
Wilson action%
\begin{equation}
S_{W}^{2}=-\beta_{2}\sum_{x\in\Lambda_{2}}\frac{1}{3}\operatorname{Re}%
TrU_{P}\label{puregauge}%
\end{equation}
with the coupling $\beta_{2}$ being $L_{0}$ times bigger than the original
coupling of $(2+1)D$ model.%

\begin{equation}
\beta_{2}=L_{0}\beta_{3}%
\end{equation}

This approximation is particularly interesting due to the fact that this model
can be analytically solved. The solution comes up to Gross and Witten
\cite{gross}, here we will sketch the main ideas and present the actual
calculation of the string tension (we will need this result in Chapter 2) in
Appendix 2.

Let us consider an infinite (or with open boundary conditions) $2D$ square
lattice as described above. Due to the nature of our $2D$ model we will use
symbolics $\hat{1}$ and $\hat{2}$ for directions $x$ and $y$, leaving $\hat
{0}$ for the time direction. For this model the following gauge may be
introduced. First, we set an arbitrary gauge matrix in the coordinate system
origin and gauge all link variables in $\hat{2}$ direction to one. Then,
starting from this line, we set all links in $\hat{1}$ direction to unity
(like in the static gauge, although we do not have time-direction now) and
have thus:%
\begin{align*}
U(x,\hat{1})  &  =1,\\
U(0\hat{1}+n\hat{2},\hat{2})  &  =1
\end{align*}

Now we introduce the following change of variables
\begin{equation}
U(x+\hat{1},\hat{2})\equiv W(x)U(x,\hat{2}),
\end{equation}
so that the action factorizes and the statistical sum becomes
\begin{align}
Z &  =\int\prod_{x}dW_{x}\exp\left[  \sum_{x}\beta\frac{1}{3}\operatorname{Re}%
TrW_{x}\right]  =\\
&  =\left(  z\right)  ^{V/a^{2}}%
\end{align}
with
\begin{equation}
z=\int dW\exp\left[  \beta\frac{1}{3}\operatorname{Re}TrW\right]
\end{equation}
being a one-link integral.

With the same argumentation one can get for the spatial Wilson loop with
extensions $R$ and $T$
\begin{equation}
W_{L}=(w)^{RT}%
\end{equation}
with
\begin{equation}
w=\int dW\frac{1}{3}TrW\exp\left[  \beta\frac{1}{3}\operatorname{Re}%
TrW\right]
\end{equation}

The string tension is then
\begin{equation}
\sigma=-\ln w=-\log\frac{\partial}{\partial\beta}z\label{sigma02}%
\end{equation}

One can immediately notice that $w$ is always smaller than one; thus string
tension is positive, and one is in the confined phase for all coupling values.
This is the property which all existing dimensionally reduced models have - no
deconfinement phase transition. This reduction is useful only as a reference
point; it is not possible to define Polyakov loops and the screening masses in
this model.

\section{Classical reduction}

The next step in our consideration is so-called classical reduction. The name
"classical" refers to it being performed on the level of action, not taking
into account the quantum fluctuations. Consider the general (we will do this
in continuum, and use different normalisation of the potential to have the
coupling constant in front of the integral) action for a pure gauge $SU(3)$
Yang-Mills theory in (2+1)D:%
\begin{equation}
S=\frac{1}{2g^{2}}\int_{0}^{1/T}dx_{0}\int d^{2}x\sum_{\mu,\nu=0}^{2}%
TrF_{\mu\nu}F_{\mu\nu}%
\end{equation}
with the field strength thensor being%
\begin{equation}
F_{\mu\nu}=\partial_{\mu}A_{\nu}-\partial_{\nu}A_{\mu}-ig[A_{\nu},A_{\mu}]
\end{equation}
For very high temperature it seems completely legitimate to ignore the
time-dependence of the fields%
\begin{equation}
\partial_{0}A_{\nu}=0
\end{equation}
Now we can rewrite $F_{\mu\nu}$ in the following symbolic way (latin-style
indexes go over spatial directions only):%
\begin{align}
F_{\mu\nu} &  =F_{km}+\partial_{k}A_{0}-ig[A_{k},A_{0}]+\partial_{m}%
A_{0}-ig[A_{m},A_{0}]\\
&  =F_{km}+D_{k}A_{0}+D_{m}A_{0}\nonumber
\end{align}
with the covariant derivative being%
\begin{equation}
D_{k}A_{0}=\partial_{k}A_{0}-ig[A_{k},A_{0}]
\end{equation}
The action then takes the form
\begin{equation}
S\simeq\frac{1}{2G^{2}}\int d^{2}x\sum_{k,m=1}^{2}TrF_{km}F_{km}+2\sum
_{k,m=1}^{2}\left(  D_{k}A_{0}\right)  ^{2}%
\end{equation}
with the new coupling%
\begin{equation}
G^{2}=g^{2}T
\end{equation}
This is the action with the same gauge group in two dimensions, coupled to the
scalar field $A_{0}$ which transforms under the adjoint representation of the
gauge group. All the fields are called static modes because they do not have
any dependence on $T.$ The latter is present only in the coupling constant.

However, as we mentioned before, this procedure is classical and ignores the
quantum fluctuations in the system. Detailed consideration performed in e.g.
\cite{landsman} shows that the complete decoupling of the non-static modes is
not supposed to be happening in this model.

\newpage

\chapter{Dimensional reduction}

\section{Perturbation Theory}

Now we will turn to the actual derivation of the reduced model. We are
interested in \textit{gauge invariant} operators, such as the Polyakov loop
correlations and the spatial Wilson loops, and those may be computed in any
gauge. We will therefore choose the following particular static gauge, which
can be realized on the periodic lattice by first setting all $A_{0}$ to be
independent of the imaginary time coordinate $x_{0}$%
\begin{equation}
A_{0}(x_{0},\vec{x})=A_{0}(\vec{x}) \label{static}%
\end{equation}

This automatically makes the Polyakov loops static operators. Remnants of the
gauge freedom are eliminated by adding the so-called Landau constraint
\begin{equation}
\sum_{x_{0}}\sum_{\hat{\imath}=1}^{2}\left[  A_{i}\left(  x\right)
-A_{i}\left(  x-a\hat{\imath}\right)  \right]  =0 \label{stalg}%
\end{equation}
Together with (\ref{static}) this condition is called Static Time-Averaged
Landau Gauge (STALG) (\cite{STALG},\cite{curci}). Spatial Wilson loops are not
static operators in this gauge.

One can split the field in the static and the non-static components,
$A_{i}^{static}$ and $A_{i}^{ns}$ which are defined by
\begin{align}
A_{i}(x)  &  =A_{i}^{static}(\vec{x})+A_{i}^{ns}(x_{0},\vec{x})\\
\sum_{x_{0}}A_{i}^{ns}(x_{0},\vec{x})  &  =0
\end{align}
The non-static modes should be integrated out, leaving only the static
components. We will do this in a perturbative fashion and will omit the
superscript \textquotedblright static\textquotedblright\ then.

It is well known that at the high temperature $Z(3)$ symmetry of the Wilson
action is broken and the average of the Polyakov loop is not zero and is close
to one of the cubic roots of one. One can eliminate this three-fold degeneracy
by rotating the phase away and keeping all gauge variables close to one. In
this case one can perform a perturbative expansion, using the Taylor series
around unity in arbitrary parametrization satisfying (\ref{stalg}) and
(\ref{static}).

We obtained the effective two-dimensional action according to the standard
perturbative techniques (\cite{appelquist},\cite{ginsparg},\cite{landsman}).
In momentum space the effective action consists of the classically reduced
$3D$ action with all fields not depending on the $0^{th}$ coordinate (static
modes only) and the one-point irreducible Feynman graphs with $n-$ external
legs and non-static internal lines, which we will denote by $\tilde{\Gamma
}_{ns}^{(n)}.$ We have for the $n-th$ order%
\begin{equation}
\tilde{S}_{eff}^{(n)}\left(  p_{1}...p_{n-1}\right)  =\tilde{S}_{0}%
^{(n)}\left(  p_{1}...p_{n-1}\right)  -\tilde{\Gamma}_{ns}^{(n)}\left(
p_{1}...p_{n-1}\right)  .
\end{equation}
It may be proven that $\tilde{\Gamma}_{ns}^{(n)}$ are analytic in momenta due
to internal lines having no singularities, so we can expand them around zero
momentum and keep only the high-temperature contributions.

Let us see which diagrams we need to calculate. We will limit ourselves to the
one-loop approximation. For superrenormalizable theory in 3 dimensions
\begin{equation}
g_{R}^{2}\Lambda=\frac{g_{0}^{2}}{a}%
\end{equation}
where $g_{R}$ is the renormalized coupling, $g_{0}$ - the bare coupling and
$\Lambda$ is the momentum cutoff parameter. Higher order contributions to
$\Gamma_{ns,\,R}^{(n)}$ get higher powers of $\Lambda,$ thus smaller powers of
$T.$ For diagrams each external leg brings in $\left(  g_{R}^{2}%
\Lambda\right)  ^{1/2}$ in the nominator.

The diagrams with three external legs, which would lead to the terms
proportional to the $TrA_{0}^{3},$ are not permitted by the time reversal
symmetry of the original action. This symmetry is discussed more in detail in
chapters 3 and 4.

The terms with four external legs are of the type $\int d^{3}k/k^{4}$ and thus
UV-finite. The restriction to the non-static modes $k_{0}\neq0$ makes them
also IR-finite. Therefore we computed the corresponding diagrams in the
continuum, which is simpler and avoids cutoff effects. The Feynman rules were
taken from \cite{rothe}. The direct calculation was compared to the adaptation
of the Nadkarni calculation \cite{nadkarni} to the $D=3$ case.

For the term proportional to the square of the "higgs" field we need the
$\tilde{\Pi}_{00}^{ns}(0),$ which is the one loop contribution of the
non-static modes to vacuum polarization at zero momenta. Two Feynman graphs
contribute, both linearly divergent with $\Lambda.$ We computed them on the
lattice (see Appendix 3), which provides the cut-off of the order of
\begin{equation}
\Lambda\simeq\frac{1}{a}=TL_{0}%
\end{equation}
at high temperature.

\section{Action}

The resulting action depends thus on the parameters of the original theory
$L_{0}$ and $\beta_{3}$ (we will set $a=1$ from now on). We make a change of
variables
\begin{equation}
A_{0}\left(  x\right)  =\phi(x)\sqrt{\frac{6}{L_{0}\beta_{3}}} \label{phi}%
\end{equation}
which will be useful later and express $S_{eff}^{2}$ as the function of this
field and the two-dimensional gauge field $U.$ It consists of the following
three parts:

\begin{itemize}
\item pure gauge part $S_{W}^{2},$ which we had already in "naive" reduction%

\begin{equation}
S_{W}^{2}=\beta_{2}\sum_{x\in\Lambda_{2}}Tr\left(  1-\frac{1}{3}%
\operatorname{Re}U(x;\hat{1})U(x+a\hat{1};\hat{2})U(x+a\hat{2};\hat{1}%
)^{-1}U(x;\hat{2})^{-1}\right)
\end{equation}

\end{itemize}%

\begin{equation}
\beta_{2}=L_{0}\beta_{3}%
\end{equation}

\begin{itemize}
\item gauge covariant kinetic term, depending both on $U$ and $\phi,$ obtained
by expanding the $S_{W}^{3}$ to the second order in $A_{0}$ (we had it already
in "classical" reduction)%

\begin{equation}
S_{U,\phi}=\sum_{x}\sum_{i=1,2}Tr\left(  D_{i}\left(  U\right)  \phi
(x)\right)
\end{equation}
with $D_{i}$ being covariant derivative
\begin{equation}
D_{i}\left(  U\right)  \phi(x)=U(x;\hat{\imath})\phi(x+a\hat{\imath}%
)U(x;\hat{\imath})^{-1}-\phi(x).
\end{equation}
Now one can easily see the reason for the normalization of the Higgs-field
(\ref{phi}) - the classical limit of the kinetic term is $Tr\left(
\partial_{i}\phi\right)  ^{2}.$

\item self-interaction term $S_{\phi}$, which we will also call
\textquotedblright Higgs potential\textquotedblright%
\begin{equation}
S_{\phi}=h_{2}\sum_{x}Tr\phi^{2}(x)+h_{4}\sum_{x}\left(  Tr\phi^{2}(x)\right)
^{2}%
\end{equation}
with the couplings%
\begin{align}
h_{2} &  =-\frac{9}{\pi L_{0}\beta_{3}}\mu,\label{h2}\\
\mu &  =\left(  \log L_{0}+\frac{5}{2}\log2-1\right)  ,\\
h_{4} &  =\frac{9}{16\pi\beta_{3}^{2}}.\label{h4}%
\end{align}

\end{itemize}

We note here that the second term in the action contains both terms of the
type $\left(  Tr\phi^{2}(x)\right)  ^{2}$ and $Tr\phi^{4}(x)$ which are
related to each other via%
\begin{equation}
\left(  Tr\phi^{2}(x)\right)  ^{2}=2Tr\phi^{4}(x)
\end{equation}
in the case of $SU(3)$ group (or, in our case, for the variables which are
related to it via multiplication by a constant factor).

The coupling $h_{4}$ comes from the diagrams with four external legs, while
the quadratic coupling is expressed via%
\begin{equation}
h_{2}=-\frac{6}{\beta_{3}}\tilde{\Pi}_{00}^{ns}(0) \label{h2pi}%
\end{equation}
which can be numerically calculated from Eq.(\ref{pi00}). We used, however,
the scaling form of this quantity (\ref{h2}), that is its asymptotic value at
large $L_{0}$ for the reason that we assume (and prove) that we work in the
scaling region of the model (see subsection \ref{subsec:scaling}). The details
of the derivation can be found in Appendix 3.

Let us now have a closer look at this action. Unlike the quadric coupling
$h_{4},$ which is positive and insures boundedness of the partition function
at large fields, the quadratic coupling $h_{2}$ is negative. The
potential\ thus has a shape typical for the gauge symmetry breaking by Higgs
mechanism. Of course, this symmetry breaking is not expected to happen in two
dimensions. In fact, the form of this coupling can be seen as the counter-term
for the logarithmic ultra-violet divergence of the two-dimensional model, here
prescribed in advance by the UV regularization of the original model. The
actual infrared\ behavior of the model is highly non-perturbative.

In the static gauge the Polyakov loop operator is given in 2 and (2+1)
dimensions by the same function of the $A_{0}$ field. According to our
normalization of the $\phi-$field it reads
\begin{equation}
L(x)=\frac{1}{3}Tr\exp\left(  i\phi(x)/\sqrt{\tau}\right)  \label{ploop2d}%
\end{equation}
with
\begin{equation}
\tau\equiv\frac{\beta_{3}}{6L_{0}}=\frac{\beta_{2}}{6L_{0}^{2}}=\frac{T}%
{g_{3}^{2}} \label{tau}%
\end{equation}
being a dimensionless temperature.

Along the line of constant physics ($\tau=const$ ) the operator used to probe
dimensional reduction remains unchanged, and it is thus consistent to compare
the correlations of the Polyakov loops measured in three dimensions at the
same fixed $\tau$ value. The continuum limit is $L_{0}\rightarrow\infty,$ and
large distance behavior can be checked within either one of the two models.

Let us now consider the effective action $S_{eff}^{2}$ as a function of the
dimensionless temperature $\tau$ and the parameter of the original lattice
$L_{0}.$ After setting
\begin{equation}
a^{2}g_{2}^{2}=\frac{6}{\beta_{2}}%
\end{equation}
and using the equations above we find the following form for the three
dimensionless couplings
\begin{equation}
a^{2}g_{2}^{2}=\frac{1}{\tau L_{0}^{2}}%
\end{equation}%
\begin{equation}
h_{2}=-\frac{3}{2\pi}\frac{c_{0}\log L_{0}+c_{1}}{\tau L_{0}^{2}}%
\end{equation}%
\begin{equation}
h_{4}=\frac{1}{64\pi}\frac{1}{\tau L_{0}^{2}}\frac{1}{\tau}%
\end{equation}
One can easily see that apart from the logarithmic corrections all three
couplings scale as $1/L_{0}^{2}$ with $L_{0}.$ As a function of $\tau,$
$h_{4}$ is $\tau$-times smaller than the two other ones.

Let us now write down the effective Lagrangian $\mathcal{L}_{eff}$ as a result
of the small $a$ expansion of the effective action $S_{eff}^{2}.$ We make a
substitution $A_{i}\rightarrow ag_{2}A_{i}$ , $\sum_{x}\rightarrow a^{-2}\int
d^{2}x$ and use $T$ and $g_{2}^{2}$ as the parameters instead of $\beta_{3}$
and $L_{0}$ and take the limit $a\rightarrow0$ everywhere but in the
logarithmic term. We get
\begin{align}
\mathcal{L}_{eff}  &  =\frac{1}{4}\sum_{c=1}^{8}F_{ij}^{c}F_{ij}^{c}+Tr\left[
D_{i}\phi\right]  ^{2}+\frac{g_{2}^{2}}{32\pi}\frac{g_{2}^{2}}{T^{2}}%
Tr\phi^{4}+\mathcal{L}_{CT}\\
D_{i}\phi &  =\partial_{i}\phi+ig_{2}\left[  A_{i},\phi\right] \\
F_{ij}  &  =\partial_{i}A_{j}-\partial_{j}A_{i}+ig_{2}\left[  A_{i}%
,A_{j}\right] \\
\mathcal{L}_{CT}  &  =\frac{3g_{2}^{2}}{2\pi}\left[  -\log aT+\frac{5}{2}%
\log2-1\right]  Tr\phi^{2}. \label{counter}%
\end{align}

This is a two-dimensional $SU(3)$ gauge invariant Lagrangian for the adjoint
scalar field $\phi$; it is far from being the most general one. The gauge
coupling $g_{2}$ has canonical dimension one in energy and sets the scale. The
non-kinetic quadratic term is the counterterm, suited to a lattice UV
regularization with spacing $a$ . The quadric self-interuction
\begin{equation}
\hat{h}_{4}=g_{2}^{2}\frac{g_{2}^{2}}{T^{2}}%
\end{equation}
goes to zero as the temperature approaches infinity and presumably plays a
marginal role.

It is easy to identify the symmetries of the reduced action. The space
inversion symmetry of the original model remains untouched, as well does the
rotational symmetry in the space-like plane. The time direction reversal
symmetry, as we pointed out above, turns now to the invariance of the action
with respect to changing the sign of the higgs fields, so we define%
\begin{equation}
R_{\tau}:\phi\rightarrow-\phi\label{rtau}%
\end{equation}

\section{Simulations}

To fulfill our aim of direct and detailed comparison, we had to simulate both
the three dimensional and the two dimensional models. For the three
dimensional part we used the Quadrics QH2 parallel computer of the Bielefeld
University. We refer to \cite{lego} for the detailed description of this
machine. The actual presented data are combined data from C.Legeland and
M.Luetgemeyer, which were crosschecked with the data obtained by the program
independently written by the author (based on a different simulation design
model of the Bielefeld Lattice Group). Both 3D programs use the
Cabbibo-Marinari pseudo-heatbath update algorithm, based on updating the
$SU(2)$ subgroups of the $SU(3)$ matrix using Kennedy-Pendelton algorithm. No
additional technique like smearing was used.

The two-dimensional project was realized on our farm of PCs. It consists of
one server and 10 diskless clients, each having AMD K6 processor at 300MHz
clock speed. The cluster was gradually upgraded to a set of AMD Athlon
processors, first to 600MHz and then to 1200MHz ones, which are (apart from
the linear scaling with the clock speed) considerably more effective in
floating point operations. Booting of the clients is done via traditional
etherboot \cite{etherboot} which resides on the network cards and boots system
from the server. The farm runs under Linux operating system from RedHat with
appropriately modified kernel. It provides Message Passing Interface (MPI)
services via Los-Alamos MPI Daemon (LAMD) from the University of Notre-Dame
\cite{lam}.

The two-dimensional program (written in collaboration with P.Bialas using
parts of the standard Bielefeld C++ Monte-Carlo code) consists of two main
parts - master and slave.

\bigskip

Master part:

\begin{itemize}
\item runs on server

\item reads and parses the configuration file

\item gives an appropriate set of parameters to all slave notes. Parameters
can be either different for exploring a certain area in coupling constants or
the same for accumulating statistics.
\end{itemize}

\bigskip

Slave part:

\begin{itemize}
\item runs on clients

\item gets the parameters

\item does the actual simulation

\item outputs the results
\end{itemize}

\bigskip

The program is realized in an object-oriented way using classes
\textit{lattice} (SU(3) group) and \textit{higgs} (SU(3) algebra).
Measurements are done via classes \textit{basic\_observables} (all local
observables), \textit{corr\_observables} (all correlations) and
\textit{swl\_observables} (non-local observables). All measurement classes are
inherited from the class \textit{observable} which does pre- and
post-processing as well as output.

For updating gauge fields we used the same algorithm as above, but
after\textit{\ each subgroup }update the resulting matrix was subjected to the
Metropolis question on the hopping term. This approach results in an
acceptance ratio of about 95\%.

Alternatively we tried the multi-hit Metropolis algorithm. An update of a
single link variable was attempted several times (typically eight) before
attempting to update the next variable. The trial matrices were generated
according to
\[
U\rightarrow\exp(i\delta A)U
\]
where $\delta A$ was a Hermitian matrix generated from the distribution
$\exp(-\epsilon TrA^{2}).$ The parameter $\epsilon$ was tuned to obtain the
acceptance ratio close to 50\%. Because generation of the trial matrices is
expensive in CPU time (for it requires matrix exponentiation), we stored the
values in the table for later use (that is we store both the matrix and its
conjugate to ensure the detailed balance). This table was updated every five
sweeps. The autocorrelation times were comparable to heat-bath but the CPU
time required to perform the metropolis update was approximately two times bigger.

The scalar fields were updated using multi-hit metropolis. At every point we
tried as before typically eight updates. New scalar fields were obtained from
\[
A\rightarrow A+\delta A
\]
where $\delta A$ was generated the same way as above, using the same desired
acceptance rate. One sweep over lattice consisted of one update of the gauge
fields followed by one update of the scalar fields.

Operators we measured were:

\begin{itemize}
\item the ''control'' variables $trA^{2}$ , $trA^{3}$ and $trA^{4}$

\item plaquette variables averages

\item Polyakov loops (defined by (\ref{ploop3d}) in 3D and (\ref{ploop2d}) in
2D) and their on-axis correlations
\begin{equation}
L(r)=\frac{1}{2}\left\langle L(x)L(x+r)+L(x+r)L(x)\right\rangle -\left\langle
L(x)\right\rangle ^{2} \label{pcorr}%
\end{equation}
where the averaging is performed over lattice and computer time. When
necessary we will use indexes $D=2+1$ or $2$ specifying the dimension considered.

\item Spatial Wilson loops $W(r_{1},r_{2})$ (defined in the same way both in
2D and 3D )\bigskip
\end{itemize}

We used the following tests to test the program:

\begin{itemize}
\item \bigskip\textbf{Gauge Sector}: remove interaction (kinetic) term in the
update routine. Then the model is reduced to the pure gauge theory with action
(\ref{puregauge}) and can be compared to the analytic solution.

\item \textbf{Higgs Sector:} set the $\beta_{3}$ and the quadric coupling to
zero. The model is reduced to the free-field model for the scalar fields which
can be exactly solved.

\item \textbf{Polyakov loop Correlations}: same as in latter case, may be
expanded and calculated in the small field limit
\end{itemize}

The analytic solution and results for the two last test are presented in the
Appendices 4 and 5.

All runs in the main simulation of the two dimensional model were performed on
$32\times32$ lattices with the parameter $L_{0}$ set to 4. The dynamics of the
gauge and Higgs sector turned out to be quite different. The integrated
autocorrelation time for the plaquette was typically of the order of one while
for the Higgs related local variables (e.g. $trA^{2}$ ) it varied between 100
and 400 sweeps when $\beta_{3}$ varied between 21.0 and 346.0. For each
$\beta$ value we collected typically one million sweeps.

\section{Polyakov loop correlations and Masses}

\subsection{Comparison with the original model}

The Polyakov loops correlations defined as in Eq.(\ref{pcorr}) were measured
every ten sweeps. While due to the large autocorrelation times this was in a
sense an \textquotedblright oversampling\textquotedblright\ - it was not too
costly in CPU time. The measurements were blocked by 50 and written out.
Afterwards the data were analyzed using the double jack-knife method with
typically 25 blocks. First, the jack-knife procedure was used in a standard
way to get the values of the $L(r)$ and the error estimates. Then these were
used to obtain the fit (see below) parameters and the errors (which were
uniform for all blocks of the second jack-knife) using the second jack-knife.
For checking consistency we also tried a different number of blocks, and the
errors, as expected, did not show any significant dependence on the number of
blocks in this region.

There are two possible statements of the dimensional reduction about the
Polyakov loops

\begin{itemize}
\item Weak: the lowest state coupled to $L(x)$ is the same in both dimensions,
so that the two functions must have the \textit{same shape} in $r $ at large
$r$

\item Strong: to the extent that the weight associated with $S_{eff}^{2}$ is a
good approximation to the integral over the non-static field of the $(2+1)D $
field (in the small $p/T$ regime at least) the averages with the respect to
the two weights of a static operator, such as $L(r),$ are equal at large $r$
\end{itemize}

We will show that the latter situation is approached at sufficiently high
temperature. Using $L_{0}=4$ fixed, we investigate the temperature dependence
of the correlations by varying $\beta_{3}$ in the range $[21,173]$. High
temperature means $\beta_{3}$ sufficiently larger than the transition point in
$(2+1)D$. These $\beta_{3}$ values correspond roughly to temperatures in the
range $T/T_{c}\in\lbrack1.43,11.73]$ \begin{figure}[ptb]
\begin{center}
\psfrag{xlabel}[cb][cb][1][0]{$r$} \psfrag{ylabel}[c][t][1][-90]{}
\psfrag{s0label}[r][r][.8][0]{(2+1)D} \psfrag{s1label}[r][r][.8][0]{2D}
\includegraphics[clip,width=14.5cm,bb=28 35 540 308]{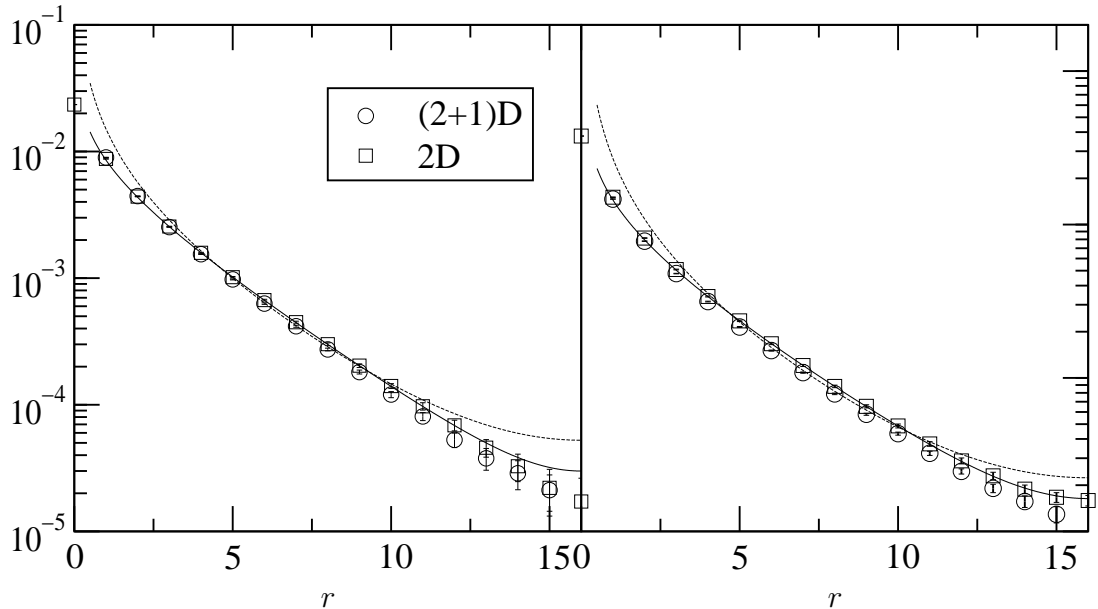}
\end{center}
\caption{Polyakov loop correlations $L_{D}(r)$, $D=2+1$ (circles) and $2$
(squares), for $\beta_{3}$ equal to $29$ (left) and $84$ (right). Distance $r$
is given in lattice units. In both cases $L_{s}=32$ and $L_{0}=4$. The $2D$
data were produced using $h_{2}$ from Eq.(\ref{h2}). The continuous and the
dashed lines result respectively from fitting formulae (\ref{pole}) and
(\ref{cut}) to the $2D$ data.}%
\label{fploop}%
\end{figure}

On the Fig.\ref{fploop} the correlations $L_{2+1}(r)$ and $L_{2}(r)$ for
$\beta_{3}=29.0$ and $84.0$ are presented. We see that the two correlations
are extremely close to each other: not only they have the same \textit{shape}
at large $r$, which is the primary prediction of dimensional reduction, but
also nearly the same \textit{normalization}. As announced, this finding
favours dimensional reduction in the strong sense.

In addition, we notice that this agreement between the two ways of computing
the Polyakov loop correlation extends down to fairly small values of $r$.
Since the normalization is set by $|\left\langle L(\vec{x})\right\rangle
|^{2}$, it shows that, although it is local, the Polyakov operator is not very
sensitive to the short wave length terms omitted in the effective action. In
Table \ref{tab1} we present the results for $\left\langle L_{2}\right\rangle $
and $\left\langle L_{2+1}\right\rangle $ to demonstrate this property.
Relatively to their distance to one (their common value at $\beta_{3}=\infty
$), their difference, already $\sim6\%$ at $\beta_{3}$=21 ($T/T_{c}=1.43)$,
decreases to only about $0.6\%$ at $\beta_{3}=173$ ($T/T_{c}=11.73)$.
\begin{table}[h]
\begin{center}%
\begin{tabular}
[c]{||l||l||l||}\hline\hline
$\vphantom{\bigg(}\beta_{3}$ & \multicolumn{1}{c||}{$|\left\langle
L_{2}\right\rangle |$} & \multicolumn{1}{c||}{$|\left\langle L_{2+1}%
\right\rangle |$}\\\hline\hline
21.0 & 0.56002 (62) & 0.53467 (16)\\\hline
29.0 & 0.67007 (25) & 0.66120 (13)\\\hline
42.0 & 0.76397 (20) & 0.76130 (12)\\\hline
84.0 & 0.87392 (11) & 0.87435 (6)\\\hline
173.0 & 0.93494 (13) & 0.93530 (11)\\\hline\hline
\end{tabular}
\end{center}
\caption{The average Polyakov loop as a function of $\beta_{3}$ in (2~+~1~)~D
and 2D.}%
\label{tab1}%
\end{table}\begin{figure}[hptbh]
\begin{center}
\psfrag{xlabel}[b][b][1][0]{$r$} \psfrag{ylabel}[c][c][1][0]{}
\includegraphics[width=12cm,clip,bb=32 70 537 507]{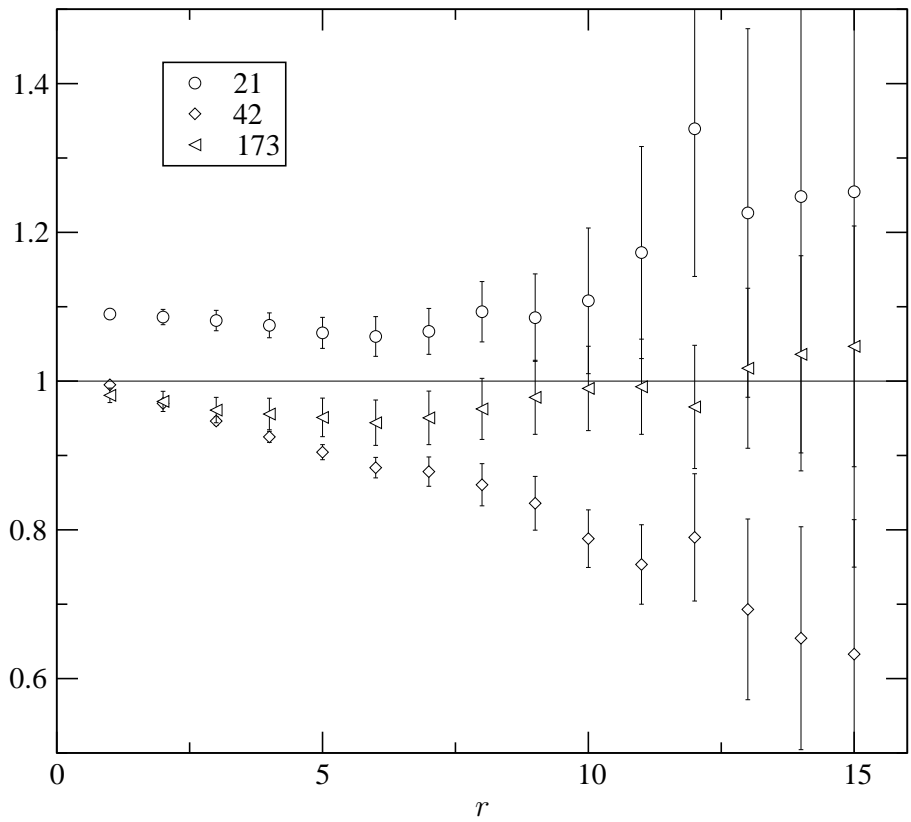}
\end{center}
\caption{Ratios $L_{2+1}(r)/L_{2}(r)$ of Polyakov loop correlations as
functions of the lattice distance $r$ for $T/T_{c}=1.4,2.8,11.7\quad(\beta
_{3}=21,42,173)$.}%
\label{fratio}%
\end{figure}

In Fig.\ref{fratio}, we illustrate more in details, for the $\beta_{3}$ values
not reported on in Fig.\ref{fploop}, how well the $2D$ and $(2+1)D$
correlations compare. Their ratio $L_{2+1}/L_{2}$ is plotted against the
distance $r$ in lattice units. These data definitely support the statement
that $L_{2+1}/L_{2}$ remains quite flat and close to one for all distances and
temperatures. Recalling that the lowest $T$ value is only $\sim1.5$ times
$T_{c}$, and that the correlation functions decrease in $r$ by about three
orders of magnitude, we conclude that the effective local 2D action reproduces
the $(2+1)D$ Polyakov loop correlations with a remarkable accuracy, soon above
the transition and down to distances even shorter than $1/T$.

\subsection{Scaling test}

\label{subsec:scaling} As for every lattice calculation, the question arises
if the lattice size is sufficient for the study, or finite size effects
influence the measurements. This is normally checked by running the same
simulation on the larger lattices and comparing results. The parameters of the
bigger lattice are defined as follows. Given the temperature, the continuum
limit is approached by taking
\begin{equation}
\tau=\frac{\beta_{3}}{L_{0}}=\frac{T}{g_{3}^{2}}\quad\text{fixed}\qquad\qquad
L_{0}=\frac{1}{aT}\quad\text{large}%
\end{equation}
\begin{figure}[ptb]
\begin{center}
\psfrag{ylabel}[c][c][1][0]{} \psfrag{xlabel}[c][c][1][0]{$r$}
\psfrag{biggerandbigger}[l][l][.8][0]{$\beta_3=58,\;L_0=8$}
\psfrag{smaller}[l][l][.8][0]{$\beta_{3}=29,\;L_{0}=4$}
\includegraphics[width=12cm]{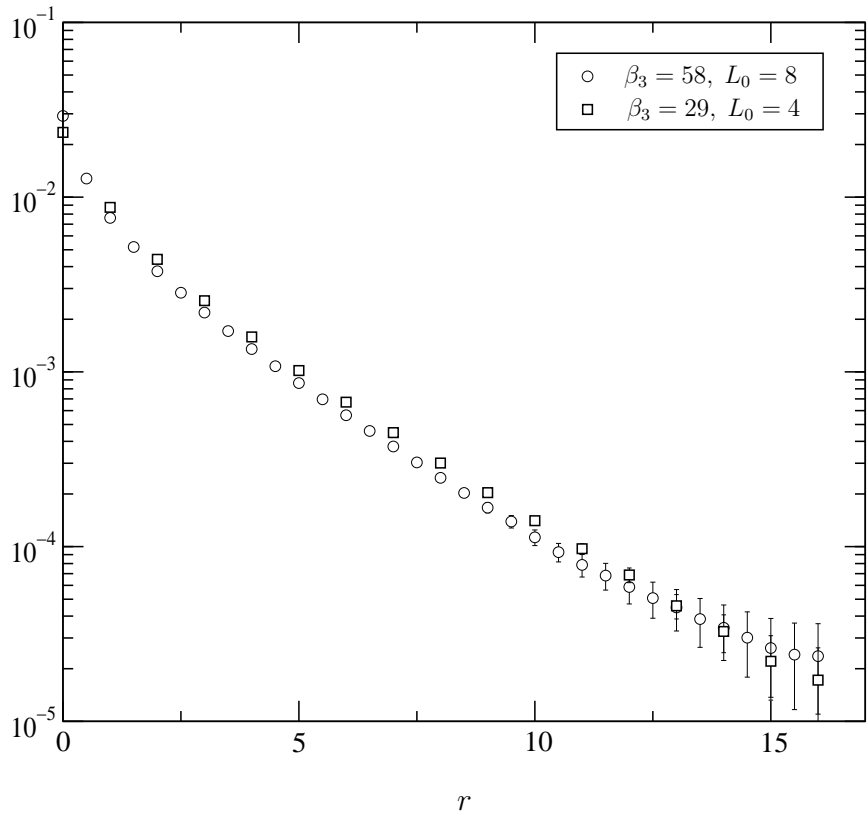}
\end{center}
\caption{Comparison of the Polyakov loop correlations in the 2D model for the
two sets of parameters corresponding to $T/T_{c}=1.97\quad$ $[\beta_{3}%
,L_{0},L_{s}]=[29,4,32]$ and $[58,8,64]$ i.e. for constant $\tau$ (\ref{tau}).
The values of $r$ for $L_{0}=8$ (circles) are scaled down by a factor two in
order to maintain the same physical scale.}%
\label{fscaling}%
\end{figure}

\begin{figure}[t]
\begin{center}
\includegraphics[width=12cm,bb=15 80 537 495]{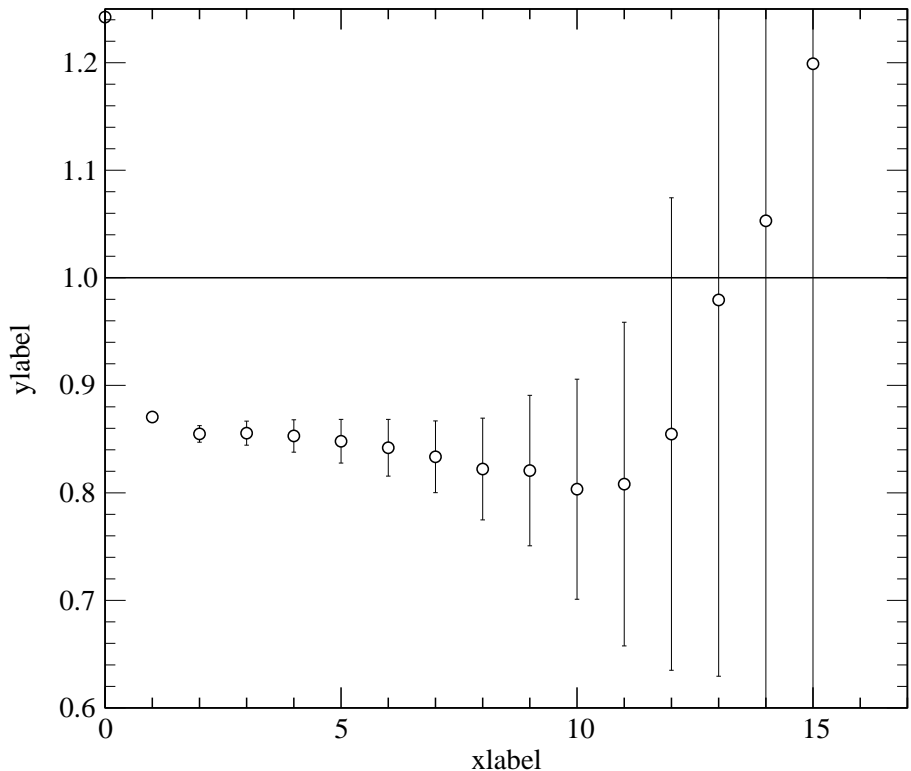}
\psfrag{xlabel}[c][c][1][0]{$r$} 
\psfrag{ylabel}[c][c][1][0]{}
\end{center}
\caption{The ratio between the two Polyakov loop correlation functions
presented in figure~\ref{fscaling}.}%
\label{fratio2}%
\end{figure}

Scaling region is then defined as the region where physics of the model at
fixed $\tau$ does not depend on $L_{0}$, if $L_{0}$ is large enough. To check
if we are in this region, we compared the Polyakov loop correlations in $2D$
for two sets of the lattice parameters, namely $[\beta_{3},L_{0}%
,L_{s}]=[29,4,32]$ and $[58,8,64]$. Given $T$, doubling $L_{0}$ means dividing
the lattice spacing $a$ by two, so that the\textit{\ physical }spatial size
$aL_{s}$ of the lattice is unchanged. The resulting correlations are presented
in Fig.\ref{fscaling}\thinspace\ showing a very similar shape as a function of
the physical distance. It is again instructive to consider their ratio, found
to be quite flat at all distances (Fig.\ref{fratio2}): within errors, there is
no sizable deviation from scaling.

The other observation is that this constant ratio is not one. And it should
not be because it is given by the corresponding values of $\left\vert
\left\langle L(\vec{x})\right\rangle \right\vert ^{2}$. They are not physical
quantities due to explicit (logarithmic) dependence of the $2D$ effective
Lagrangian on $a$ via its counterterm Eq.(\ref{counter}). 

Hence scaling is verified in the range of interest. It justifies keeping
$L_{0}=4$, which is less expensive in computer time and allowed us to use the
existing data of \cite{lego} in $(2+1)D$.

\subsection{Screening Lengths}

\label{subsec:pscreening}The fast decay of the correlation functions with $r$
is consistent with the existence of a finite spatial correlation length,
$\xi_{S}$, associated with the quantum numbers of the Polyakov loop operator,
a color singlet scalar. A mass can be defined as usual by $M_{S}=\xi_{S}^{-1}$.

For the (3+1) dimensional model D'Hoker advocated the scenario (based on
perturbation theory) that the screening mass $M_{S}$ is $2M_{E}$, twice the
so-called electric screening mass, because the lowest \textquotedblleft
state\textquotedblright\ coupled to the loop is a two electric gluon state
\cite{dhoker}. In this case the $L(r)$ correlation is expected to be
proportional at large $r$ to the square of the correlation function for a one
particle state of mass $M_{E}$:%
\begin{equation}
L_{D}^{(2m_{E})}(r)\simeq c^{\prime}\,\left(  \frac{1}{[m_{E}r]^{1/2}%
}e^{-m_{E}\,r}+\frac{1}{[m_{E}(L_{s}-r)]^{1/2}}e^{-m_{E}(L_{s}-r)}\right)
^{2}. \label{cut}%
\end{equation}

In the present case, however, the infrared sector is more complicated and
perturbation theory is doubtful. One of the options is that there exists in
the Polyakov loop channel an \textit{independent} screening mass $M_{S}$,
associated with a true excitation of the $2D$ system. In this case we
parametrize the data according to
\begin{equation}
L_{D}^{(m_{S})}(r)\simeq c\,\left(  \frac{1}{[m_{S}r]^{1/2}}e^{-m_{S}%
\,r}+\frac{1}{[m_{S}(L_{s}-r)]^{1/2}}e^{-m_{S}(L_{s}-r)}\right)  .\label{pole}%
\end{equation}
The second term in both cases arises from the finite size of the lattice and
its form is determined by the periodicity. It actually does improve fitting
process significantly. The $m$ symbols denote masses in lattice units, i.e.
$m\equiv aM$. These parametrizations respect the lattice symmetry
$r\rightarrow L_{s}-r$. Since we have no direct access to $m_{E}$, the two
expressions above differ in shape through the prefactors, $r^{-1/2}$ and
$r^{-1}$ respectively. We checked that for $r>1$ lattice artefacts are
negligible in the mass range considered. This was achieved by comparing
Eq.(\ref{pole}) to the lattice propagator
\begin{align}
L_{Latt}(m_{S},r) &  =\frac{1}{L_{s}^{2}}\sum_{p_{1},p_{2}}\cos(p_{1}%
r)\widetilde{L}_{Latt}(m_{S},\vec{p}),\\
\widetilde{L}_{Latt}^{-1}(m_{S},\vec{p}) &  =\widehat{p}^{2}+4\sinh(m_{S}%
^{2}/4).
\end{align}
We carefully analyzed our numerical data, and \textit{we find that the ansatz}
(\ref{pole}) \textit{is by far the best one}, giving a good fit of the data
down to small $r$ values. For an illustration, the continuous curves of
Fig.\ref{fploop} are fits of the form (\ref{pole}) to $L_{2}(r),r\geq r_{min}%
$, with $r_{min}=4$, and changing $r_{min}$ from 3 to 6 does not change the
resulting $m_{S}$ within errors. At the shorter distances a noticeable
deviation exists, which can be either due to the presence of the more massive
contribution or just be an artefact of the lattice UV regularization. It is
anyway too small to be reliably analyzed. On the contrary fits to the same
data of the form Eq.(\ref{cut}) lead in Fig.\ref{fploop} to the dashed curves,
which are clearly not acceptable. \begin{figure}[ptb]
\begin{center}
\psfrag{xlabel}[c][c][1][0]{$g^2_3/T$}
\psfrag{ylabel}[c][c][1][0]{$M_S/\sqrt
{Tg^2_3}$} \includegraphics[width=12cm]{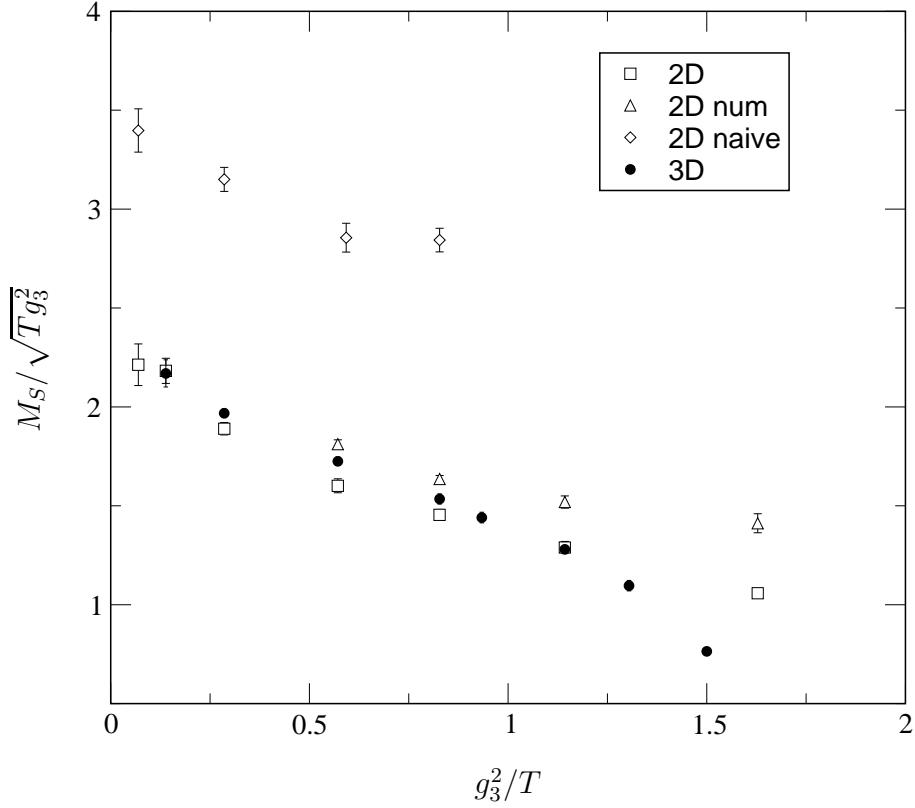}
\end{center}
\caption{Physical screening masses $M_{S}$ in units of $g_{3}\sqrt{T}$ versus
$g_{3}^{2}/T$, in $(2+1)D$ (black points) and $2D$ (squares). Also shown are
the masses obtained with the numerical value of $h_{2}$ from Eq.(\ref{h2pi})
(triangles) instead of its asymptotic expression (\ref{h2}), and with the
classically\ reduced (labeled naive) action: no Higgs potential, $h_{2}%
=h_{4}=0$. Masses of the original model vanish at critical point, while in
reduced one they do not}%
\label{fscreen}%
\end{figure}The results of systematic fits of expression (\ref{pole}) to all
available correlations $L_{2}(r)$ and $L_{(2+1)}(r)$ are presented in
Fig.\ref{fscreen}. We define the corresponding dimensionless quantities%
\begin{align}
M_{S}/(g_{3}\sqrt{T}) &  \equiv m_{S}\sqrt{L_{0}\beta_{3}/6}\\
g_{3}^{2}/T &  \equiv6L_{0}/\beta_{3}%
\end{align}
and plot them one versus the other.

\bigskip It is also interesting to check to which extent the calculated
coefficients are important for the simulations and if they have only minor
effect. Thus we also plot the masses resulting from a few simulations made
with the classically reduced effective action ($h_{2}=h_{4}=0$). They are far
away from the former, showing the important effect of taking into account the
non-static degrees of freedom. One may also want to check if the simulation
which uses the so-called scaling form (\ref{h2}) of $h_{2}$ differs much from
the simulation with the value obtained numerically from Eqs. (\ref{h2pi},
\ref{pi00}). The results are similar, especially at high temperature, but
anyway distinguishable within our statistical accuracy. This shows a great
sensitivity of the static properties to the quadratic counterterm
(\ref{counter}), an interesting feature \textit{per se}.

The scaling properties observed in the previous subsection reflect themselves
of course in the mass values. For the two cases compared there, we find
$M_{S}a$=0.331(6) and 0.169(3): the lattice spacing is reduced exactly by two
within errors. Of course the fact that dimensional reduction works well for
masses directly follows from its success for correlations. We preferred to
illustrate it first on directly measured quantities, as done in
Fig.(\ref{fscaling}). This comparison does not depend on any interpretation of
the nature of the observed screening lengths.

\section{Spatial Wilson loop}

The spatial Wilson loops are measured on a lattice via%
\begin{align}
\tilde{W}(R,R^{\prime},x,y,m)  &  =\prod_{i=0}^{R-1}U(x+i,y,\hat{0})_{m}%
\prod_{j=0}^{R^{\prime}-1}U(x+R,y+j,\hat{1})_{m}\times\\
&  \times\prod_{i=R-1}^{0}U^{\dagger}(x+i,y+R^{\prime},\hat{0})_{m}%
\prod_{j=R^{\prime}-1}^{0}U^{\dagger}(x,y+j,\hat{1})_{m}%
\end{align}%
\begin{equation}
W(R,R^{\prime})=\frac{1}{N_{m}}\sum_{m=1}^{N_{m}}\frac{1}{V}\sum
_{x,y=0}^{L_{s}-1}\tilde{W}(R,R^{\prime})
\end{equation}
where $N_{m}$ is the number of measurements, and index $m$ refers to the
particular measurement.

In our simulations the Wilson loops were measured every ten sweeps and blocked
by 50. Integral autocorrelation time for the measurements of $14\times14$
Wilson loops was, in the worst case, of the order of one and was negligible in
the majority of cases. The measurement routine is quite expensive and consumed
almost half of the CPU time.

In order to extract the string tension, local potentials were first extracted
from the ratios~:
\begin{equation}
V(R,R^{\prime})=\log\frac{\left\langle W(R,R^{\prime})\right\rangle
}{\left\langle W(R,R^{\prime}+1)\right\rangle }.
\end{equation}
By definition, the potential is
\begin{equation}
V(R)=\lim\limits_{R^{\prime}\rightarrow\infty}V(R,R^{\prime}).
\end{equation}
For each given $R$, $V(R,R^{\prime})$ was found to decay exponentially to a
constant in $R^{\prime}$. In practice, the constant regime is reached within
errors above $R^{\prime}=3$ and the potential was fitted in the range
$R^{\prime}\in\left[  4,12\right]  $. The string tension $\sigma$ was then
obtained from the ansatz%
\begin{equation}
V\left(  R\right)  =V_{0}+\sigma R. \label{vfit}%
\end{equation}

The errors on the string tension were calculated using the triple jack-knife
method as follows:

\begin{enumerate}
\item split the data into $N$ blocks

\item omitting one block at time, produce $N$ sets, each consisting of $(N-1)
$ blocks

\item in each of the sets repeat this procedure, so that we get $N\times(N-1)
$ sets each consisting of $(N-2)$ blocks

\item use these in traditional jack-knife way to obtain $N\times(N-1)$ sets of
$V(R,R^{\prime})$ with errors

\item repeat it to obtain $N$ copies of the $V\left(  R\right)  $ with errors
(which we need to do the weighted fitting)

\item fit to the formula (\ref{vfit}) get string tension and errors
\begin{figure}[t]
\begin{center}
\includegraphics[width=12cm]{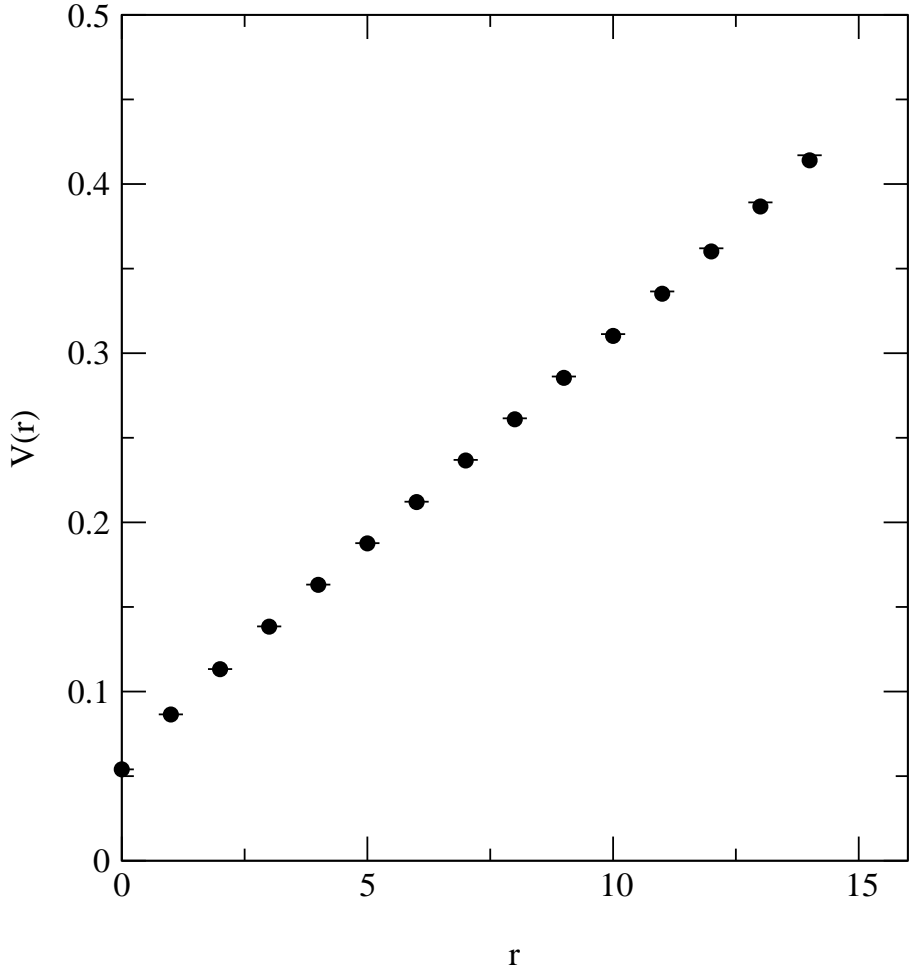}
\end{center}
\caption{Potential extracted from the spatial Wilson loops, $\beta_{3}=42$}%
\label{fpotential}%
\end{figure}
\end{enumerate}

\begin{figure}[t]
\begin{center}
\includegraphics
[width=12cm]{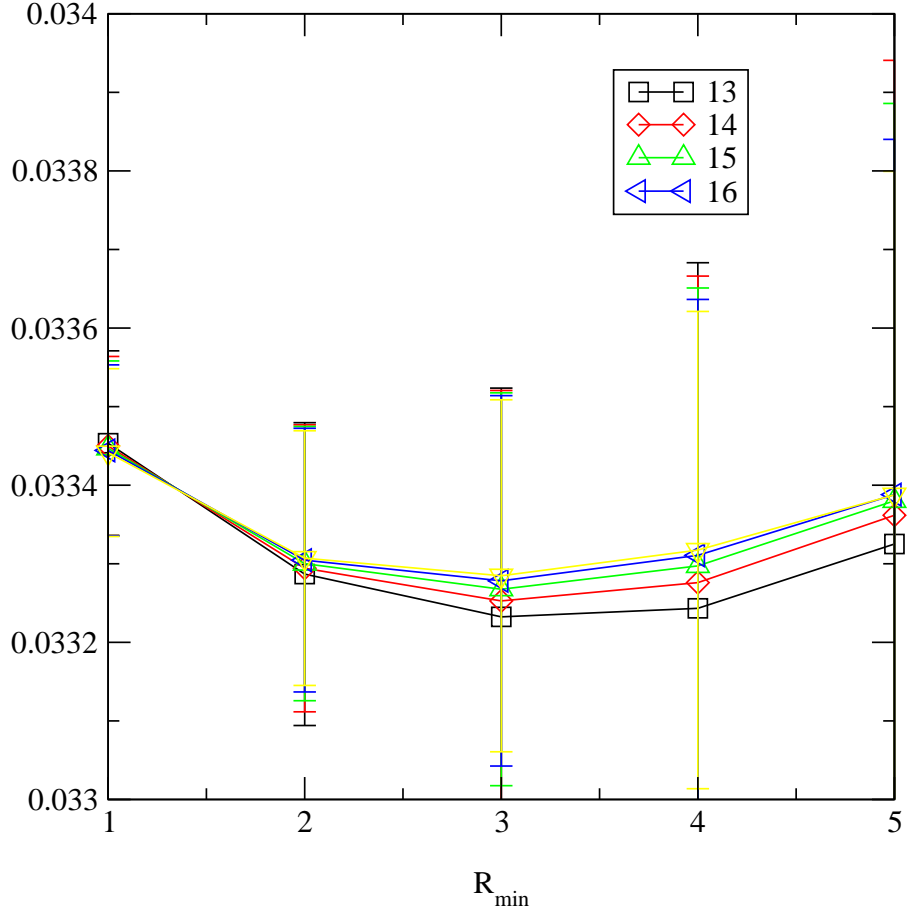}
\end{center}
\caption{ We plotted string tension for 4 different upper cutoff values
against lower cutoff value}%
\label{fstability}%
\end{figure}

A typical form of the potential is presented in Fig.\ref{fpotential}. A fit
using the linear potential (\ref{vfit}) is sufficient to obtain a stable value
of $\sigma$ within errors. To prove that, we produced for every $\beta_{3}$ a
fit from $R_{min}$ to $R_{max}$. The $R_{max}$ values, depending upon the
value of $\beta_{3}$, varied from $10$ to $14$, the $R_{min}$ varied from 1 to
4. To judge about stability of the fit we plotted the resulting value of
$\sigma$ with errors as the function of $R_{min},$ $R_{max}.$ As it may be
seen from Fig.\ref{fstability} the fit is indeed stable. This is in good agreement 
with fluctuating flux tube model \cite{Caselle:mt} - the Luescher $1/R$ term 
is not expected to be present above the phase transition.

The $\chi^{2}$ value in the 3D case was of the order of one and
about ten times less in the 2D case, likely due to the large correlation
between points.

\begin{figure}[t]
\begin{center}
\psfrag{ylabel}[c][c][1][0]{$\sqrt{{\sigma}/{Tg_{3}^{2} }}$}
\psfrag {xlabel}[c][c][1][0]{${g_{3}^{2}}/{T}$} \includegraphics
[width=12cm]{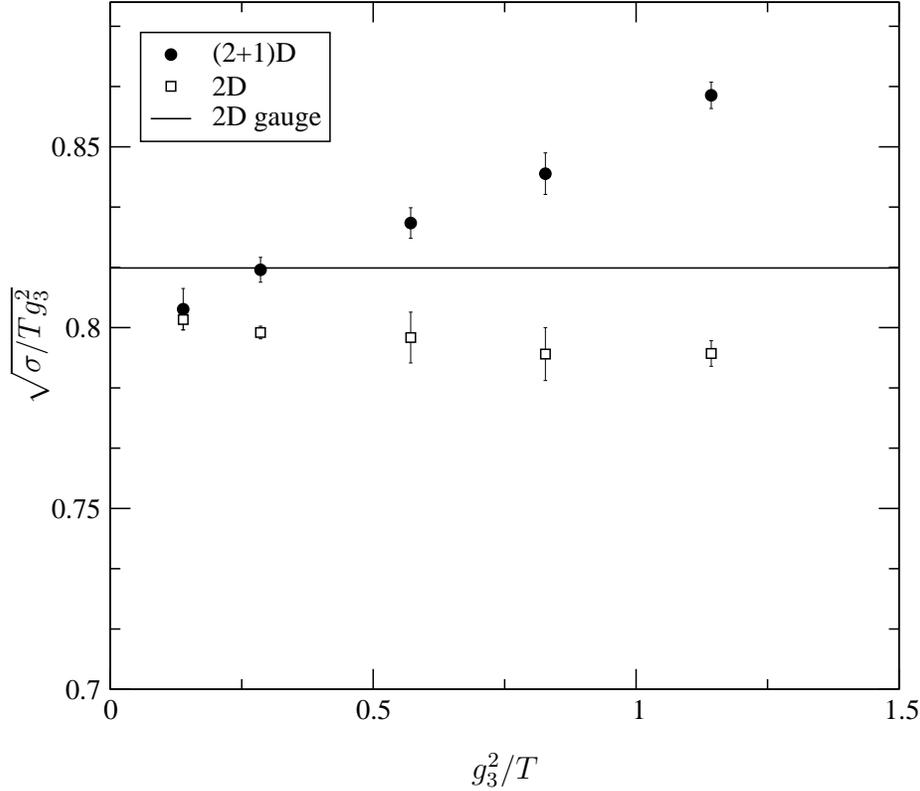}
\end{center}
\caption{The square root of the physical string tension in units of
$g_{3}\sqrt{T}$ as a function of $g_{3}^{2}/T$ in (2+1)D (filled circles) and
2D (squares). The line denotes the scaling limit $\sqrt{2/3}$ in 2D pure gauge
theory (\ref{sigma02}). }%
\label{ftension}%
\end{figure}For the dimensionally reduced treatment the case of the spatial
Wilson loop $W$ in 3D is more difficult than the Polyakov loop correlations.
It is not a static operator, and thus by no means is predicted to be
reproduced well in the dimensionally reduced model in the high temperature
region. However, there are the following considerations:

\begin{itemize}
\item[i)] a non-zero spatial string tension $\sigma_{2+1}$ is known to exist
\cite{lego} above the deconfinement temperature

\item[ii)] the pure gauge 2D theory is confining and produces a finite string
tension $\sigma_{2}^{0}$ \cite{gross}

\item[iii)] in two dimensions we expect confinement and thus a finite string
tension $\sigma_{2}^{\phi}$ to survive when the Higgs field $\phi$ is turned on.
\end{itemize}

It is thus interesting to compare these three quantities as a function of the
temperature of the $(2+1)D$ model. For this comparison we take $a^{2}%
\sigma_{2+1}$ and $a^{2}\sigma_{2}^{\phi}$ from the analysis of $W$ data
provided by \cite{lego} and by our simulation, whereas $a^{2}\sigma_{2}^{0}$
is computed analytically in Appendix 2:
\[
a^{2}\sigma_{2}^{0}=\frac{4}{\beta_{2}}+\frac{7}{\beta_{2}^{2}}+\mathcal{O}%
(\beta_{2}^{-3}).
\]
The next terms in this expansion are easy to derive using the computer algebra
system. However, they are not required for our purpose. This expression can be
equivalently rewritten
\[
\frac{\sigma_{2}^{0}}{g_{3}^{2}T}=\frac{2}{3}+\frac{7}{36}\frac{g_{3}^{2}}%
{T}(aT)^{2}+\mathcal{O}\left[  (\frac{g_{3}^{2}}{T})^{2}(aT)^{4}\right]  ,
\]
showing that up to scaling violations of order
\begin{equation}
(aT)^{2}=\frac{1}{L_{0}^{2}}%
\end{equation}
$\sigma_{2}^{0}$ scales as $2g_{3}^{2}T/3$ at finite $T$ . The quantities
reported in Fig.\ref{ftension} versus $g_{3}^{2}/T$ are the values of
$\sqrt{\sigma}$ in units of $g_{3}\sqrt{T}$, which we thus compare to
$\sqrt{2/3}$. The numerical values of ${\sigma_{2}^{\phi}/(g_{3}^{2}T)}$ and
$\sigma_{2+1}/(g_{3}^{2}T)$ were obtained from the simulations for $L_{0}=4.$

All three string tensions are quite close to each other; they differ by at
most $10\%$ within our temperature range. From the two-dimensional point of
view, we see that introducing the $\phi$ field modifies the picture quite
weakly: no sizable $g_{3}^{2}/T$ dependence around a value close to
$\sqrt{2/3}$. The behavior for $\sigma_{2+1}$ is different, with a sizable
slope in $g_{3}^{2}/T$. The difference observed between $\sigma_{2+1}$ and
$\sigma_{2}^{\phi}$ may be understood as a consequence of $W$ not being a
static operator: the average of its non-static modes with the $(2+1)D$ weight
are missing in its calculation with the effective action. However, scaling
violations may also contribute differently to these two $\sigma$'s, so that
definite conclusions require complementary simulations.

Another possible reason for this disagreement are the heavy space correlations
in the Wilson loops measurements, which may lead to systematic errors for the
$\sigma$ measurements. Our data, presented in Fig.\ref{ftension}, do not lead
to the definite conclusion whether or not the infinite $T$ limit of the string
tension in the two- and three-dimensional models is the same, and if it is
equal exactly to $2/3$.

\newpage

\chapter{Reduced Model Per Se}

\label{chap:PerSe}

\section{ Phase structure of the model}

\label{sec:phase}

As we have seen in the previous chapter, the reduced model reproduces the
$(2+1)D$ model very well; it is interesting thus to study the properties of
the reduced model as such. As we pointed out in Chapter 2, the wanted
deconfinement phase transition is absent in this model; one can see it either
from the fact that screening masses do not vanish, or via direct test. The
reason for that is the explicitly broken $Z(3)$ symmetry in the perturbative
approach. One of the remaining symmetries is the $R_{\tau}$ (see \ref{rtau}) -
the imaginary time reversal symmetry of the original model. As it was pointed
out in the studies of the $(3+1)D$ case, there exist two phases, symmetric and
broken with respect to this symmetry. The same scenario happens also in our
model. The test runs on the small lattices have shown the possibility of the
very different dynamics of the system. Indeed, we were able to see a strong
first order phase transition happening. The order parameter for this
transition is
\begin{equation}
O_{1}=\left\langle trA^{3}\right\rangle
\end{equation}
It is equal to zero in all the simulations discussed previously and is
different from zero in the broken phase.

We performed extensive simulations for lattice sizes $8\times8$ and $6\times6$
with parameter $L_{0}=4.$ To see how the transition point moves with scaling
we also simulated on $8\times8$, $10\times10$ and $12\times12$ lattices with
$L_{0}=8$. Couplings $\beta_{2}$ and $h_{4}$ were fixed at three--dimensional
values corresponding to $\beta_{3}=29$ ($\beta_{2}=116,$ $h_{4}=0.954930$) and
$\beta_{3}=58$ ($\beta_{2}=464,$ $h_{4}=3.819718$) respectively. The operators
of interest this time were $TrA^{2}$ and $TrA^{3}$. To map the transition
point we performed a large number of measurements, which varied from
$3.000.000$ to $30.000.000$ depending upon coupling. The typical
autocorrelation time was of the order of several thousands due to closeness to
the phase transition.

\begin{figure}[h]
\begin{center}
\includegraphics[clip,width=12cm,height=12cm,angle=-90]{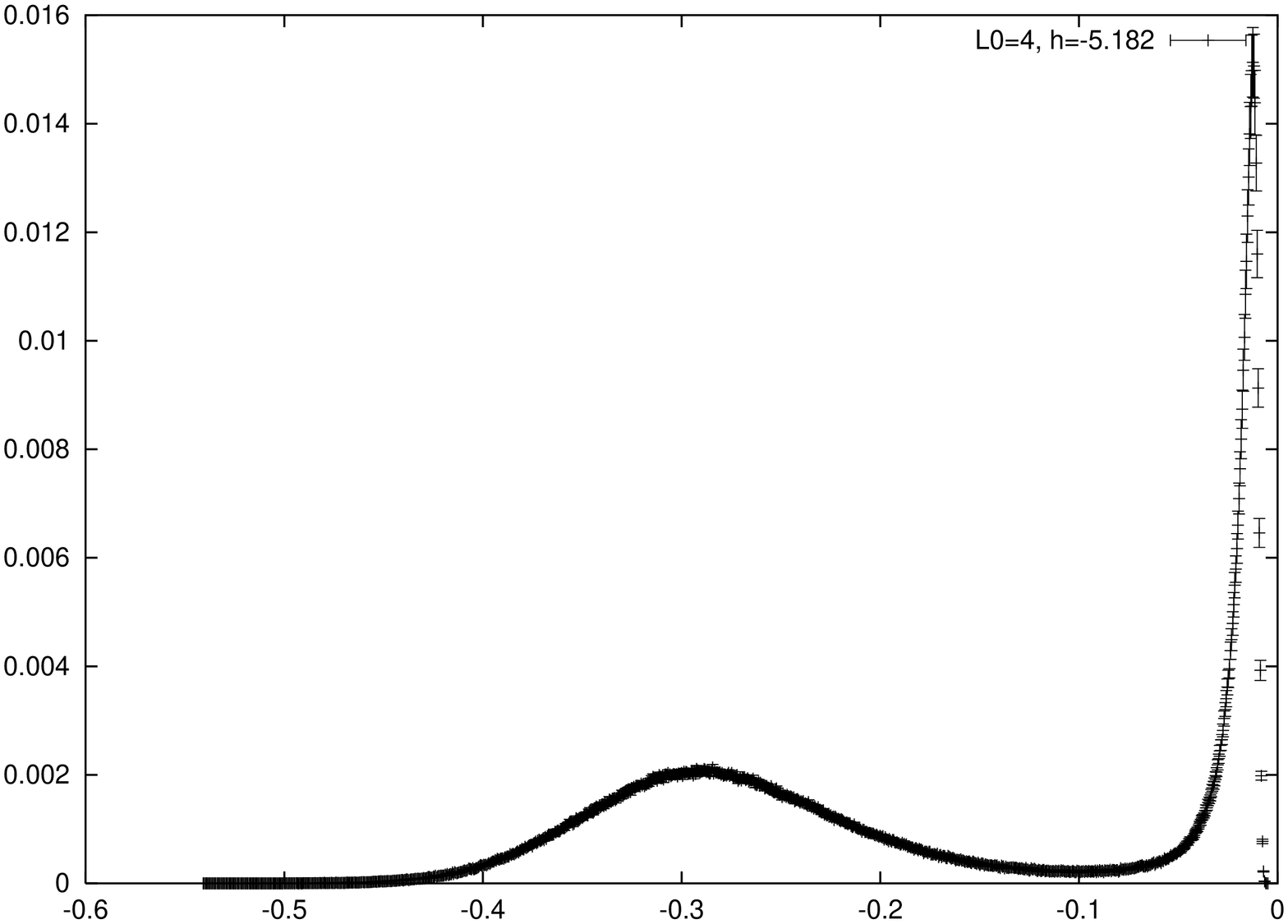}
\end{center}
\caption{Histogram, $Tr A^{2}$,$6\times6$ lattice, $L_{0}=4$}%
\label{A24}%
\end{figure}

\begin{figure}[h]
\begin{center}
\includegraphics[clip,width=12cm,height=12cm,angle=-90]{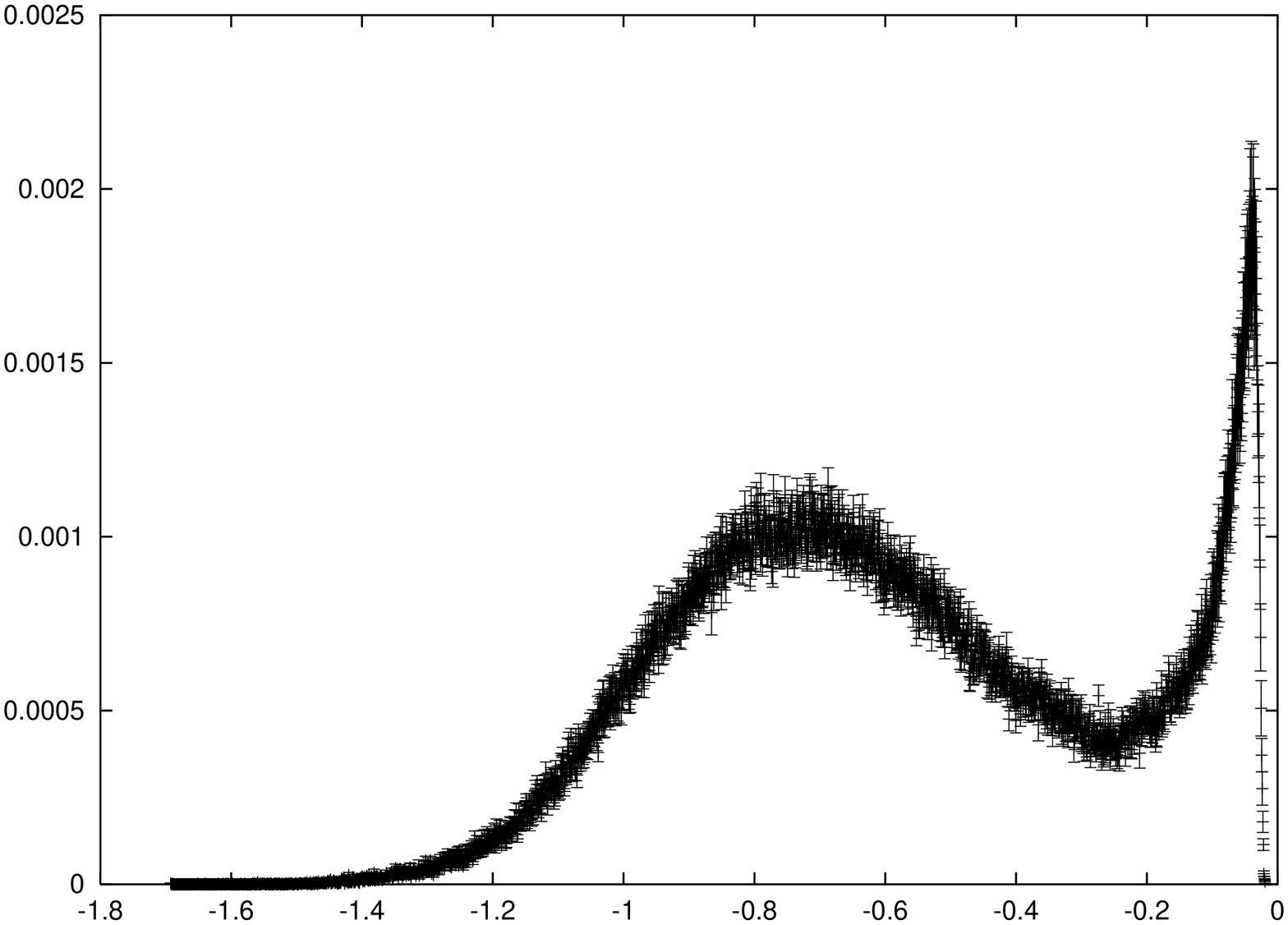}
\end{center}
\caption{Histogram, $TrA^{2}$, $8\times8$ lattice, $L_{0}=8$}%
\label{A28}%
\end{figure}

We used the histogram technique to find out the position of the phase
transition. Unlike in the previous chapter, the data were not blocked to
increase the statistics for histogramming. Typical histograms for $TrA^{2}$
are presented on Fig.\ref{A24} and Fig.\ref{A28} for $L_{0}$=4 and 8
respectively. They have a clear two-peak structure, which is the traditional
picture for the phase transition of the first order. Behavior of the system
was significantly different in these cases: for $L_{0}=4$ already at
$6\times6$ lattice a very sharp two--peak structure was present, and two
phases are separated clearly. For $L_{0}=8$ the transition is much softer, the
relative height of the peaks is smaller compared to the dip. This is
compatible with the fact that we decreased the lattice spacing twice when
increasing $L_{0}$.

\begin{figure}[h]
\begin{center}
\includegraphics[clip,width=12cm,height=12cm,angle=-90]{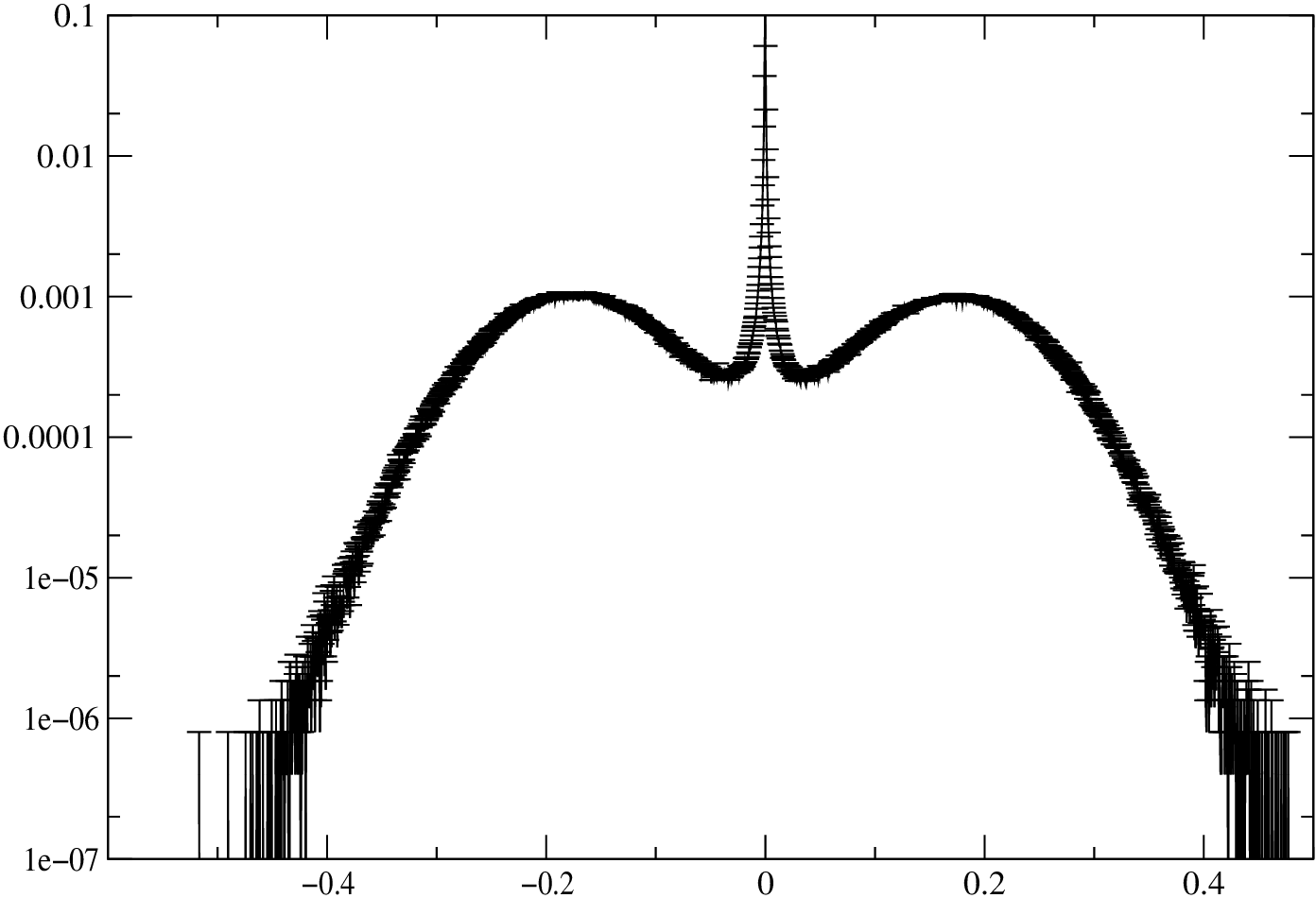}
\end{center}
\caption{Histogram, $TrA^{3}$, $8\times8$ lattice, $L_{0}=4$}%
\label{A34}%
\end{figure}

A typical histogram for $O_{1}$ is presented on Fig.\ref{A34} - a clear
three-peak structure is present; the scaling behavior also turned out to be
similar. This histogram shows that in fact there are not two but three
different phases, the symmetric one and two broken phases with positive and
negative value of the order parameter. For the data analysis we used the Ferrenberg-Swendsen multiple
histogram reweighting algorithm. It allows to use the data from several
histograms, effectively accumulating statistics for a more precise
determination of the critical point. Details of the algorithm may be found in
\cite{swendsen}. The single histogram reweighting method was used for the cross-checking.

There are several criteria which one may use in case of the first order phase
transition, among them the equal weights criteria and the susceptibility
criteria. To employ the equal weights criteria we need to take into account
that there are three phases. Thus the criteria should be formulated as such
that broken phases weight is $2/3$. After scanning the $h_{2}$ space, we
spotted the approximate point where transition happens and performed about a
dozen of long runs in that area. Then we used the Ferrenberg-Swendsen
reweighting to produce the table of the pairs $(h_{2},W_{b})$ where $W_{b}$ is
the weight of the broken phase (i.e. number of measured values which are on
the left side of the minima, divided by the overall number of measurements)
with a very small step in $h_{2}.$ Afterwards the $h_{2}$ corresponding to the
closest to $2/3$ value of $W_{b}$ was selected to be an estimator for the
critical point. This estimator was used in jack-knife like treatment to obtain
the errors on the critical value of $h_{2}.$

\begin{figure}[ptb]
\begin{center}
\includegraphics[clip,width=12cm]{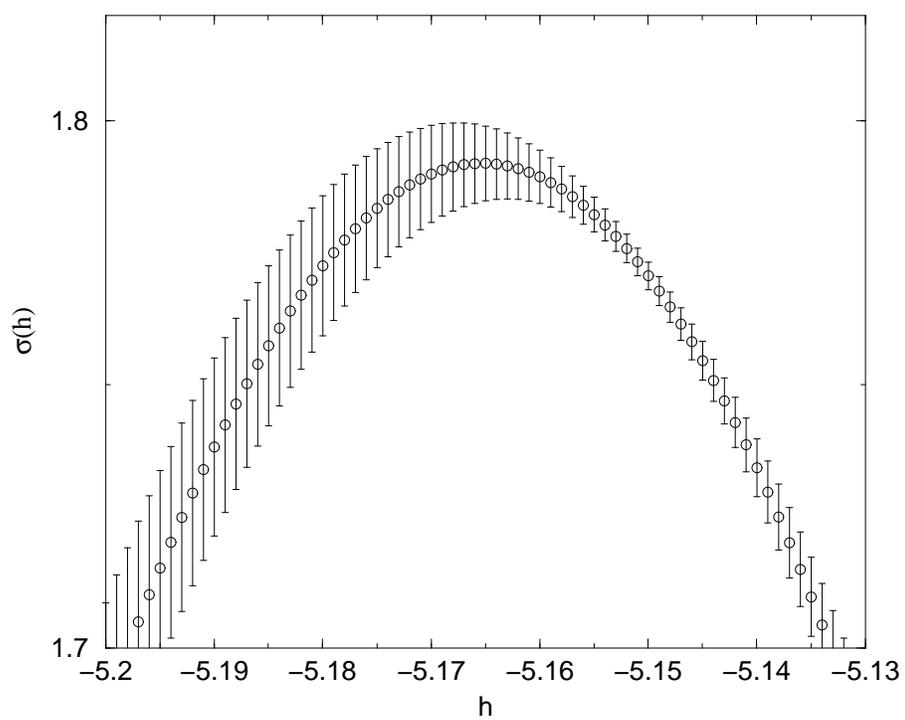}
\end{center}
\caption{Susceptibility maximum criteria, $L_{0}=4$}%
\label{suscmax}%
\end{figure}

A typical picture for the susceptibility is presented on Fig.\ref{suscmax}
where susceptibility is plotted versus $h_{2}$ for $L_{0}=4$. To estimate the
$h_{critical},$ we find the absolute maximum for every jack-knife bin,
ignoring the actual errors on susceptibility. As it can be seen from table
\ref{tabL4} below, the susceptibility maximum criteria estimate gets closer to
the equal weights criteria one when increasing volume and, moreover, coincides
within errors for the bigger value of the parameter $L_{0}$.

We also provide in the table the reduced values (labeled 3D) and two
extrapolated values, an \textquotedblleft infinite volume\textquotedblright%
\ value and the $32\times32$ value, which we need for making the conclusion
about where we are on the actual lattice we work on. The scaling behavior of
the following type was assumed
\begin{equation}
h_{V}=h_{\inf}+\frac{C_{1}}{V}+O\left(  \frac{1}{V^{2}}\right)
\end{equation}
To check the validity of this ansatz we compared the predicted from the two
points value for the third point with the actual value, and they agree within errors.

\begin{figure}[ptb]
\begin{center}
\includegraphics[clip,width=12cm]{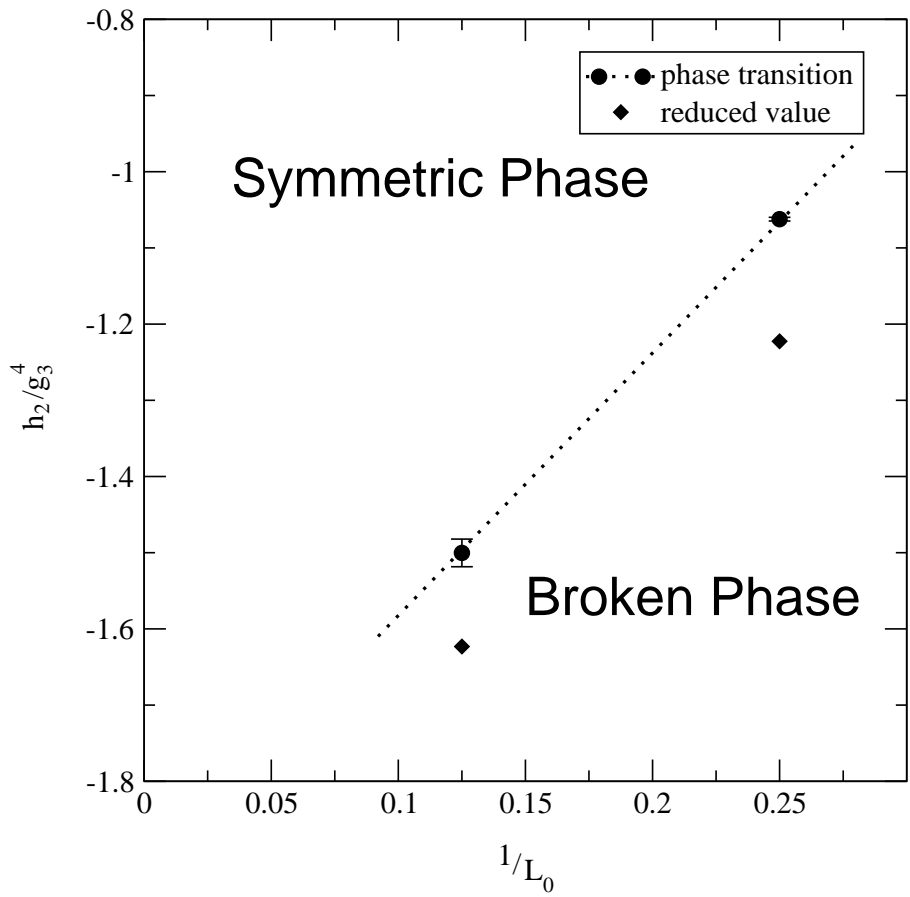}
\end{center}
\caption{Phase Diagram of the Model}%
\label{phase}%
\end{figure}

The phase diagram is sketched on Fig.\ref{phase}, the dotted line is there to
guide the eye. The reduced values are obviously in the broken phase, and
situation does not get any better with increasing $L_{0}.$ The Polyakov loops
and the string tension are not reproduced by the simulations in that phase.
However, due to it being a very strong first order phase transition - it never
(on a scale of several gigaflops-monthes) happens on the $32\times32$ lattice
which we used in the previous chapter. Starting from a weak field
configuration, we are in the symmetric phase and always stay in it.

\begin{table}[ptbptb]%
\begin{tabular}
[c]{|l|l|l|l|l|}\hline
\label{tabL4} & $6\times6$ & $8\times8$ & $\infty$ & $3D$\\\hline
equal weights & $-5.186(2)$ & $-5.225(4)$ & $-5.275(12)$ & $-6.071$\\\hline
susceptibility & $-5.165(3)$ & $-5.208(4)$ & $-5.263(13)$ & $-6.071$\\\hline
\end{tabular}
\par%
\begin{tabular}
[c]{|l|l|l|l|l|l|}\hline
\label{tabL8} & $8\times8$ & $10\times10$ & $12\times12$ & $\infty$ &
$3D$\\\hline
equal weights & $-7.28(2)$ & $-7.33(2)$ & $-7.37(2)$ & $-7.45(9)$ &
$-8.06$\\\hline
susceptibility & $-7.28(2)$ & $-7.34(2)$ & $-7.39(4)$ & $-7.44(11)$ &
$-8.06$\\\hline
\end{tabular}
\caption{Critical couplings for $L_{0}=4$ and $L_{0}=8$}%
\label{tphase}%
\end{table}

One could perform here a more detailed analysis following Ref.
\cite{engelsFSS} as done in the determination of the critical exponents and
transition temperature in \cite{karsch2p1} along with the study of the larger
lattices. Latter would require some advanced simulation method (e.g.
multi-canonical simulations). However, as long as the transition is of a
strong first order we conclude that for the small values of $L_{0}$ our
lattices are big enough to make a statement that the reduced values are in the
broken phase.

\section{The spectrum of the model}

\label{sec:spectrum}In this section we will study the lowest states of the
spectrum of the reduced model \textit{per se}. We restrict ourselves to the
operators directly related to the static operator $L(x)$ because only they are
directly relevant for the $(2+1)D$ system.

Let us now expand the $L(x)$ in powers of $A(x)$ and study the operators
$A^{n}(x)$ and connected correlations of their traces. We define
\begin{align}
A_{n}(x)  &  \equiv TrA^{n}(x),\label{an}\\
A_{n,m}(x)  &  \equiv\left\langle A_{n}(x)\,A_{m}(0)\right\rangle
\,-\,\left\langle A_{n}\right\rangle \left\langle \,A_{m}\right\rangle
.\label{anm}\\
A_{n}  &  \equiv\left\langle A_{n}(x)\right\rangle
\end{align}
Any operator $A_{n}(x)$ is gauge invariant, even or odd under the $R_{\tau}%
$-symmetry of the $2D$ action for $n$ even or odd respectively. We will
continue working in the unbroken $R_{\tau}$ phase only, where $A_{2p+1}=0$.

Using the same action as in Chapter 2 we have performed a numerical simulation
of the two dimensional model for lattice size $L_{S}=32$ till the end of the
chapter. The $\beta_{3}$ values are 29, 42, 84 and 173, while $L_{0}=4$. They
correspond to the values of $T/T_{c}$ equal approximately to 1.97, 2.85, 5.70
and 11.73 respectively. We were able to extract information from the operators
and correlations corresponding to $n=2$ to 5. The cases $n=2$ and 3 for the
$3D$ reduced model were investigated in \cite{kajantie2}.

Here we will present our results for the $A_{n,m}$ correlations measured and
describe them for each temperature in terms of two states, which we will call
$S$ and $P$, respectively even and odd under $R_{\tau}.$ They appear for $n,m$
both even and both odd. Their physical masses will be denoted $M_{S}$ and
$M_{P}$.

We will use in this and the next section either the bare lattice parameters
$r$ and $\beta_{3}$ or the physical quantities $R$ and $T,$ which are related
to each other by the following equations: \newline%
\begin{equation}
RT=r/L_{0}%
\end{equation}
\begin{figure}[ptb]
\label{fAcorr}
\par
\begin{center}
\includegraphics[width=12cm]{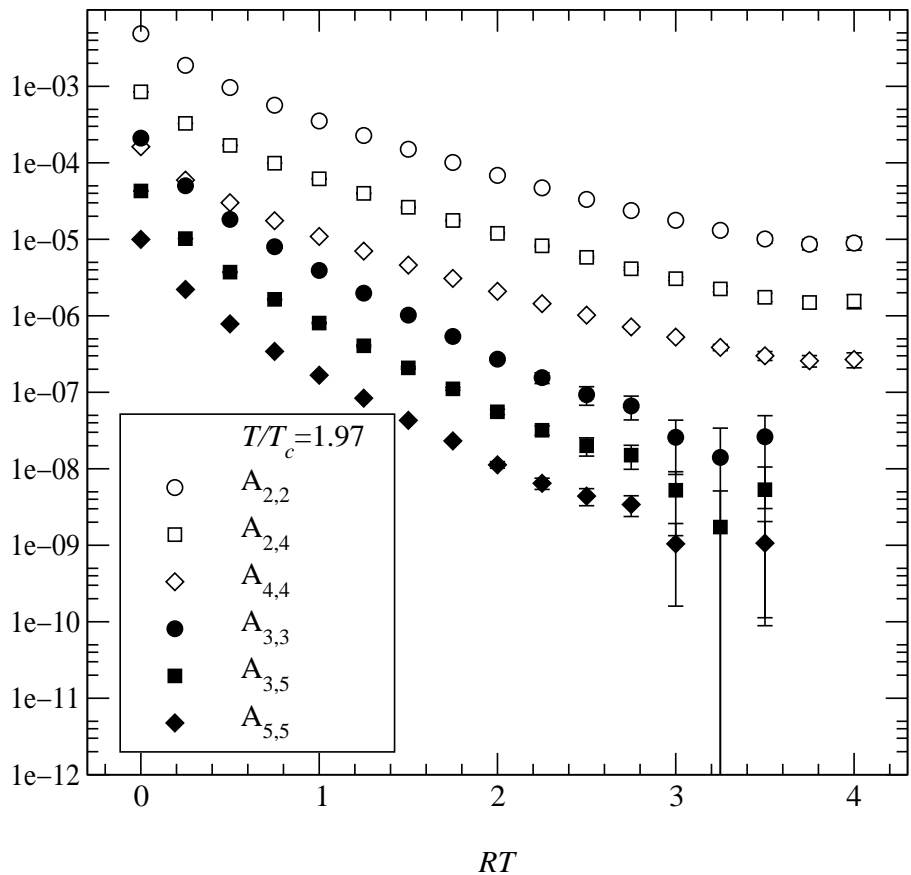}
\end{center}
\caption[1]{The on-axis correlations $A_{n,m}(r)$ at $T/T_{c}=1.97$
($\beta_{3}=29$), versus the distance in units of $1/T$. The even cases
$[n,m]=[2,2],[2,4]$ and $[4,4]$ all have the same shape, and the odd cases
$[3,3],[3,5]$ and $[5,5]$, again similar with each other in shape, are
steeper.}%
\label{fAnm}%
\end{figure}

We show in Fig.\ref{fAnm} the on-axis correlations $A_{n,m}(r),\,\,n\leq
m\in\lbrack2,5]$ at $T/T_{c}=1.97\,(\beta_{3}=29)$. They are plotted against
$RT$, that is the physical distance in units of the inverse temperature. As
expected from general considerations one can see two types of behavior, for
the even and odd channels. In the even cases, the three correlations all have
the same shape, and they decrease by about one order of magnitude each time
two more powers of $A$ are involved. The same is true for the odd cases, with
a common decay of the correlations steeper than in the former case (smaller
correlation length in lattice units). The overall situation is similar for
$\beta_{3}$ higher. For $n,m$ larger than 5, as well as for $\beta_{3}$ very
large, the signal/noise ratio becomes very small. This can be understood at
the qualitative level by noting that the rescaling Eq.(\ref{phi}) of the
$A$-fields normalizes the kinetic term for the $\phi$-fields to the standard,
parameter independent form $1/2Tr(D_{i}\phi)^{2}\,$. Hence, if the field
renormalization by the interactions is weak, the $\phi$ correlations should
depend only weakly on $\beta_{3}$, which means that $A_{n,m}$ scales like
$\beta_{3}^{-(n+m)/2}$. This will be illustrated more quantitatively in the
next subsection. Due to this scale factor, the $A$-fields remain
\textquotedblleft small\textquotedblright\ in practice down to quite low
values of $\beta_{3}$, which a posteriori explains why the perturbative
reduction may still work at a temperature as low as 1.5 $T_{c}$. In fact, we
checked that the Polyakov loop correlations are actually fully reconstructed
within errors by keeping $\{n,m\}$ up to $\{5,5\}$ only in their expansion in
$A_{n,m}$'s obtained from the small $A$ expansion of (\ref{pcorr}).

Now we want to analyze these $A_{n,m}$ data quantitatively in terms of the
lowest states of the spectrum. Let $m_{i}=a\,M_{i}$ be the lowest mass with
quantum number $i=S,P$, in lattice units. For particle $i$ with momentum $p$
we introduce a lattice propagator $\widetilde{\Delta}_{Latt}(m_{i},p)$ in
momentum space:%

\begin{align}
\widetilde{\Delta}_{Latt}^{-1}(m_{i},p)  &  =\widehat{p}^{2}+4\sinh(m_{i}%
^{2}/4),\label{proplatt}\\
\widehat{p}^{2}  &  =4\sin^{2}(p_{1}/2)+4\sin^{2}(p_{2}/2).\nonumber
\end{align}
The corresponding contribution to $A_{n,m}(r)$ then reads
\begin{equation}
A_{n,m}(m_{i},r)=g_{n,m}^{i}\frac{1}{L_{s}^{2}}\sum_{p_{1},p_{2}}\cos
(p_{1}r)\widetilde{\Delta}_{Latt}(m_{i},p), \label{onaxis}%
\end{equation}
where $g_{n,m}^{i}$ measures the residue of $A_{n,m}$ at the pole of
(\ref{proplatt}), which on large enough lattices sits at $p^{2}\sim\widehat
{p}^{2}\sim-m_{i}^{2}$. With our definitions, $g_{n,m}^{i}$ is non-zero only
for $i=S$ if $n$ and $m$ are even, and for $i=P$ if $n$ and $m$ are odd.

There is a simple result which one may get directly from Eq.(\ref{onaxis}).
For the case of the single-particle propagation between two spinless states
the residue $g_{n,m}^{i}$ factorizes (see, for example, \cite{martin}):%
\begin{equation}
g_{n,m}^{i}=\gamma_{n}^{i}\gamma_{m}^{i}%
\end{equation}
This property we can probe directly on the correlations since, as $r$ becomes
large, it implies
\begin{equation}
X_{n}\equiv\frac{A_{n,n}(r)\,A_{n+2,n+2}(r)}{A_{n,n+2}^{2}(r)}\rightarrow1.
\label{factor}%
\end{equation}
We demonstrate this fact for the temperatures $T/T_{c}=1.97$ and 5.7
($\beta_{3}=29$ and 84) on Fig.\ref{fx2_29}-\ref{fx3_84}.The displayed
quantities $X_{2}$ and $X_{3}$ (symbols $\diamond$) indeed approach one for
the large distances in all cases. However, the quality of the data is poorer
for $X_{3} $ due to the correlations involving $A_{5}$ getting very small.
Similar results are obtained for other values of $T/T_{c}$. We thus conclude
at this point that a single particle propagation accounts very well for the
largest correlation length occurring in each of the two channels. Most of the
observed deviations of $X_{n}$ from one will be interpreted in the next
section in terms of the two particle state contributions (symbols $\circ$ in
the same figures).

We now proceed to assign values to the two lowest masses $M_{S}$ and $M_{P}$
expected from the above findings. This we do by various ways in order to
further enforce the statement that the correlations do have the
characteristics associated with the pole structure of Eq.(\ref{proplatt}).
Down to $r\sim1$, an excellent approximation to the on-axis correlation
(\ref{onaxis}) is given by
\begin{equation}
A_{n,m}(m_{i},r)\simeq c\,\left(  \frac{1}{[m_{i}r]^{1/2}}e^{-m_{i}\,r}%
+\frac{1}{[m_{i}(L_{s}-r)]^{1/2}}e^{-m_{i}(L_{s}-r)}\right)  , \label{poleA}%
\end{equation}
where $c$ is constant in $r$. The second term takes into account the finite
size of the lattice.

We performed fits of this formula to all our $A_{n,m}(r)$ data taken at
$r>r_{min}$. These fits are stable with respect to $r_{min},$ provided it is
larger than about 4, and the values found for $m_{i}$ in different
correlations are always consistent with each other. The smallest errors were
obtained by using fits to $A_{2,2}$ and $A_{3,3}$ for the reasons described above.

An alternative way for the extracting the effective masses without any fitting
is the use of the $0-$momentum correlations. They defined for a generic
$x-$space correlation $C(x_{1},x_{2})$ by
\begin{equation}
C^{0}(r)=\frac{1}{L_{s}}\sum_{x_{2}}\,C(r,x_{2}). \label{proj}%
\end{equation}
If the lowest mass in $C$ is $m$, the ansatz (\ref{proplatt}) gives
\[
C^{0}(r)\propto\cosh\left(  m(L_{s}/2-r)\right)  ,
\]
and $m$ can be extracted at any $r$ by inverting this relation:
\begin{align}
m  &  =\log\left(  Y(r)+\sqrt{Y^{2}(r)-1}\right)  ,\label{meff}\\
Y(r)  &  =\frac{C^{0}(r+1)+C^{0}(r-1)}{2C^{0}(r)}.\nonumber
\end{align}
As an overall consistency check, we have extracted an effective mass
$m^{eff}(r)$ from $0-$momentum Polyakov loops correlations (\ref{pcorr}) and
compared it to the $m_{S}$ values obtained by our fits to $A_{2,2}$. We find
that $m^{eff}(r)$ is indeed nearly constant, in fact slowly decreasing towards
a value compatible with $m_{S}$, due to smaller and steeper contributions to
(\ref{pcorr}) of the heavier particle $P$.

We also investigated the D'Hoker's perturbative scenario, as we have done in
Section \ref{subsec:pscreening}. Again the largest correlation length in
$A_{n,n}$ should be $n$ times shorter than the \textquotedblleft Debye
screening length\textquotedblright, the inverse of a mass $m_{E}$ associated
with \textquotedblleft electric\textquotedblright\ gluons of the initial
(2+1)D model (the scalars of the reduced model). If such was the case, the
on-axis correlations $A_{n,n}$ should rather look like%
\begin{equation}
A_{n,n}(n\,m_{E},r)\propto\left(  \frac{1}{[m_{E}r]^{1/2}}e^{-m_{E}\,r}%
+\frac{1}{[m_{E}(L_{s}-r)]^{1/2}}e^{-m_{E}(L_{s}-r)}\right)  ^{n},
\label{cutA}%
\end{equation}
which differs in shape from (\ref{poleA}), as was illustrated in the previous
chapter for the Polyakov loop correlations. We, nevertheless, tried fits with
(\ref{cutA}), but got a definitely worse agreement in the range of
temperatures, which we have investigated, i.e. up to $12T_{c}$. Hence, this
scenario is ruled out by the data in this temperature range; if a mass can be
defined for the electric gluon in high temperature $QCD_{3},$ it is most
probably larger than both $m_{S}$/2 and $m_{P}$/3. In the \textquotedblright
constituent gluon\textquotedblright\ picture, as advocated in Ref.
\cite{buchmuller}, one would have bound states instead of a cut. One would,
however, expect $m_{P}/m_{S}\approx3/2$.

Our final results for the $S$ and $P$ masses in units of the scale
$\sqrt{g_{3}^{2}T}$ are collected in Table \ref{tbl:msmp} for the values of
$T/T_{c} $ investigated. They are taken from fits to $A_{2,2}$ and $A_{3,3}$
respectively. The values for $M_{S}$ agree with those obtained before from the
Polyakov loop correlations. As can be seen from the tables, the ratios
$M_{P}/M_{S}$ vary with $T/T_{c}$. There is, however, no clear tendency in the
region we have investigated, the ratios being 1.8, 2.0, 1.7, 1.6 in order of
increasing temperature. This ratio may, of course, go to 1.5 at still higher
temperatures. \begin{table}[ptb]
\begin{center}
\begin{tabular}{||r||l|l||l|l||}
\hline\hline
$T/T_c$ & $M_S/\sqrt{g_3^2T}$ & $M_P/\sqrt{g_3^2T}$ \\\hline\hline
1.97 & 1.39(3) & 2.51(13) \\\hline
2.85 & 1.48(5) & 3.02(16) \\\hline
5.70 & 1.89(5) & 3.22(15) \\\hline
11.73& 2.12(8) & 3.44(11) \\\hline\hline
\end{tabular}

\end{center}
\caption{Masses in units of $\sqrt{g_{3}^{2}T}$ for the $S$ and $P$ states, as
measured from fits to $A_{2,2}$ and $A_{3,3}$ respectively, for different
values of $T/T_{c}$.}%
\label{tbl:msmp}%
\end{table}

\section{Weak \textquotedblleft Strong\textquotedblright\ Interactions Between
Colorless States}

\label{sec:weak}Here we will show that even at quite short distances ($r$
small compared to $m_{S}^{-1}$) all the condensates $A_{n}\equiv\left\langle
A_{n}(x)\right\rangle $ and correlations $A_{n,m}(r)$ can be reconstructed to
a good accuracy from the data for $A_{2}$, $A_{2,2}$ and $A_{3,3}$. The
assumption is that the elementary fields $A^{\alpha}(x)$ (Greek superscripts
are color indices) interact only through $S$ and $P$ exchanges between the
non-interacting composite $A_{2}(x)$ and $A_{3}(x)$, the scale of the fields
being fixed by the size of $A_{2}$, while $A_{3}=0$. The details of the
derivations are presented in Appendix 3, while here we limit ourselves to the
simplest applications and give the results, starting with the local condensates.

\subsection{Weak Residual Interactions: The $A$-fields condensates}

\label{subsec:condens}Since $SU(3)$ has rank 2, any $A_{n}(x)$ can be reduced
to a polynomial in $A_{2}(x)$ and $A_{3}(x)$. For $n$ odd $A_{n}$ is zero by
$R_{\tau}$ symmetry. For $n$ even, we apply Wick contraction to all pairs of
$A^{\alpha}$ elementary fields, followed by the meanfield-like substitution
\begin{equation}
A^{\alpha}(x)\,A^{\beta}(x)\,\,\rightarrow\,\,\frac{1}{4}\delta_{\alpha,\beta
}\,A_{2}(x). \label{subst}%
\end{equation}
As an illustration consider $A_{4}$. With the definitions of Section
\ref{sec:spectrum}, we have
\begin{equation}
2A_{4}(x)=A_{2}^{2}(x)=\frac{1}{2^{2}}\,\sum_{\alpha,\beta=1}^{8}A^{\alpha
}(x)A^{\alpha}(x)A^{\beta}(x)A^{\beta}(x). \label{suma4}%
\end{equation}
There we apply (\ref{subst}) and then replace $A_{2}(x)$ by its average
$A_{2}$. The $A^{\alpha}A^{\alpha}$ and $A^{\beta}A^{\beta}$ contractions give
$(8\times A_{2}/4)^{2}$, and the additional contributions from $\alpha=\beta$
give $2\times8(A_{2}/4)^{2}$. Noting that
\begin{equation}
A_{2,2}(0)=\left\langle A_{2}^{2}(x)\right\rangle -A_{2}^{2}%
\end{equation}
$\,$ (see (\ref{anm})), the net result can be put into two equivalent forms
\begin{align}
2A_{4}  &  =\frac{5}{4}\,A_{2}^{2},\label{a4}\\
A_{2,2}(0)  &  =\frac{1}{4}\,A_{2}^{2}. \label{a22}%
\end{align}
This prediction is remarkably well verified in all cases. At $\beta_{3}=29$,
the left and right hand sides of (\ref{a22}) are respectively
$4.86(1)\,10^{-3}$ and $4.825(10)\,10^{-3}$. They are $6.093(13)\,10^{-4}$ and
$6.069(1)\,10^{-4}$ at $\beta_{3}=84$. The formula for the $A_{3,3}(0)$
correlations
\begin{equation}
A_{3,3}(0)=\frac{5}{64}A_{2}^{3}. \label{a33}%
\end{equation}
leads to the similar correspondence. In this case the left and right hand
sides are measured to be $2.105(7)\,10^{-4}$ and $2.095(6)\,10^{-4}$ for
$\beta_{3}=29$, $9.365(30)\,10^{-6}$ and $9.344(3)\,10^{-6}$ for $\beta
_{3}=84$. Hence, the effects of residual interactions via higher orders
effective couplings in $A_{2}(x)$ and $A_{3}(x)$ are less than the percent in
the correlations at zero distance.

\begin{figure}[ptb]
\begin{center}
\includegraphics[width=12cm,angle=-90]{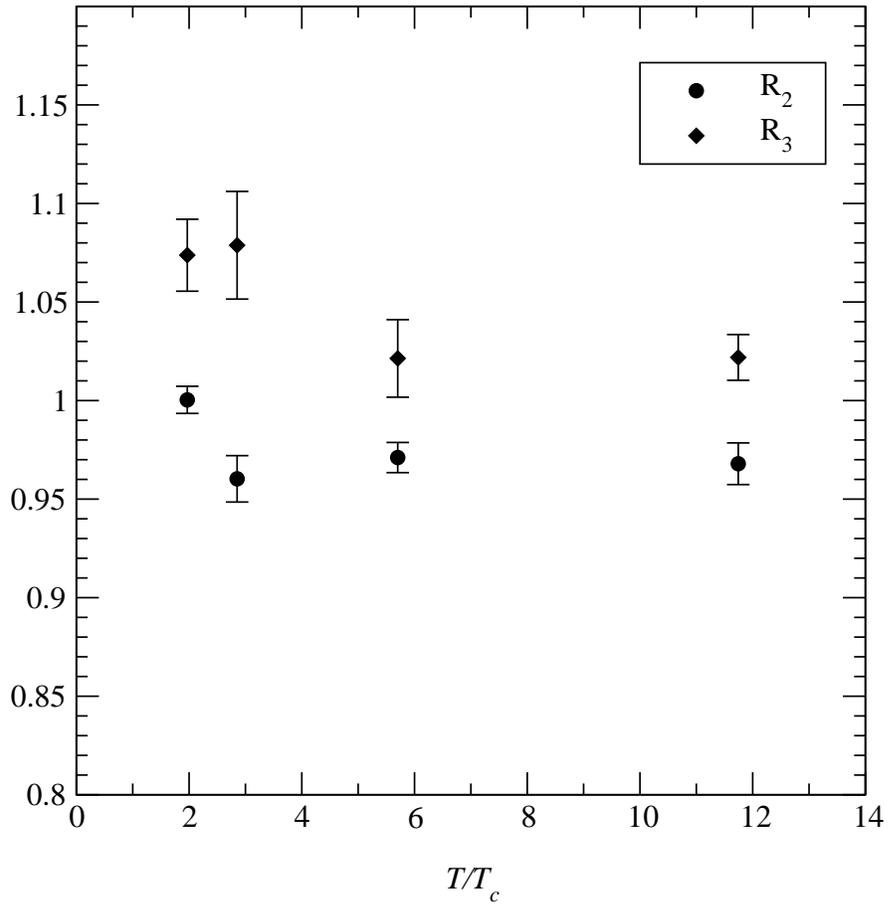}
\end{center}
\caption[6]{The residues $R_{i}$ defined by Eq. (\ref{residue}) stay close to
one in the whole temperature range.}%
\label{fresidue}%
\end{figure}

Before going to the correlations at non-zero distance, let us discuss their
normalization, as measured by the values of $A_{2,2}(0)$ and $A_{3,3}(0)$
described just above. At the beginning of the previous section, we argued that
the behavior in $\beta_{3},\,n,\,m$ observed for $A_{n,m}$ could follow from
the absence of a large renormalization of the $\phi$-fields (defined by Eq.
(\ref{phi})) by the interactions . Here we note that in the confined phase the
effective degrees of freedom are the \textit{massive} composites $\phi
_{i}=Tr\phi^{i},\,i=2,\,3$, so that in the limit where they are considered as
free fields, one may write (see (\ref{proplatt}))
\begin{align}
\left\langle \phi_{i}(0)\,\phi_{i}(0)\right\rangle  &  \simeq R_{i}\int
d^{2}p\,\frac{1}{\widehat{p}^{2}+4\sinh(m_{i}^{2}/4)},\label{residue}\\
\widehat{p}^{2}  &  =4\sin^{2}(p_{1}/2)+4\sin^{2}(p_{2}/2),\nonumber
\end{align}
the residue $R_{i}$ being one if neither $\phi$ nor the composites get
renormalized. We computed $R_{i}$ as the ratio of the l.h.s. of (\ref{residue}%
), directly measured, to the integral in the r.h.s, evaluated numerically on
the lattice for the mass values fitted to the correlation data. The result is
shown in Fig.~\ref{fresidue}: in the whole temperature range, both residues in
the even and odd channels remain uniformly very close to one.

\subsection{Weak Residual Interactions: Properties of the Correlations}

\begin{figure}[ptb]
\begin{center}
\includegraphics[width=12cm]{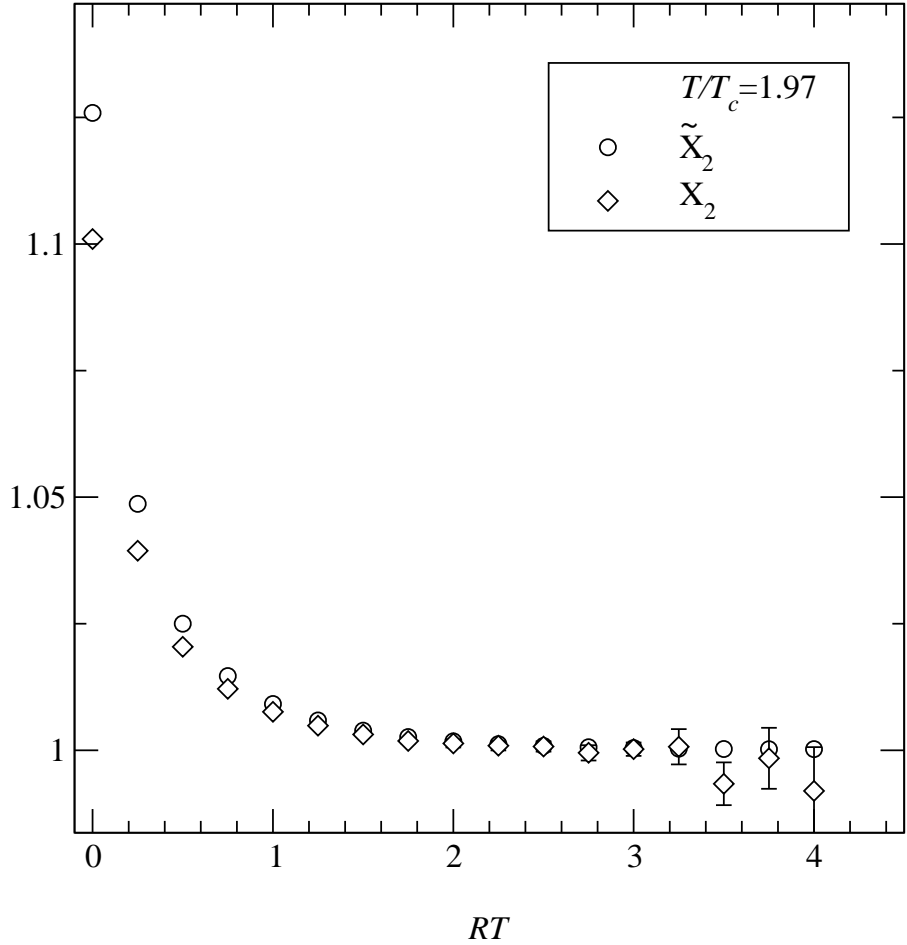}
\end{center}
\caption[2]{Residue factorization: data for the quantity $X_{2}$, Eq.
(\ref{factor}) at $T/T_{c}=1.97$ ($\beta_{3}=29$) versus the distance in units
of $1/T$. It approaches one at large distances. The quantity $\widetilde
{X}_{2}$ corresponds to our interpretation (Eq. (\ref{correct2})) of the
deviation from one of $X_{2}$ at shorter distances.}%
\label{fx2_29}%
\end{figure}

\begin{figure}[ptb]
\begin{center}
\includegraphics[width=12cm]{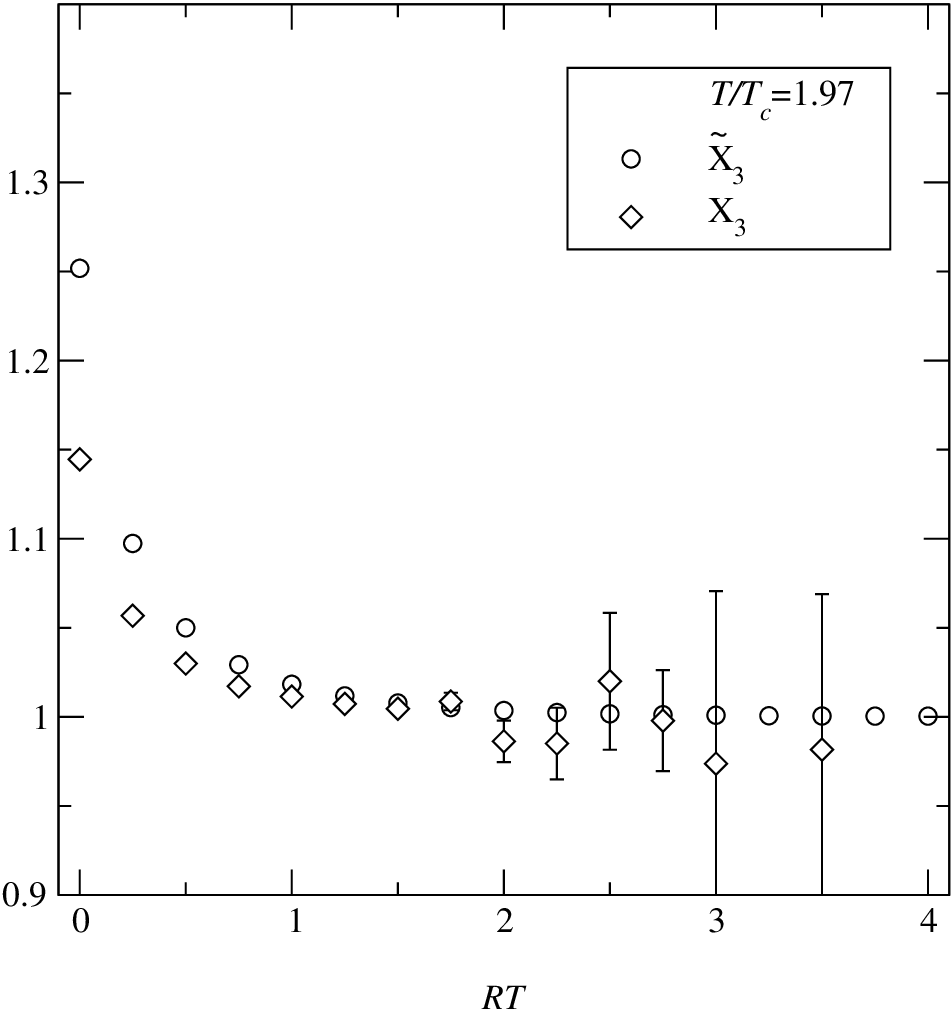}
\end{center}
\caption[3]{Residue factorization: data for the quantity $X_{3}$, Eq.
(\ref{factor}) $T/T_{c}=1.97$ ($\beta_{3}=29$) versus the distance in units of
$1/T$. It approaches one at large distances. The quantity $\widetilde{X}_{3}$
corresponds to our interpretation (Section 4, Eq. (\ref{correct3})) of the
deviation from one of $X_{3}$ at shorter distances.}%
\label{fx3_29}%
\end{figure}

\begin{figure}[ptb]
\begin{center}
\includegraphics[width=12cm]{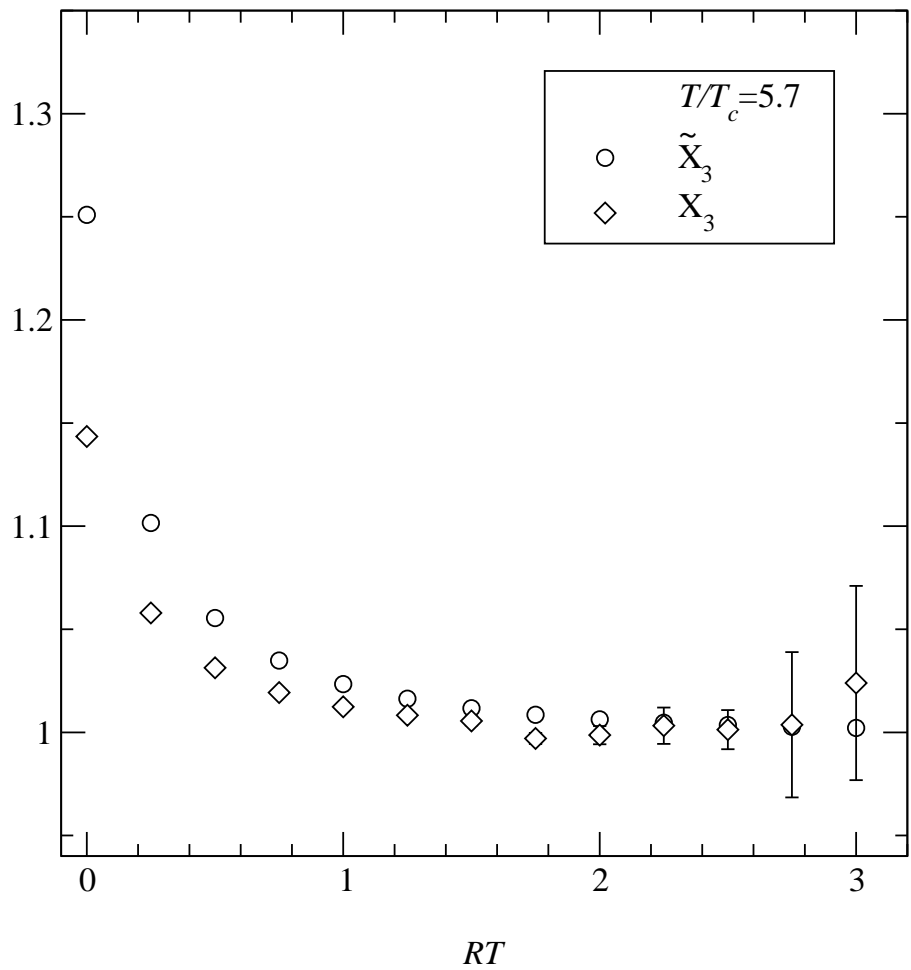}
\end{center}
\caption[5]{Same as in Fig. \ref{fx3_29} at $T/T_{c}=5.7$ ($\beta_{3}=84$).}%
\label{fx3_84}%
\end{figure}

As we have seen in section \ref{sec:spectrum} (see Fig.~\ref{fAcorr}), the
different $A_{n,m}(r)$'s corresponding to the same channel have very similar
shapes. Their analysis in terms of one particle exchange was successful,
confirmed by the agreement with residue factorization. Nevertheless, although
the quantity $X_{n}$ of Eq. (\ref{factor}) does go to one at large distances,
it is significantly different from one at medium and short distances (see
Figs.~\ref{fx2_29}-\ref{fx3_84} ). We will now show that two particle exchange
is responsible for most of this lack of factorization. 

\begin{figure}[ptb]
\begin{center}
\includegraphics[width=12cm]{paper.plot.X2.X2d.b084.eps}
\end{center}
\caption[4]{Same as in Fig. \ref{fx2_29} at $T/T_{c}=5.7$ ($\beta_{3}=84$).}%
\label{fx2_82}%
\end{figure}

\begin{figure}[ptbptb]
\begin{center}
\includegraphics[width=12cm]{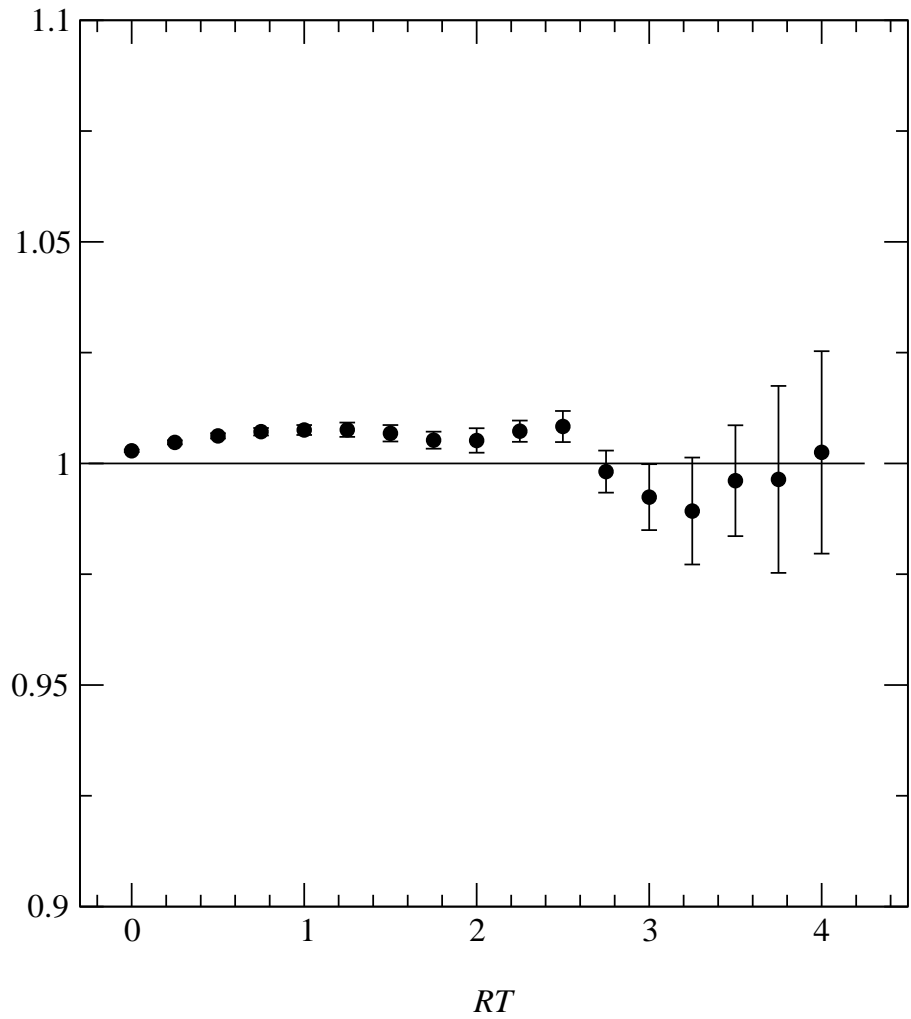}
\end{center}
\caption[7]{ Plot of ${4A_{24}}/{5A_{2}A_{22}}$ at $T/T_{c}=1.97$ ($\beta
_{3}=29$). This quantity is one if Eq. (\ref{wicka24}) exactly holds.}%
\label{fAratio}%
\end{figure}

The simplest consequence of our assumptions for correlations at finite $r$ is,
using Eq.(\ref{a4}),
\begin{equation}
A_{2,4}(r)=\frac{5}{4}\,A_{2}\,A_{2,2}(r), \label{wicka24}%
\end{equation}
which is very well verified at any distance as shown in Fig.\ref{fAratio} for
$T/T_{c}=1.97$ ($\beta_{3}~=~29.0$). A similar agreement is found for the
relation
\begin{equation}
A_{3,5}(r)=\frac{35}{24}\,A_{2}\,A_{3,3}(r), \label{wicka35}%
\end{equation}
derived in the appendix. A new situation arises when we consider $A_{4,4,}$ or
$A_{5,5}$ where both the initial and final states may couple to a two particle
state, $(S\,S)$ or $(S\,P)$ respectively. Then the intermediate state in a
connected correlation between $0$ and $r$ may consist of either one or two
particles. For example, to compute $A_{4,4,}(r)$, we apply the substitution
rule (\ref{subst}) to the sum (\ref{suma4}), and then average using the
definitions of $A_{2,2}$ and $A_{2}$. One finds
\[
4A_{4,4}(r)=\left(  \frac{5}{4}\right)  ^{2}\,\left[  4A_{2}^{2}%
\,A_{2,2}(r)\,+2\,A_{2,2}^{2}(r)\right]  .
\]
and
\[
A_{5,5}(r)=\left(  \frac{35}{24}\right)  ^{2}\,\left[  A_{2}^{2}%
\,A_{3,3}(r)\,+\,A_{2,2}(r)\,A_{3,3}(r)\right]  .
\]
These both expressions contain two parts. The first term in the square
brackets is the contribution of the single particle propagation. The second
term is a product of two propagators in space, and is that of a two-particle
intermediate state. It provides a correction to exact factorization.


From the definitions Eq.(\ref{factor}) of $X_{2}$ and $X_{3}$ one actually
gets the following estimates:
\begin{align}
X_{2}(r)  &  \simeq\widetilde{X}_{2}(r)\equiv1+\frac{A_{2,2}(r)}{2A_{2}%
^{2}(r)},\label{correct2}\\
X_{3}(r)  &  \simeq\widetilde{X}_{3}(r)\equiv1+\frac{A_{2,2}(r)}{A_{2}^{2}%
(r)}. \label{correct3}%
\end{align}
The estimates $\widetilde{X}_{2}(r),\,\widetilde{X}_{3}(r)$ are displayed in
Figs.\ref{fx2_29}-\ref{fx3_29} (resp.\ref{fx2_82}-\ref{fx3_84}), for
comparison with the measured values $X_{2}(r),\,X_{3}(r)$ at $T/T_{c}=1.97$
(resp. 5.7), i.e. $\beta_{3}=29.0$ (resp. 84.0). We see that the corrections
to factorization implied by the two-particle propagation provide a reasonable
explanation of the behavior observed for the $X$'s at intermediate and short
distances. This is especially true in the case of $X_{2}$, showing that there
is very little room for contributions from direct non-quadratic couplings in
$A_{2}(x)$ in the full effective action (that resulting from integration over
the gauge fields). This justifies our statement that the residual interactions
between the colorless boundstates of the adjoint scalars are very weak.

\newpage

\chapter{ Z(3) symmetric action}

\label{chap:Z3sym}In this chapter we describe an attempt to overcome the
absence of the deconfinement phase transition in the model. As we mentioned
before, this phenomena is not possible due to the fact that we broke the
$Z(3)$ symmetry explicitly when doing the perturbation expansion. One may try
to restore this symmetry by hand and derive an effective action possessing it.
We will use $A_{0}$ variables throughout this chapter, them being a more
natural choice for our consideration. The starting point is then:%

\[
S_{eff}^{2}\left(  U,A_{0}\right)  =S_{W}^{2}\left(  U\right)  +S_{int}\left(
U,A_{0}\right)  +S_{pot}\left(  A_{0}\right)
\]
where $S_{W}^{2}\left(  U\right)  $ is the standard Wilson action in two
dimensions as defined in Eq.(\ref{puregauge}),%
\[
S_{int}\left(  U,A_{0}\right)  =\frac{L_{0}\beta_{3}}{6}\sum_{x}\sum
_{i}Tr\left[  \left(  D_{i}A_{0}\right)  ^{2}\right]
\]%
\[
D_{i}A_{0}=U\left(  x;i\right)  A\left(  x+ai\right)  U\left(  x;i\right)
^{-1}-A\left(  x\right)
\]%
\[
S_{pot}\left(  A_{0}\right)  =\sum_{x}h_{2}TrA_{0}^{2}\left(  x\right)
+h_{4}\left(  TrA_{0}^{2}\left(  x\right)  \right)  ^{2}%
\]
with couplings
\[
h_{2}=\frac{9}{16\pi\beta_{3}}%
\]%
\[
h_{4}=\frac{9}{\pi L_{0}\beta_{3}}\mu
\]%
\[
\mu=\ln L_{0}+\frac{5}{2}\ln2-1
\]
We may now formulate the basic requirements for this action:

\begin{description}
\item[i)] should be $Z_{3}$ symmetric

\item[ii)] should have also the basic gauge symmetry

\item[iii)] should reproduce the results from the \textquotedblright
old\textquotedblright\ model at the large value of $\beta_{3}/L_{0},$ because
we need it still for describing the $(2+1)$-dimensional model
\end{description}

One would also like that such a model \textit{predicts} $Z_{3}$-symmetry
restoration (reconfinement) as $\beta_{3}/L_{0}$ (temperature) is lowered. The
Polyakov loop should be the order parameter. Close to one (up to a $Z_{3}$
phase $\exp(\pm2\pi i/3))$ at high temperature, it is zero (on the lattice it
means - becomes small) in the symmetric phase.

It is obvious that there are infinitely many ways to create the desired
action. The most reasonable for us seems the following form:%
\begin{equation}
S_{Z_{3}}=S_{W}^{2}\left(  U\right)  +S_{U,V}+S_{V} \label{new}%
\end{equation}

The plaquette action $S_{W}^{2}\left(  U\right)  $ remains untouched. The
kinetic term \vspace{7mm}$S_{U,V}$ we rewrite using the Wilson Lines
\begin{equation}
V(x)=U_{0}^{L_{0}}(x)
\end{equation}
where $U_{0}$ is the element of the $SU(3)$ group, which we place on every
site of our lattice. It relates to the $A_{0}$ field as usually via
$U_{0}=\exp(iA_{0}).$ The simplest choice for the kinetic term is%
\begin{equation}
S_{U,V}\;\sim\;\frac{\beta_{3}}{L_{0}}\;\sum_{x,\mu}\left(  1-\frac{1}{3}%
\Re\,TrU(x;\mu)V(x+a\widehat{\mu})U(x;\mu)^{-1}V^{-1}(x)\right)
\end{equation}
There are two invariants for the $SU(3)$ group, so one can choose any two
independent quantities as the basic variables for writing up the potential. We
made our choice in favour of the following quantities:%
\begin{equation}
O(x)=\frac{1}{3}TrV(x)\,\,\qquad\,O^{\ast}(x)=\frac{1}{3}TrV^{\ast}(x)
\end{equation}
that is the Polyakov loop and its conjugate. This choice of variables leads to
the following considerations

\begin{itemize}
\item gauge invariance implies that $S_{V}^{R}$ depends only on the set of all
pairs $\{O(x),O^{\ast}(x)\}$

\item by $Z_{3}$ invariance, the lowest degree \textit{local} monomial in
$O,O^{\ast}$ is $O_{2}(x)=O(x)O^{\ast}(x)$ (degree two).

\item next terms should be $O^{3}(x)$ and $O^{\ast3}(x)$ \textit{with degree
three,} there is no reason for them to be absent. They are independent degrees
of freedom

\item reality implies that at order 3 in $O$ the local action depends only on
\[
O_{3}(x)=\frac{1}{2}(O^{3}+O^{\ast3}).
\]
and terms proportional to it are expected on very general grounds in the full
effective Polyakov loop potential.
\end{itemize}

The simplest form of such potential is then%
\begin{equation}
S_{V}=\sum_{x\in\Lambda_{2}}\left(  \lambda_{2}\,|O(x)|^{2}\;+\;\lambda
_{3}\,\operatorname{Re}(O^{3}(x))\right)  \label{S_cub}%
\end{equation}
The corresponding couplings $\lambda_{2}$ and $\lambda_{3}$ have to be fixed
by again matching the small $A_{0}$ expansion of (\ref{new}) to $S_{A_{0}}$.
Although this proposal still contains a substantial amount of arbitrariness,
the following may be stated:\newline

\begin{description}
\item[i)] All desired symmetries are implemented,

\item[ii)] All independent degrees of freedom are involved,

\item[iii)] In the region of phase space where the $O$'s are close to one, the
action looks like the old one. The hope is that it is the relevant region at
\textit{large} $\beta_{3}/L_{0}$ values,

\item[iv)] In the region of phase space where the $O$'s are close to zero, the
action is truncated to the two lowest degree monomials in $O,O^{\ast}$, that
is two and three, not two and four. The hope is that it is the relevant region
at \textit{low} $\beta_{3}/L_{0}$ values.\newline
\end{description}

The other option would be to ignore the requirements ii) and iv) and use
instead of $\Re(O^{3}(x))$ the term with $O^{4}(x).$ This way it is also
possible to match all the terms of the potential (see below).

One wants to be working in the unbroken phase, and assume that $trA^{3}=0.$
The obvious choice for the kinetic term is%
\[
S_{I}=\frac{\beta_{3}}{6L_{0}}\sum(1-\operatorname{Re}Tr\left\{  U\left(
x,-\mu\right)  O\left(  x+\mu\right)  U^{\dagger}\left(  x,\mu\right)
O^{\ast}\left(  x+\mu\right)  \right\}
\]

Expanding the action (\ref{S_cub}) one easily gets:
\begin{align*}
\lambda_{2}^{c}  &  =\allowbreak\frac{9}{16}\frac{20\mu-1}{\pi L_{0}^{2}}\\
\lambda_{3}^{c}  &  =\allowbreak-\frac{3}{8}\frac{12\mu-1}{\pi L_{0}^{2}}%
\end{align*}

Here is the table for the coefficients for $\beta_{3}=29$:

\begin{center}%
\begin{tabular}
[c]{|l|l|l|}\hline
$L_{0}$ & $\lambda_{2}^{c}$ & $\lambda_{3}^{c}$\\\hline
4 & 0.4631026079 & -0.182256888\\\hline
8 & 0.1545592536 & -0.06107766263\\\hline
\end{tabular}

\end{center}

\bigskip The model described above represents a good candidate for the
improved effective model, which can allow us to calculate the
three-dimensional quantities closer to the phase transition. Our test runs
indicate that this model does have the second order phase transition in
$\beta_{2}$ with Polyakov loop being an order parameter. However, it still
requires a detailed study, which goes beyond the scope of the current work.

\newpage

\chapter{Conclusions and outlook}

This work is the detailed investigation of the dimensional reduction for SU(3)
gauge theory in (2+1)-dimensions at high temperature. We have produced an
effective reduced model for the static variables following \cite{su2,su3,qcd}.
This means that the integration over the non-static modes has been performed
perturbatively, and the effective couplings have been kept to one loop order
for the two point and four point functions. Higher order contributions,
neglecting which was a part of our approximation, are unimportant at large
distances and high temperatures.

We investigated the validity of this approximation for the correlation
function between Polyakov loops and corresponding screening masses, as well as
the spatial string tension. For the (2+1)-dimensional theory and the
corresponding 2-dimensional reduced adjoint Higgs-gauge model we were able to
obtain very precise numerical results. Therefore, we could make a detailed
comparison, including a scaling analysis, which strongly supports the
assumption that our lattices correspond to a large enough value of the
temporal extent $L_{0}$.

For the correlation between Polyakov loops the dimensional reduction works
very well down to temperatures of the order of $1.5T_{c},$ where $T_{c}$ is
the critical temperature of the deconfinement phase transition of the original
model. This is true even at distances down to or below $1/T$ . And it does
even better than in the (3+1)-dimensional case. The correlation between
Polyakov loops is well described by a simple pole in momentum space, even at
high temperatures. This situation is not manifested in (3+1)-dimensional
theory where at high temperatures the data favour the exchange of two Debye
screened gluons \cite{gao,petereisz,frules,kacz}. The disagreement may be
related to the stronger infrared divergences in (2+1) dimensions, which
invalidate the perturbatively resumed Debye screening picture. A further
investigation of this behavior in (3+1) dimensions would certainly be very interesting.

The spatial Wilson loops in (2+1) dimensions do not correspond to a static
operator and is not reproduced at all by itself. However, for the string
tension, which is extracted from Wilson loops of extent larger than $1/T$, one
can hope that the non-static corrections are small. Our comparison favours
this situation. In fact, the Higgs sector seems to have little influence on
the value of the string tension; there is a fairly good agreement also with
the analytically solvable pure two-dimensional gauge theory where confinement
is given by the two-dimensional Coulomb potential. The differences between the
string tension in the three cases considered are of the order of a few
percent. To make a definite statement if these differences are real continuum
effects, one must further study finite size and scaling corrections.

Our work shows that it may be possible to explain the non-perturbative
features of (2+1)-dimensional QCD in the deconfined phase with a relatively
simple two-dimensional model. The model possesses, however, a non-physical
phase transition associated with the symmetry of the Higgs field sign reversal
$R_{\tau},$ which corresponds to the time reversal of the original model. The
reduces values of the Higgs potential are in the non-physical broken phase.
However, we are able to simulate with these values in the metastable phase,
and never drop into the broken one on sufficiently large lattices due to this
transition being strongly first order.

\bigskip We studied the reduced model in the region of the unbroken symmetry
to identify two boundstates $S$ and $P$, respectively, even and odd under
$R_{\tau}$ and thus coupled to monomials, respectively, of degree $2n$ and
$2n+1$ in the higgs field. The $S$ signal coincides with that previously
obtained from the Polyakov loops correlations, where, however, the P-state
contributions could not be disentangled. These results came from the
measurement of three even--even and three odd--odd distinct correlations, as
functions of the on-axis lattice distance $r$. We very carefully analyzed
their shape in $r$, with the result that in all cases the signal found was
that expected from the occurrence of genuine poles in momentum space. Our data
invalidated the D'Hoker's scenario where the decay of such correlations with
the distance reflects the propagation in $3D$ of $p=2n$ or $2n+1$
\textquotedblleft electric gluons\textquotedblright. In other words, the
perturbative result that the correlation length equals to $1/p$ times the
\textquotedblleft Debye screening length\textquotedblright\ is inadequate in
the present case.

By comparing the size of the three different correlations measured for each of
the $S$ and $P$ sectors, we were able to show that residue factorization
holds, as expected on general grounds when one particle propagates between
different states. The agreement with factorization was expectedly found to be
particularly good at large distances. We were also able to demonstrate that
deviations at shorter distances are to a large extent compatible with
propagation of two particles, namely two $S$ or $S$ and $P$, respectively, in
the $S$ or $P$ channel. We thus make a conclusion that the scalar sector of
the reduced model at large distance can be described by the two weakly
interacting colorless states.

The perturbative approach used to derive the reduced model explicitly broke
the $Z(3)$ symmetry of the (2+1) dimensional system, thus we do not have the
phase transition which would correspond to the deconfinement phase transition
of the original model. We presented here a candidate for the effective model,
based on the original spatial gauge variables and Wilson Lines. This model is
likely not only to be able to reproduce the full model at high temperature but
also exhibits a deconfinement-like phase transition with Polyakov loop being
an order parameter. Careful study of this model is needed to make a definite statement.

Farther interesting topics for investigations are of course more real-world
models. The similarly detailed analysis for pure gauge QCD in (3 + 1)D would
be interesting, with extending to the full QCD with dynamic fermions.
Furthermore, the study of thermodynamical quantities like pressure and entropy
may be of interest, as well as study of the operators which are even and odd
under other symmetries of the model - space reversal and charge conjugation.
\newpage
\chapter{Acknowlegements}
At this point I would like to thank everybody, who during these years
supported me at my work and made it possible. At first, of course, my
supervisor Bengt Petersson, who with infinite patience guided me through the
process of learning, while giving me as much freedom as I could digest. I am
really happy to have you as a supervisor, Bengt.

I also would like to thank the Bielefeld particle physics group for having
so much of different activities going on and providing a real atmosphere of
high-end physics, which is very motivating and inspiring. Also thanks go to
our collaborators: Andre Morel, who made me more disciplined in presentation
of the results and supported my interest in analytic calculations; Thomas
Reisz and Piotr Bialas - for having an immense influence on me in analytical
and numerical sides of physics respectively. 

It would be impossible not to mention Gena Zinovjev, the supervisor of my
diploma work - who invested a lot of effort to make it all happened and
supported me in my scientific career.

The last but not the least I would like to thank our secretaries - Susi von
Reder, and most of all - Gudrun Eickmeyer who made my life here as
uncomplicated as it could be, helping me with all the administrative things
and guiding me through the dark forests of german, ukrainian and american
bureaucracy.

I dedicate this work to my parents - my mother Maria and my father Vladimir
- for combination of freedom and support in whatever I was doing, which is
really invaluable and my wife Jenny - for becoming such despite my doing
this Ph.D. thesis. 

Finally I would like to thank DFG and DAAD for supporting this work
financially.

\newpage\appendix

\chapter{Appendix}

%\appendix

\section{General SU(3) formulae and definitions}

In this Appendix we present the formulae and definitions which will be used in
the following calculations.

The Gell-Mann matrices are:%
\begin{align}
\lambda_{1}  &  =\left(
\begin{array}
[c]{ccc}%
0 & 1 & 0\\
1 & 0 & 0\\
0 & 0 & 0
\end{array}
\right)  \quad\lambda_{2}=\left(
\begin{array}
[c]{ccc}%
0 & -i & 0\\
i & 0 & 0\\
0 & 0 & 0
\end{array}
\right)  \quad\lambda_{3}=\left(
\begin{array}
[c]{ccc}%
1 & 0 & 0\\
0 & -1 & 0\\
0 & 0 & 0
\end{array}
\right) \label{lambda}\\
\lambda_{4}  &  =\left(
\begin{array}
[c]{ccc}%
0 & 0 & 1\\
0 & 0 & 0\\
1 & 0 & 0
\end{array}
\right)  \quad\lambda_{5}=\left(
\begin{array}
[c]{ccc}%
0 & 0 & -i\\
0 & 0 & 0\\
i & 0 & 0
\end{array}
\right)  \quad\lambda_{6}=\left(
\begin{array}
[c]{ccc}%
0 & 0 & 0\\
0 & 0 & 1\\
0 & 1 & 0
\end{array}
\right) \nonumber\\
\lambda_{7}  &  =\left(
\begin{array}
[c]{ccc}%
0 & 0 & 0\\
0 & 0 & -i\\
0 & i & 0
\end{array}
\right)  \quad\lambda_{8}=\frac{1}{\sqrt{3}}\left(
\begin{array}
[c]{ccc}%
1 & 0 & 0\\
0 & 1 & 0\\
0 & 0 & -2
\end{array}
\right) \nonumber
\end{align}

The actual generators of the $SU(3)$ group/algebra we take as the half of the
lambda-matrices
\begin{equation}
T_{a}\equiv\frac{1}{2}\lambda_{a} \label{gen}%
\end{equation}
They satisfy (for any $SU(N)$ group) the following commutation/anticommutation
relations:
\begin{equation}
\left[  T_{a}T_{b}\right]  =if_{abc}T_{c} \label{gencomm}%
\end{equation}%
\begin{equation}
\left\{  T_{a}T_{b}\right\}  =\frac{4}{N}\delta_{ab}I_{N}+4d_{abc}T_{c}
\label{genacomm}%
\end{equation}
with $f_{abc}$ being the absolutely antisymmetric and $d_{abc}$ - absolutely
symmetric thensors of the third rank. Traces of the generators and their
products satisfy the following relations:
\begin{equation}
TrT_{a}=0 \label{gentrace1}%
\end{equation}%
\begin{equation}
TrT_{a}T_{b}=\frac{1}{2}\delta_{ab} \label{gentrace2}%
\end{equation}%
\begin{equation}
TrT_{a}T_{b}T_{c}=\frac{1}{4}\left(  d_{abc}+if_{abc}\right)
\label{gentrace3}%
\end{equation}%
\begin{equation}
TrT_{a}T_{b}T_{a}T_{c}=-\frac{1}{4N}\delta_{bc} \label{gentrace4}%
\end{equation}
Also%
\begin{equation}
T_{a}T_{b}=\frac{1}{2}\left[  \frac{1}{N}\delta_{ab}+(d_{abc}+if_{abc}%
)T_{c}\right]
\end{equation}%
\begin{equation}
T_{a}^{ij}T_{a}^{kl}=\frac{1}{2}\left(  \delta_{il}\delta_{kj}-\frac{1}%
{N}\delta_{ij}\delta_{kl}\right)  \label{genmul}%
\end{equation}
Structure constants of the group satisfy the Jacobi identities
\begin{align}
f_{abc}f_{ecd}+f_{cbe}f_{aed}+f_{dbe}f_{ace}  &  =0\label{struc1}\\
f_{abc}f_{ecd}+f_{cbe}d_{aed}+f_{dbe}d_{ace}  &  =0 \label{struc2}%
\end{align}
From the antisymmetric nature of the $f_{abc}$ directly follows
\[
f_{abb}=0
\]
for any $b.$ Similar self-contraction relation holds for $d_{abc},$ where we
put the explicit summation to distinguish it from the previous formula.
\[
\sum_{b}d_{abb}=0
\]
The contraction relations for $f$ and $d$ are%
\begin{equation}
f_{acd}f_{bcd}=N\delta_{ab} \label{contr1}%
\end{equation}%
\begin{equation}
d_{acd}d_{bcd}=\frac{N^{2}-4}{N}\delta_{ab} \label{contr2}%
\end{equation}
\newpage

\section{One Link Integral}

We will use here the Gell-Mann parametrization of the arbitrary $SU(3)$
matrix:%
\begin{align}
U  &  =e^{i\frac{\alpha_{i}\lambda_{i}}{2}}\quad\\
U^{\dagger}  &  =e^{-i\frac{\alpha_{i}\lambda_{i}}{2}}%
\end{align}
where $\lambda_{i}$ are the Gell-Mann matrices (\ref{lambda}). For small
$\alpha_{i}$ we can expand%
\begin{align}
U  &  \sim1+i\frac{\alpha_{i}\lambda_{i}}{2}-\frac{\alpha_{i}\alpha_{j}%
\lambda_{i}\lambda_{j}}{8}-i\frac{\alpha_{i}\alpha_{j}\alpha_{k}\lambda
_{i}\lambda_{j}\lambda_{k}}{8\cdot3!}\\
U^{\dagger}  &  \sim1-i\frac{\alpha_{i}\lambda_{i}}{2}-\frac{\alpha_{i}%
\alpha_{j}\lambda_{i}\lambda_{j}}{8}+i\frac{\alpha_{i}\alpha_{j}\alpha
_{k}\lambda_{i}\lambda_{j}\lambda_{k}}{8\cdot3!}\nonumber
\end{align}

Let us calculate the measure first. Following the standard procedure we
calculate differentials:%
\begin{align*}
dU  &  \sim i\frac{d\alpha_{i}\lambda_{i}}{2}-\frac{\left(  d\alpha_{i}%
\alpha_{j}+\alpha_{i}d\alpha_{j}\right)  \lambda_{i}\lambda_{j}}{8}%
-i\frac{\left(  d\alpha_{i}\alpha_{j}\alpha_{k}+\alpha_{i}d\alpha_{j}%
\alpha_{k}+\alpha_{i}\alpha_{j}d\alpha_{k}\right)  \lambda_{i}\lambda
_{j}\lambda_{k}}{8\cdot3!}=\\
&  =i\frac{d\alpha_{i}\lambda_{i}}{2}-\frac{d\alpha_{i}\alpha_{j}\left(
\lambda_{i}\lambda_{j}+\lambda_{j}\lambda_{i}\right)  }{8}-i\frac{d\alpha
_{i}\alpha_{j}\alpha_{k}\left(  \lambda_{i}\lambda_{j}\lambda_{k}+\lambda
_{j}\lambda_{i}\lambda_{k}+\lambda_{k}\lambda_{j}\lambda_{i}\right)  }%
{8\cdot3!}=\\
&  i\frac{d\alpha_{i}F_{i}}{2}-\frac{d\alpha_{i}\alpha_{j}F_{ij}}{8}%
-i\frac{d\alpha_{i}\alpha_{j}\alpha_{k}F_{ijk}}{8\cdot3!}%
\end{align*}%
\[
dU^{\dagger}=-i\frac{d\alpha_{i}F_{i}}{2}-\frac{d\alpha_{i}\alpha_{j}F_{ij}%
}{8}+i\frac{d\alpha_{i}\alpha_{j}\alpha_{k}F_{ijk}}{8\cdot3!}%
\]

Now we calculate the \textquotedblright metric tensor\textquotedblright\ in
this parametrization:%
\begin{align}
TrdUdU^{\dagger}  &  \sim Tr\left(  i\frac{d\alpha_{i}F_{i}}{2}-\frac
{d\alpha_{i}\alpha_{j}F_{ij}}{8}-i\frac{d\alpha_{i}\alpha_{j}\alpha_{k}%
F_{ijk}}{8\cdot3!}\right)  \times\nonumber\\
&  \times\left(  -i\frac{d\alpha_{a}F_{a}}{2}-\frac{d\alpha_{a}\alpha
_{b}F_{ab}}{8}+i\frac{d\alpha_{a}\alpha_{b}\alpha_{c}F_{abc}}{8\cdot3!}\right)
\nonumber\\
&  =Tr\allowbreak\frac{d\alpha_{i}F_{i}}{2}\frac{d\alpha_{a}F_{a}}{2}%
+\frac{d\alpha_{i}\alpha_{j}F_{ij}}{8}\frac{d\alpha_{a}\alpha_{b}F_{ab}}%
{8}-2\frac{d\alpha_{i}F_{i}}{2}\frac{d\alpha_{a}\alpha_{b}\alpha_{c}F_{abc}%
}{8\cdot3!}\nonumber\\
&  \equiv g_{ia}d\alpha_{i}d\alpha_{a}%
\end{align}

with%
\begin{align}
g_{ia}  &  \equiv\frac{TrF_{i}F_{a}}{2^{2}}+\frac{\alpha_{j}\alpha_{b}%
TrF_{ij}F_{ab}}{2^{2}\left(  2\times2!\right)  ^{2}}-2\frac{\alpha_{b}%
\alpha_{c}TrF_{i}F_{abc}}{2^{2}\times2^{2}\cdot3!}=\nonumber\\
&  =\frac{\delta_{ia}}{2}+\frac{\alpha_{j}\alpha_{b}TrF_{ij}F_{ab}}{64}%
-\frac{\alpha_{b}\alpha_{c}TrF_{i}F_{abc}}{8\cdot3!}%
\end{align}%
\begin{equation}
q_{il}=2g_{il}=1+\alpha_{j}\alpha_{k}\left(  \frac{TrF_{ij}F_{lk}}{32}%
-\frac{TrF_{i}F_{lkj}}{24}\right)
\end{equation}
\qquad\qquad\qquad\qquad\qquad\qquad\qquad\qquad\qquad\qquad\qquad\qquad

According to \cite{Creutz} the invariant measure is then
\begin{equation}
d\mu=\sqrt{\det g}\prod_{i}d\alpha_{i}%
\end{equation}

Using the following representation of the determinant
\[
\det q_{il}=\exp Tr\log q_{il}%
\]
$\qquad$%
\begin{equation}
\log q_{il}=\alpha_{j}\alpha_{k}\left(  \frac{TrF_{ij}F_{lk}}{32}%
-\frac{TrF_{i}F_{lkj}}{24}\right)
\end{equation}%
\begin{equation}
Tr\log q_{il}=\sum_{i}\alpha_{j}\alpha_{k}\left(  \frac{TrF_{ij}F_{ik}}%
{32}-\frac{TrF_{i}F_{ikj}}{24}\right)  \equiv\alpha_{j}\alpha_{k}T_{jk}%
\end{equation}

where we defined (and calculated)%
\[
T_{jk}=\sum_{i}\left(  \frac{TrF_{ij}F_{ik}}{32}-\frac{TrF_{i}F_{ikj}}%
{24}\right)  =-\frac{1}{4}\delta_{jk}%
\]

So the expression for the measure takes the following simple form%
\begin{equation}
d\mu=\sqrt{\det g}\prod_{i}d\alpha_{i}=Ce^{-\frac{1}{8}\alpha_{i}\alpha_{i}%
}\prod_{i}d\alpha_{i}%
\end{equation}

We also define%
\[
G_{ijkl}=Tr\lambda_{i}\lambda_{j}\lambda_{k}\lambda_{l}%
\]

Explicit calculation leads to:%
\[
G_{ijkl}\alpha_{i}\alpha_{j}\alpha_{k}\alpha_{l}=2\sum_{i,j}a_{i}^{2}a_{j}^{2}%
\]

Now let us define the following average%
\begin{equation}
\left\langle A\right\rangle _{\gamma}=\int_{-\infty}^{\infty}\prod d\alpha
_{i}A\exp\left(  -\gamma a_{i}a_{i}\right)
\end{equation}

We will need these averages:
\[
\left\langle 1\right\rangle _{\gamma}=\frac{\pi^{4}}{\gamma^{4}}%
;\quad\left\langle a_{i}^{2}a_{j}^{2}\right\rangle _{\gamma}=\frac{\pi^{4}%
}{4\gamma^{6}};\quad\left\langle a_{i}^{4}\right\rangle _{\gamma}=\frac
{3\pi^{4}}{4\gamma^{6}}%
\]%
\begin{equation}
\left\langle \sum_{i,j}a_{i}^{2}a_{j}^{2}\right\rangle _{\gamma}%
=(8\ast8-8)\left\langle a_{i}^{2}a_{j}^{2}\right\rangle _{\gamma
}+8\left\langle a_{i}^{4}\right\rangle _{\gamma}=20\frac{\pi^{4}}{\gamma^{6}}
\label{aver}%
\end{equation}

\bigskip Now we can calculate the statistical sum:%
\begin{align}
Z  &  =\int d\mu e^{\frac{\beta}{2N}Tr\left(  U+U^{\dagger}\right)  }%
=e^{\beta}\int d\mu e^{-\frac{\beta}{2N}\left(  Tr\left(  1-U\right)
+Tr\left(  1-U^{\dagger}\right)  \right)  }=\nonumber\\
&  =Ce^{\beta}\int e^{\alpha_{j}\alpha_{k}T_{jk}}e^{-\frac{\beta}{2N}\left(
\frac{\alpha_{i}\alpha_{i}\delta_{ij}}{2}-\frac{\alpha_{i}\alpha_{j}\alpha
_{k}\alpha_{l}Tr\lambda_{i}\lambda_{j}\lambda_{k}\lambda_{l}}{8\cdot
4!}\right)  }=\nonumber\\
&  =Ce^{\beta}\int e^{-\left(  \alpha_{i}\alpha_{i}\left[  \frac{\beta}%
{4N}\delta_{ij}+\frac{1}{8}\delta_{ij}\right]  -\frac{\beta}{2N}\frac
{\alpha_{i}\alpha_{j}\alpha_{k}\alpha_{l}G_{ijkl}}{8\cdot4!}\right)  }=
\end{align}

Defining the new coupling%
\begin{equation}
\gamma=\left(  \frac{\beta}{4N}+\frac{1}{8}\right)
\end{equation}

and expanding the integrand we get%
\begin{equation}
Z=Ce^{\beta}\int e^{-\alpha_{i}\alpha_{i}\gamma\delta_{ij}}\left[
1+\frac{\beta}{2N}\frac{G_{ijkl}}{8\cdot4!}\alpha_{i}\alpha_{j}\alpha
_{k}\alpha_{l}\right]
\end{equation}

Now we will use the formulae for the averages (\ref{aver}) and get%
\begin{equation}
Z=Ce^{\beta}\frac{\pi^{4}}{\left(  \frac{\beta}{4N}+\frac{1}{8}\right)  ^{4}%
}\left[  1\allowbreak+\frac{\beta}{2N}\frac{20}{96\left(  \frac{\beta}%
{4N}+\frac{1}{8}\right)  ^{2}}\right]
\end{equation}

At this point we set $N=3$ (in fact our derivation is not for arbitrary $N$ at
this point, but we kept it variable for transparency) and obtain for the
string tension%
\begin{align}
\sigma &  =-\ln\frac{\partial\ln\left(  Ce^{\beta}\frac{\pi^{4}}{\left(
\frac{\beta}{12}+\frac{1}{8}\right)  ^{4}}\left[  1\allowbreak+\frac{\beta}%
{6}\frac{20}{96\left(  \frac{\beta}{12}+\frac{1}{8}\right)  ^{2}}\right]
\right)  }{\partial\beta}=\nonumber\\
&  =-\ln\frac{\partial\left(  \beta-4\ln\left(  2\beta+3\right)
+\allowbreak\ln\left(  1+20\frac{\beta}{\left(  2\beta+3\right)  ^{2}}\right)
\right)  }{\partial\beta}\nonumber\\
&  =-\ln\left(  1-\frac{8}{2\beta+3}+\frac{\frac{20}{\left(  2\beta+3\right)
^{2}}-80\frac{\beta}{\left(  2\beta+3\right)  ^{3}}}{1+20\frac{\beta}{\left(
2\beta+3\right)  ^{2}}}\right)
\end{align}
and expanding to the second order in $1/\beta$ we arrive to the final answer:
\begin{equation}
\sigma=\frac{4}{\beta}+\frac{7}{\beta^{2}}+O(\frac{1}{\beta^{3}})
\end{equation}

\section{One-loop Feynman diagrams for the vacuum polarization}

\begin{figure}[ptb]
\begin{center}
\epsfig{file=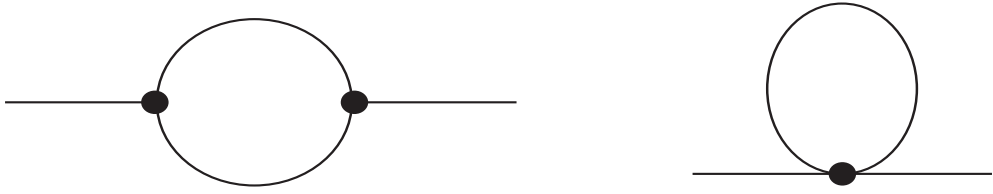, width=150mm, height=40mm}
\end{center}
\caption{Feynman diagramms for the second order term in the potential}%
\label{diag1}%
\end{figure}Here we calculate the Feynman diagrams for the $\tilde{\Pi}%
_{00}^{ns}$. We define the following variables:
\begin{equation}
\hat{q}_{\mu}=2\sin\frac{q_{\mu}}{2}%
\end{equation}%
\begin{equation}
c_{q_{\mu}}=\cos\frac{q_{\mu}}{2}%
\end{equation}%
\begin{equation}
\mathbf{\hat{q}}^{2}=\hat{q}^{2}+\hat{q}_{0}^{2}=\sum_{m=0}^{3}\hat{q}_{\mu
}\hat{q}_{\mu}%
\end{equation}%
\begin{equation}
\hat{q}^{2}=\sum_{m=1}^{3}\hat{q}_{\mu}\hat{q}_{\mu}%
\end{equation}

Feynman rules we need are \cite{curci}:

\begin{description}
\item[1.] 3-line vertex: $\left(  a,p,\mu\right)  \left(  b,q,\nu\right)
\left(  c,k,\lambda\right)  $ gives factor
\begin{equation}
\qquad igf^{abc}\left[  \delta_{\lambda\nu}\Hat{\left(  k-q\right)  }_{\mu
}c_{p_{\nu}}+\delta_{\lambda\mu}\left(  p-k\right)  _{\nu}c_{q_{\lambda}%
}+\delta_{\nu\mu}\left(  q-p\right)  _{\lambda}c_{k_{\mu}}\right]
\end{equation}

\item[2.] \bigskip propagator $(a,i)[q](b,j)$ gives factor
\begin{equation}
D_{ij}^{ab}\left(  q\right)  =\frac{1}{\hat{q}^{2}}\left(  \delta_{ij}%
-\frac{\hat{q}_{i}\hat{q}_{j}}{\mathbf{\hat{q}}^{2}}\right)  +\frac{\hat
{q}_{i}\hat{q}_{j}}{\mathbf{\hat{q}}^{2}}\left(  \frac{1}{\hat{q}_{0}^{2}%
}\left(  1-\delta_{q_{0},0}\right)  +\alpha\frac{\delta_{q_{0},0}%
}{\mathbf{\hat{q}}^{2}}\right)
\end{equation}%
\begin{equation}
D_{00}^{ab}\left(  q\right)  =\frac{\delta^{ab}\delta_{q_{0},0}}%
{\mathbf{\hat{q}}^{2}}%
\end{equation}

\end{description}

Non-static part is%

\begin{equation}
D_{ij}^{ab}\left(  q\right)  =\frac{1}{\hat{q}^{2}}\left(  \delta_{ij}%
-\frac{\hat{q}_{i}\hat{q}_{j}}{\mathbf{\hat{q}}^{2}}\right)  +\frac{\hat
{q}_{i}\hat{q}_{j}}{\mathbf{\hat{q}}^{2}}\left(  \frac{1}{\hat{q}_{0}^{2}%
}\right)
\end{equation}%
\begin{equation}
D_{00}^{ab}\left(  q\right)  =0
\end{equation}

\bigskip\textbf{First diagram}

The first diagram (Fig.\ref{diag1}, left) has two internal lines and two
vertices, that is

$\left(  a,q,\mu\right)  \left(  d,q^{\prime}-k,\lambda\right)  \left(
c,-q^{\prime},\rho\right)  $ and $\left(  c,q^{\prime},\sigma\right)  \left(
d,q-q^{\prime},\tau\right)  \left(  b,-q,\nu\right)  $

Hence, according to the Feynman rules above we have
\begin{align}
Q_{1}\left(  q\right)   &  =\frac{1}{L^{2}L_{0}}\sum\left(  \frac{i}%
{\sqrt{2c_{f}}}\right)  ^{2}\tilde{D}_{\lambda\tau}^{dd}\left(  q^{\prime
}-q\right)  \tilde{D}_{\rho\sigma}^{cc}\left(  q^{\prime}\right)
\times\nonumber\\
&  \times f^{adc}\left[  \delta_{\lambda\rho}\left(  -q^{\prime}+q-q^{\prime
}\right)  _{\mu}c_{q_{\lambda}}+\delta_{\mu\rho}\left(  q+q^{\prime}\right)
_{\lambda}c_{\left(  q^{\prime}-q\right)  _{\rho}}+\delta_{\lambda\mu}\left(
q^{\prime}-2q\right)  _{\rho}c_{q_{\mu}^{\prime}}\right]  \times\nonumber\\
&  \times f^{cdb}\left[  \delta_{\tau\nu}\left(  -q+q^{\prime}-q\right)
_{\sigma}c_{q_{\tau}^{\prime}}+\delta_{\sigma\nu}\left(  q+q^{\prime}\right)
_{\tau}c_{\left(  q^{\prime}-q\right)  _{\nu}}+\delta_{\sigma\tau}\left(
q-2q^{\prime}\right)  _{\nu}c_{q_{\sigma}}\right]
\end{align}
Now we set $\mu=\nu=0,$ and we calculate for $q=0$%
\begin{align}
Q_{1}\left(  0\right)   &  =\frac{1}{L_{s}^{2}L_{0}}\sum\left(  \frac{i}%
{\sqrt{2c_{f}}}\right)  ^{2}\tilde{D}_{\lambda\tau}^{dd}\left(  q^{\prime
}\right)  \tilde{D}_{\rho\sigma}^{cc}\left(  q^{\prime}\right)  \times
\nonumber\\
&  \times f^{adc}\times\left[  \delta_{\lambda\rho}\left(  -2q^{\prime
}\right)  _{0}+\delta_{0\rho}\left(  q^{\prime}\right)  _{\lambda}%
c_{q_{0}^{\prime}}+\delta_{\lambda0}\left(  q^{\prime}\right)  _{\rho}%
c_{q_{0}^{\prime}}\right] \nonumber\\
&  \times f^{cdb}\left[  \delta_{\tau0}\left(  q^{\prime}\right)  _{\sigma
}c_{q_{0}^{\prime}}+\delta_{\sigma0}\left(  q^{\prime}\right)  _{\tau}%
c_{q_{0}^{\prime}}+\delta_{\sigma\tau}\left(  -2q^{\prime}\right)
_{0}\right]
\end{align}
Only the first term in the first bracket and the last term in the second
survive%
\begin{equation}
Q_{1}\left(  0\right)  =\frac{1}{L_{s}^{2}L_{0}}\sum\left(  \frac{i}%
{\sqrt{2c_{f}}}\right)  ^{2}f^{adc}f^{cdb}(\tilde{D}_{ij}^{dd}\left(
q^{\prime}\right)  \tilde{D}_{ij}^{cc}\left(  q^{\prime}\right)  \left(
-2q^{\prime}\right)  _{0}\left(  -2q^{\prime}\right)  _{0}%
\end{equation}
Now we insert the propagators and omit the prime%
\begin{equation}
Q_{1}\left(  0\right)  =\frac{1}{L_{s}^{2}L_{0}}\sum\left(  \frac{i}%
{\sqrt{2c_{f}}}\right)  ^{2}f^{adc}f^{cdb}\left(  \frac{1}{\hat{q}^{2}}\left(
\delta_{ij}-\frac{\hat{q}_{i}\hat{q}_{j}}{\mathbf{\hat{q}}^{2}}\right)
+\frac{\hat{q}_{i}\hat{q}_{j}}{\mathbf{\hat{q}}^{2}}\frac{1}{\hat{q}_{0}^{2}%
}\right)  ^{2}\frac{1}{4}\sin^{2}q_{0}%
\end{equation}
Using identity (\ref{contr1}) and setting $a,b=0$
\begin{equation}
Q_{1}\left(  0\right)  =\frac{N_{c}}{L_{s}^{2}L_{0}}\sum_{\vec{k},k_{0}\neq
0}\sum_{ij=1}^{D-1}\left(  \frac{i}{\sqrt{2c_{f}}}\right)  ^{2}\left(
\frac{1}{\hat{q}^{2}}\left(  \delta_{ij}-\frac{\hat{q}_{i}\hat{q}_{j}%
}{\mathbf{\hat{q}}^{2}}\right)  +\frac{\hat{q}_{i}\hat{q}_{j}}{\mathbf{\hat
{q}}^{2}}\frac{1}{\hat{q}_{0}^{2}}\right)  ^{2}\frac{1}{4}\sin^{2}q_{0}%
\end{equation}

\begin{align}
&  =-\frac{N_{c}}{2L_{s}^{2}L_{0}}\sum_{\vec{q},q_{0}\neq0}\sum_{ij=1}%
^{D-1}\frac{\left(  \hat{q}_{i}\hat{q}_{j}+\delta_{ij}\hat{q}_{0}^{2}\right)
^{2}}{\left(  \hat{q}^{2}+\hat{q}_{0}^{2}\right)  ^{2}\hat{q}_{0}^{4}}\frac
{1}{4}\sin^{2}q_{0}\nonumber\\
&  =-\frac{N_{c}}{2L_{s}^{2}L_{0}}\sum_{\vec{q},q_{0}\neq0}\frac{1}{4}\sin
^{2}q_{0}\sum_{ij=1}^{D-1}\frac{q_{i}^{2}q_{j}^{2}+2q_{i}q_{j}\delta_{ij}%
q_{0}^{2}+\delta_{ij}^{2}q_{0}^{4}}{\left(  \hat{q}^{2}+\hat{q}_{0}%
^{2}\right)  ^{2}\hat{q}_{0}^{4}}\nonumber\\
&  =-\frac{N_{c}}{2L_{s}^{2}L_{0}}\sum_{\vec{q},q_{0}\neq0}\frac{4c_{0}^{2}%
}{\hat{q}_{0}^{2}}+2\frac{N_{c}}{L^{2}L_{0}}\sum_{\vec{q},q_{0}\neq0}%
\frac{c_{0}^{2}q_{0}^{2}}{\left(  \hat{q}^{2}+\hat{q}_{0}^{2}\right)  ^{2}%
}\nonumber\\
&  =-2\frac{N_{c}}{L_{s}^{2}L_{0}}\sum_{\vec{q},q_{0}\neq0}\frac{c_{0}%
^{2}q_{0}^{2}}{\left(  \hat{q}^{2}+\hat{q}_{0}^{2}\right)  ^{2}}-\frac{N_{c}%
}{L_{0}}\sum_{q_{0}\neq0}\frac{2}{\hat{q}_{0}^{2}}+\frac{N_{c}\left(
L_{0}-1\right)  }{2L_{0}}%
\end{align}

\textbf{Second diagram}

\bigskip Now let us calculate the second diagram on (Fig.\ref{diag1}, right).
For Feynman rules see \cite{rothe}%
\[
Q_{2}(0)=\frac{N_{c}}{L^{2}L_{0}}\sum_{k_{0}^{\prime}\neq0,k^{\prime}}\left(
\frac{1}{2c_{f}}\right)  D_{\rho\tau}^{cc}(k^{\prime})2\sum_{e}f_{mce}%
f_{cne}\left[
\begin{array}
[c]{c}%
\delta_{0\tau}\delta_{0\rho}c_{k_{\rho}^{\prime}}c_{k_{0}^{\prime}}%
+\delta_{\rho\tau}c_{2k_{0}^{\prime}}-\frac{1}{6}\delta_{0\tau}k_{0}^{\prime
}k_{\rho}^{\prime}\\
-\frac{1}{6}\delta_{0\rho}k_{0}^{\prime}k_{\tau}^{\prime}+\frac{1}{12}%
\delta_{0\tau}\delta_{0\rho}\left(  k^{\prime}\right)  ^{2}%
\end{array}
\right]  =
\]

\begin{equation}
=\frac{N_{c}}{L^{2}L_{0}}\sum_{k_{0}^{\prime}\neq0,k^{\prime}}D_{\rho\rho
}^{cc}(k^{\prime})c_{2k_{0}^{\prime}}\equiv\frac{N_{c}}{L^{2}L_{0}}\sum
_{k_{0}\neq0,k}D_{\rho\rho}^{cc}(k)c_{2k_{0}}%
\end{equation}

\begin{equation}
c_{2k_{0}}=\cos k_{0}=\cos^{2}\frac{k_{0}}{2}-\sin^{2}\frac{k_{0}}{2}%
=1-2\sin^{2}\frac{k_{0}}{2}=1-\frac{1}{2}k_{0}^{2}%
\end{equation}

\[
Q_{2}(0)=-\frac{N_{c}}{L^{2}L_{0}}\sum_{k,k_{0}\neq0}\frac{1}{2}+\frac{N_{c}%
}{L^{2}L_{0}}\sum_{k,k_{0}\neq0}\frac{1-\frac{1}{2}\hat{k}_{0}^{2}}{\hat
{k}^{2}+\hat{k}_{0}^{2}}+\frac{N_{c}}{L^{2}L_{0}}\sum_{k,k_{0}\neq0}\frac
{1}{\hat{k}_{0}^{2}}%
\]

\begin{equation}
=\frac{N_{c}}{L^{2}L_{0}}\sum_{k,k_{0}\neq0}\frac{1-\frac{1}{2}\hat{k}_{0}%
^{2}}{\hat{k}^{2}+\hat{k}_{0}^{2}}+\frac{N_{c}}{L_{0}}\sum_{k_{0}\neq0}%
\frac{1}{\hat{k}_{0}^{2}}-\frac{1}{2}\frac{N_{c}}{L_{0}}\left(  L_{0}%
-1\right)
\end{equation}
In addition to these two diagrams we need the corresponding ghost diagrams and
the diagram for the measure. All of such were calculated as in
\cite{petereisz} and amount to the term
\begin{equation}
Q_{0}=\frac{N_{c}}{L_{0}}\sum_{k_{0}\neq0}\frac{1}{\hat{k}_{0}^{2}}%
\end{equation}
Adding all these we get $\tilde{\Pi}_{00}^{ns}$
\begin{equation}
\tilde{\Pi}_{00}^{ns}(0)=\frac{3}{L_{0}L_{s}^{2}}\sum_{k_{0}\neq0}\sum
_{k}\left\{  \frac{2c_{k_{0}}^{2}\hat{k}_{0}^{2}}{\left(  \hat{k}^{2}+\hat
{k}_{0}^{2}\right)  ^{2}}-\frac{1-\hat{k}_{0}^{2}/2}{\hat{k}^{2}+\hat{k}%
_{0}^{2}}\right\}  \label{pi00}%
\end{equation}%
\begin{equation}
\tilde{\Pi}_{00}^{ns}(0)=\frac{3}{L_{0}L_{s}^{2}}\sum_{k_{0}\neq0}\sum
_{k_{1},k_{2}}\left\{  \frac{2\cos^{2}\frac{k_{0}}{2}4\sin^{2}\frac{\hat
{k}_{0}}{2}}{\left(  \cos^{2}k_{1}+\cos^{2}k_{2}+\cos^{2}k_{0}\right)  ^{2}%
}-\frac{1-2\sin^{2}\frac{\hat{k}_{0}}{2}}{\cos^{2}k_{1}+\cos^{2}k_{2}+\cos
^{2}k_{0}}\right\}
\end{equation}
The infinite volume limit is obtained by replacement
\begin{equation}
\left(  \frac{2\pi}{L_{s}}\right)  ^{2}\sum_{\vec{k}}\rightarrow\int_{-\pi
}^{\pi}d^{2}k
\end{equation}%
\begin{equation}
\tilde{\Pi}_{00}^{ns}(0)=\frac{3}{L_{0}}\sum_{k_{0}\neq0}\int_{-\pi}^{\pi
}\frac{d^{2}k}{4\pi^{2}}\left\{  \frac{8\cos^{2}\frac{k_{0}}{2}\sin^{2}%
\frac{\hat{k}_{0}}{2}}{\left(  \cos^{2}k_{1}+\cos^{2}k_{2}+\cos^{2}%
k_{0}\right)  ^{2}}-\frac{1-2\sin^{2}\frac{\hat{k}_{0}}{2}}{\cos^{2}k_{1}%
+\cos^{2}k_{2}+\cos^{2}k_{0}}\right\}
\end{equation}
We define the following%

\begin{align*}
t  &  =\sin^{2}\left(  \frac{\pi n_{0}}{L_{0}}\right) \\
S\left(  t\right)   &  =\frac{1}{4\pi^{2}}\int_{-\pi}^{\pi}\frac{dk_{1}dk_{2}%
}{2\left(  1+t\right)  -\cos\left(  k_{1}\right)  -\cos\left(  k_{2}\right)  }%
\end{align*}
and $\tilde{\Pi}_{00}^{ns}$ can be presented in the more convenient for the
practical calculation form
\begin{equation}
\tilde{\Pi}_{00}^{ns}(0)=\frac{3}{L_{0}}\sum_{n_{0}=1}^{L_{0}-1}\left(
-\frac{1}{2}\left(  1-2t\right)  S(t)-t\left(  1-t\right)  \frac{2S\left(
t\right)  }{dt}\right)  ,\qquad t=\sin^{2}\left(  \frac{\pi n_{0}}{L_{0}%
}\right)
\end{equation}
In a more compact way it can be expressed as
\begin{align}
W  &  \equiv\tilde{\Pi}_{00}^{ns}(0)=-\frac{3}{L_{0}}\sum_{n_{0}=1}^{L_{0}%
-1}\frac{d}{dx}G(x),\qquad x=\frac{\pi n_{0}}{L_{0}},\\
G(x)  &  \equiv\sin x\cos xS(\sin^{2}x)
\end{align}
In the integral defining $S$ we replace the inverse denominator in a following
way%
\begin{equation}
S(t)=\int_{-\pi}^{\pi}dk_{1}dk_{2}\int_{0}^{\infty}dy\exp[-y\left(  2\left(
1+t\right)  -\cos k_{1}-\cos k_{2}\right)  ]
\end{equation}
and identify two modified Bessel functions $I_{0}$ to get%
\begin{equation}
S\left(  t\right)  =\int_{0}^{\infty}dy\exp[-y\left(  2\left(  1+t\right)
\right)  ]I_{0}^{2}(y).
\end{equation}
This can be expressed via hyper-geometric function \cite{abram} as%
\[
S\left(  t\right)  =\frac{\left(  -z\right)  ^{1/2}}{2}F\left(
1/2;1/2;1;z\right)  ,\qquad z=\frac{1}{t\left(  t+2\right)  }.
\]
This ansatz relates the behaviour of $S\left(  t\right)  $ at low $t$ to that
of $F$ at large negative $z.$ Using \cite{abram} again one finds%
\begin{equation}
S\left(  t\right)  =\frac{1}{2\pi^{2}}\sum_{n=0}^{\infty}\left[  \frac
{\Gamma\left(  n+1/2\right)  }{\Gamma\left(  n+1\right)  }\right]  ^{2}%
z^{-n}\left[  \log\left(  -z\right)  +2\Psi\left(  n+1\right)  -2\Psi\left(
n+1/2\right)  \right]  \label{sasymp}%
\end{equation}
A first approximation $W_{0}$ to $W$is obtained from the standard relationship
between a sum and an integral:%

\begin{align}
W_{0}  &  =\frac{L_{0}}{2\pi}\int_{\pi/L_{0}}^{\pi-\pi/L_{0}}dG(x)+\frac{1}%
{4}\left[  \frac{dG(x)}{dx}\biggl \vert_{x=\frac{\pi}{L_{0}}}+\frac{dG(x)}%
{dx}\biggl \vert_{x=\pi-\frac{\pi}{L_{0}}}\right] \label{summation}\\
&  =-\frac{L_{0}}{\pi}G(\frac{\pi}{L_{0}})+\frac{1}{2}\frac{dG(x)}%
{dx}\biggl \vert_{x=\frac{\pi}{L_{0}}},
\end{align}
where use has been made of the symmetry of $\frac{dG(x)}{dx}$ under
$x\leftrightarrow\pi-x$. Up to terms which vanish as $\pi/L_{0}\rightarrow0$,
Eq. (\ref{sasymp}) gives:
\[
W_{0}\simeq-\frac{1}{2\pi}\left[  \frac{3\log2}{2}+1-\log\pi+\log
L_{0}\right]  .
\]

For that part of $G(x)$ which is analytic at $x=0$, this is the right answer.
A correction to the constant term in $L_{0}$ however arises from the
logarithmic singularity of $G(x)$ (Euler-Mac-Laurinformula,\cite{abram}). As
$x\rightarrow0$, we have $\frac{dG}{dx}\simeq-\frac{1}{\pi}\log\frac{n_{0}%
}{L_{0}}$ up to analytic terms, whose contribution $W^{log}$ to $W$ from Eq.
(\ref{sasymp}) can be computed using
\[
\sum_{n_{0}=1}^{L_{0}/2}\log\frac{n_{0}}{L_{0}}=\log\left(  \frac{(L_{0}%
/2)!}{L_{0}^{L_{0}/2}}\right)  \simeq\log\frac{\sqrt{\pi L_{0}}}%
{(2e)^{L_{0}/2}}%
\]
and found to be
\[
W^{log}=-\frac{1}{2\pi}\left[  \log(2L_{0}\pi)-L_{0}(1+\log2)\right]  .
\]
This we compare to the contribution $W_{0}^{log}$ to $W_{0}$ of the same
singular part of $\frac{dG}{dx}$, namely
\[
W_{0}^{log}=-\frac{1}{2\pi}\left[  \log(2L_{0})-L_{0}(1+\log2)\right]  ,
\]
so that in the limit $L_{0}\rightarrow\infty$ the final result is
\begin{align}
W  &  =W_{0}-W_{0}^{log}+W^{log}\\
&  =-\frac{1}{2\pi}\left[  \frac{5}{2}\log2-1+\log L_{0}\right]  .
\end{align}
\newpage

\section{Free field theory}

We present here a solution of the free-field theory, used in Chapter 2 for
checking simulation validity for the Higgs sector of the 2D model. Statistical
sum is
\begin{equation}
Z_{0}=\int e^{S_{0}}%
\end{equation}
with action being%
\begin{equation}
S_{0}=-\frac{\beta}{N}\sum_{x}\left(  \sum_{i=1}^{2}\operatorname{Tr}%
A_{0}(x)A_{0}(x+\hat{\imath})-2\operatorname{Tr}A_{0}^{2}(x)-\frac{h}%
{2}\operatorname{Tr}A_{0}^{2}(x)\right)
\end{equation}%
\begin{equation}
x=(i,j)\quad i=0,\ldots,n_{x}-1,\quad j=0,\ldots,n_{y}-1,
\end{equation}
Fields belong to the $SU(3)$ algebra and can be expressed via generators%
\begin{equation}
A_{0}(x)=i\sum_{a}A_{0}^{a}T^{a},\qquad A_{0}^{a}(x)\;\;\text{real}%
\end{equation}
Using the tracelessness of the generators, we get
\begin{equation}
\operatorname{Tr}A_{0}(x)A_{0}(y)=-\frac{1}{2}\sum_{a}A_{0}^{a}(x)A_{0}^{a}(y)
\end{equation}

We may rewrite the action in the following form%
\begin{align}
S_{0}  &  =\frac{1}{2}\frac{\beta}{N}\sum_{a}\sum_{x}\left(  \sum_{i=1}%
^{2}A_{0}^{a}(x)A_{0}^{a}(x+\hat{\imath})-2A_{0}^{a}(x)^{2}-\frac{h}{2}%
A_{0}(x)^{2}\right) \\
&  =-\frac{1}{4}\frac{\beta}{N}\sum_{a}\sum_{x}\left(  \sum_{i=1}^{2}\left(
A_{0}^{a}(x)-A_{0}^{a}(x+\hat{\imath})\right)  ^{2}+hA_{0}(x)^{2}\right)
\end{align}
and diagonalize it using the Fourier transform:
\begin{equation}
A(x)=\frac{1}{n_{x}n_{y}}\sum_{k}A(k)e^{ik\cdot x}%
\end{equation}%
\begin{equation}
k=(\frac{2\pi}{n_{x}}q,\frac{2\pi}{n_{y}}r)\quad q(r)=0,\ldots,n_{x}(n_{y})
\end{equation}
In Fourier space%
\begin{equation}
\sum_{x}A(x)^{2}=\frac{1}{n_{x}n_{y}}\sum_{k}\left\vert A(k)\right\vert ^{2}%
\end{equation}%
\[
\sum_{x}\left(  A(x+\hat{\imath})-A(x\right)  )^{2}=\frac{1}{n_{x}^{2}%
n_{y}^{2}}\sum_{x}\sum_{k,k^{\prime}}A(k)(e^{ik\cdot\hat{\imath}}-1)e^{ik\cdot
x}A(k^{\prime})(e^{ik^{\prime}\cdot\hat{\imath}}-1)e^{ik^{\prime}\cdot x}%
\]%
\begin{equation}
\sum_{x}e^{i(k+k)^{\prime}\cdot x}=n_{x}n_{y}\delta_{k,-k^{\prime}}%
\end{equation}%
\begin{equation}
A(-k)=\bar{A}(k),\qquad-k=2\pi-k
\end{equation}%
\begin{align}
\sum_{x}\left(  A(x+\hat{\imath})-A(x)\right)  ^{2}  &  =\frac{1}{n_{x}n_{y}%
}\sum_{k}|A(k)|^{2}(e^{ik\cdot\hat{\imath}}-1)(e^{-ik^{\prime}\cdot\hat
{\imath}}-1)\nonumber\\
&  =\frac{2}{n_{x}n_{y}}\sum_{k}|A(k)|^{2}\left(  1-\cos(k\cdot\hat{\imath
})\right)
\end{align}%
\begin{equation}
S_{0}=\frac{1}{n_{x}n_{y}}\sum_{a}\sum_{k}\left[  -\frac{1}{2}\frac{\beta}%
{N}|A_{0}^{a}(k)|^{2}\left(  \sum_{i}\left(  1-\cos(k\cdot\hat{\imath
})\right)  +\frac{h}{2}\right)  \ \right]
\end{equation}

We define
\begin{equation}
K_{mn}=\frac{1}{n_{x}n_{y}}\frac{\beta}{N}\left(  \sum_{i}\left(
1-\cos(k\cdot\hat{\imath})\right)  +\frac{h}{2}\right)  \delta_{mn}%
\end{equation}
and the action takes the final form
\begin{equation}
S_{0}=\ \sum_{a}\sum_{k}\left[  -\frac{1}{2}\ |A_{0}^{a}(k)|^{2}%
K_{mn}\ \right]
\end{equation}
Taking the integral over the Higgs fields, we get
\begin{equation}
Z_{0}=\prod_{k}\prod_{a}2\sqrt{\frac{\pi N}{\beta}}\left(  \frac{1}{2\sum
_{i}\left(  1-\cos(k\cdot\hat{\imath})\right)  +h}\right)  ^{\frac{1}{2}}%
\end{equation}

We are interested in the following estimate:
\begin{equation}
\sum_{x}\langle\operatorname{Tr}A_{0}^{2}(x)\rangle=\frac{2N}{\beta}%
\frac{\partial\log Z_{0}}{\partial h}%
\end{equation}%
\[
\log Z_{0}=-\frac{1}{2}(N^{2}-1)\left(  \sum_{k}\log\left(  2\sum_{i}\left(
1-\cos(k\cdot\hat{\imath})\right)  +h\right)  +\frac{1}{2}V\log\beta\right)
+\mathcal{C}%
\]%
\begin{align*}
\frac{\partial\log Z_{0}}{\partial h}  &  =-\frac{1}{2}(N^{2}-1)\sum_{k}%
\frac{1}{2\sum_{i}\left(  1-\cos(k\cdot\hat{\imath})\right)  +h}\\
&  -\frac{1}{2}(N^{2}-1)\sum_{q=0}^{n_{x}-1}\sum_{r=0}^{n_{y}-1}\frac
{1}{2\left(  1-\cos(\frac{2\pi}{n_{x}}q)\right)  +2\left(  1-\cos(\frac{2\pi
}{n_{y}}r)\right)  +h}%
\end{align*}%
\begin{equation}
\langle\operatorname{Tr}A_{0}^{2}(x)\rangle=-\frac{(N^{2}-1)N}{n_{x}n_{y}%
\beta}\sum_{q=0}^{n_{x}-1}\sum_{r=0}^{n_{y}-1}\frac{1}{2\left(  1-\cos
(\frac{2\pi}{n_{x}}q)\right)  +2\left(  1-\cos(\frac{2\pi}{n_{y}}r)\right)
+h}%
\end{equation}

This quantity can be computed numerically for any given lattice and is very
suitable to test the simulation of the Higgs sector of the model. \newpage

\section{Weak field expansion for Polyakov loops}

\bigskip One may also test the measurement of the Polyakov loops correlations
using the small field expansion%
\begin{equation}
P_{0,x}=\left\langle Tr\exp\left(  A_{0}\left(  0\right)  L_{0}\right)
Tr\exp\left(  A_{0}\left(  x\right)  L_{0}\right)  \right\rangle
\end{equation}%
\begin{equation}
P_{0,x}=\int\prod_{x}dATr\exp\left(  A_{0}\left(  0\right)  L_{0}\right)
Tr\exp\left(  A_{0}\left(  x\right)  L_{0}\right)  \exp S_{0}%
\end{equation}

We expand Polyakov loops, assuming that $A_{0}$ is small:
\begin{equation}
\exp\left(  A_{0}\left(  0\right)  L_{0}\right)  =I+A_{0}\left(  0\right)
L_{0}+\frac{1}{2}A_{0}\left(  0\right)  ^{2}L_{0}^{2}+..
\end{equation}%
\begin{align}
P_{0,x}  &  =\left\langle Tr\left(  I+A_{0}\left(  0\right)  L_{0}+\frac{1}%
{2}A_{0}\left(  0\right)  ^{2}L_{0}^{2}\right)  Tr\left(  I+A_{0}\left(
x\right)  L_{0}+\frac{1}{2}A_{0}\left(  x\right)  ^{2}L_{0}^{2}\right)
\right\rangle \\
&  =\left\langle TrI\right\rangle +\frac{1}{2}L_{0}^{2}\left\langle Tr\left(
A_{0}\left(  0\right)  ^{2}\right)  \right\rangle +\frac{1}{2}L_{0}%
^{2}\left\langle Tr\left(  A_{0}\left(  x\right)  \right)  ^{2}\right\rangle
\\
&  +\frac{1}{4}L_{0}^{4}\left\langle Tr\left(  A_{0}\left(  0\right)
^{2}\right)  Tr\left(  A_{0}\left(  x\right)  ^{2}\right)  \right\rangle
\end{align}%
\begin{equation}
\operatorname{Tr}A_{0}(0)A_{0}(x)=-\frac{1}{2}\sum_{i}A_{0}^{i}(0)A_{0}^{i}(x)
\end{equation}%
\begin{equation}
F(0,x)=\left\langle Tr\left(  A_{0}\left(  0\right)  ^{2}\right)  Tr\left(
A_{0}\left(  x\right)  ^{2}\right)  \right\rangle =-\frac{1}{4}\left\langle
\sum_{a}A_{0}^{a}(0)^{2}\sum_{b}A_{0}^{b}(x)^{2}\right\rangle
\end{equation}

To calculate such correlation function we proceed in a standard manner,
introducing the sources in the action%
\begin{equation}
S[J]=-\frac{1}{2}\frac{\beta}{N}\frac{1}{n_{x}n_{y}}\sum_{k}\left(  \sum
_{a,b}A_{0}^{a}(k)\Delta_{ab}^{-1}(k)A_{0}^{b}(-k)-\sum_{a}\frac{\beta}%
{N}\frac{1}{n_{x}n_{y}}\operatorname{Re}J^{a}(k)\bar{A}_{0}^{a}(k)\right)
\end{equation}
with
\begin{equation}
\Delta_{ab}(k)=\left(  \left(  1-\cos(k\cdot\hat{\imath})\right)  +h\right)
^{-1}\delta_{a,b}%
\end{equation}
New partition function is
\begin{equation}
Z[J]=Z_{0}\exp\left(  \frac{1}{2}\frac{\beta}{N}\frac{1}{n_{x}n_{y}}\sum
_{k}\sum_{a,b}J^{a}(k)\Delta_{ab}(k)J^{b}(-k)\right)
\end{equation}
and the required correlation can be calculated taking the derivatives with
respect to the sources and then setting them to zero
\begin{align*}
F(0,x)  &  =-\frac{1}{4}\left(  \frac{1}{n_{x}n_{y}}\right)  ^{2}%
\sum_{a,b,p,p^{\prime}}\exp\left(  ip^{\prime}R\right)  \left(  A_{0}%
^{a}(p)\right)  ^{2}\left(  A_{0}^{b}(p^{\prime})\right)  ^{2}\\
&  \times\exp\left(  \frac{1}{2}\frac{\beta}{N}\frac{1}{n_{x}n_{y}}\sum
_{k}\sum_{a,b}J^{a}(k)\Delta_{ab}(k)J^{b}(-k)\right) \\
&  =-\frac{1}{4}\left(  \frac{1}{n_{x}n_{y}}\right)  ^{2}\sum_{c,d,p,p^{\prime
}}\exp\left(  ip^{\prime}R\right) \\
&  \times\frac{\partial^{2}}{J^{c}\left(  p^{\prime}\right)  ^{2}}%
\frac{\partial^{2}}{J^{d}\left(  p\right)  ^{2}}\exp\left(  \frac{1}{2}%
\frac{\beta}{N}\frac{1}{n_{x}n_{y}}\sum_{k}\sum_{a,b}J^{a}(k)\Delta
_{ab}(k)J^{b}(-k)\right)
\end{align*}%
\begin{equation}
F(0,x)=-\frac{\beta^{2}}{4N^{2}\left(  n_{x}n_{y}\right)  ^{4}}\sum
_{c,d,p,p^{\prime}}\exp\left(  ip^{\prime}R\right)  \left[  \left(
\Delta_{cc}(p^{\prime})\right)  \left(  \Delta_{dd}(p)\right)  +\delta
_{p,p^{\prime}}\left(  2\Delta_{cd}(p^{\prime})\right)  ^{2}\right]
\end{equation}
\newpage

\section{Factorization of Matterfields}

First, we have to derive some identities which will be used to prove the
factorization of the higher-order averages. Let $A$ be an arbitrary matrix
belonging to the $SU(3)$ algebra. Such matrix can always be transformed to the
diagonal form. Then on diagonal will be its eigenvalues, which we will call
$\gamma_{1},$ $\gamma_{2}$ and $\gamma_{3}.$ Tracelessness implies $\gamma
_{3}=-\gamma_{1}-$ $\gamma_{2}.$ Our consideration \cite{bauer} is based on
the general expression for the determinant%
\begin{equation}
\log Det(1-tA)=Tr\log(1-tA)
\end{equation}
Let us calculate the left-hand side of it:%
\begin{align*}
Det(1-tA)  &  =\left(  1-t\gamma_{1}\right)  \left(  1-t\gamma_{2}\right)
\left(  1+t\left(  \gamma_{1}+\gamma_{2}\right)  \right) \\
&  =1-t^{2}\left(  \gamma_{1}^{2}+\gamma_{2}^{2}+\gamma_{1}\gamma_{2}\right)
+t^{3}\left(  \gamma_{1}^{2}\gamma_{2}+\allowbreak\gamma_{1}\gamma_{2}%
^{2}\right)
\end{align*}
The quantities which we are looking for are:%
\begin{align*}
A_{2}  &  \equiv TrA^{2}=\gamma_{1}^{2}+\gamma_{2}^{2}+\left(  \gamma
_{1}+\gamma_{2}\right)  ^{2}\\
&  =\allowbreak2\left(  \gamma_{1}^{2}+\gamma_{2}^{2}+\gamma_{1}\gamma
_{2}\right)
\end{align*}%
\begin{align*}
A_{3}  &  \equiv TrA^{3}=\gamma_{1}^{3}+\gamma_{2}^{3}-\left(  \gamma
_{1}+\gamma_{2}\right)  ^{3}=\allowbreak\\
&  =-3\left(  \gamma_{1}^{2}\gamma_{2}+\gamma_{1}\gamma_{2}^{2}\right)
\end{align*}
and we have%
\[
Det(1-tA)=1-\frac{A_{2}}{2}t^{2}-\frac{A_{3}}{3}t^{3}%
\]

Now we shall expand logarithms on the both sides of the identity. As long as
this expansion should be valid for arbitrary values of $t,$ we can match the
coefficients near any power of it. Left-hand side:%
\[
\ln Det(1-tA)=\allowbreak-\frac{1}{2}A_{2}t^{2}-\frac{1}{3}A_{3}t^{3}-\frac
{1}{8}A_{2}^{2}t^{4}-\frac{1}{6}A_{2}A_{3}t^{5}-\frac{1}{18}\allowbreak
A_{3}^{2}t^{6}-\frac{1}{24}A_{2}^{3}t^{6}+O(t^{7})
\]
Right-hand side:%
\[
Tr\log(1-tA)=-\frac{1}{2}t^{2}TrA^{2}-\frac{1}{3}t^{3}TrA^{3}-\frac{1}{4}%
t^{4}TrA^{4}-\allowbreak\frac{1}{5}t^{5}TrA^{5}-\frac{1}{6}t^{6}%
TrA^{6}+O(t^{7})
\]
and we immediately get:%
\begin{equation}
A_{2}^{2}=2TrA^{4} \label{appa4}%
\end{equation}%
\begin{equation}
\frac{5}{6}A_{2}A_{3}=TrA^{5} \label{appa5}%
\end{equation}%
\begin{equation}
\frac{1}{3}\allowbreak A_{3}^{2}+\frac{1}{4}A_{2}^{3}=TrA^{6} \label{appa6}%
\end{equation}

\textbf{Mean Field Calculation for A$_{22}:$}

Here we will work within the mean field approximation:
\[
A_{\alpha}(x)A_{\beta}(x)\rightarrow\frac{1}{4}\delta_{\alpha\beta}A_{2}(x)
\]%
\[
2A_{4}(x)=\frac{1}{2^{2}}\sum A_{\alpha}(x)A_{\alpha}(x)A_{\beta}(x)A_{\beta
}(x)
\]
Using Wick's theorem we get:
\begin{align*}
\left\langle 2A_{4}(x)\right\rangle  &  =\frac{1}{2^{2}}\sum_{\alpha,\beta
}\left\langle A_{\alpha}(x)A_{\alpha}(x)A_{\beta}(x)A_{\beta}(x)\right\rangle
=\frac{1}{2^{2}}\sum_{\alpha,\beta}\left\langle A_{\alpha}(x)A_{\alpha
}(x)\right\rangle \left\langle A_{\beta}(x)A_{\beta}(x)\right\rangle \\
&  +\frac{1}{2^{2}}2\sum_{\alpha,\beta}\left\langle A_{\alpha}(x)A_{\beta
}(x)\right\rangle \left\langle A_{\alpha}(x)A_{\beta}(x)\right\rangle \\
&  =\frac{1}{2^{2}}\left(  \sum_{\alpha}\frac{1}{4}\delta_{\alpha\alpha}%
A_{2}(x)\right)  ^{2}+\frac{1}{2^{2}}2\sum_{\alpha,\beta}\left(  \frac{1}%
{4}\delta_{\alpha\beta}A_{2}(x)\right)  ^{2}\\
&  =\frac{1}{2^{2}}\left(  8\frac{1}{4}A_{2}\right)  ^{2}+\frac{1}{2^{2}}%
2\ast8\left(  \frac{1}{4}A_{2}\right)  ^{2}=\frac{5}{4}A_{2}^{2}%
\end{align*}%
\[
\left\langle A_{n}\left(  x\right)  A_{m}\left(  0\right)  \right\rangle
=A_{nm}\left(  x\right)  +\left\langle A_{n}\left(  x\right)  \right\rangle
\left\langle A_{m}\left(  x\right)  \right\rangle
\]

And we arrive to the final result%
\[
\left\langle A_{2}^{2}\left(  x\right)  \right\rangle =A_{2,2}(0)+A_{2}^{2}%
\]

\textbf{Calculation of $A_{33}$}%
\[
\left\langle A_{3}\left(  x\right)  A_{3}\left(  0\right)  \right\rangle
=A_{33}\left(  x\right)  +\left\langle A_{3}\left(  x\right)  \right\rangle
\left\langle A_{3}\left(  x\right)  \right\rangle
\]
In our phase
\begin{equation}
\left\langle A_{3}\left(  x\right)  \right\rangle =0 \label{appsym}%
\end{equation}
so%
\[
A_{33}\left(  0\right)  =\left\langle A_{3}\left(  x\right)  ^{2}%
\right\rangle
\]%
\[
TrA^{3}=\sum_{ijk}A_{i}A_{j}A_{k}TrT_{i}T_{j}T_{k}=\sum_{ijk}A_{i}A_{j}%
A_{k}\frac{1}{4}\left(  d_{ijk}+if_{ijk}\right)  =\frac{1}{4}\sum_{ijk}%
d_{ijk}A_{i}A_{j}A_{k}%
\]
where we used the antisymmetric properties of $f_{ijk}.$
\[
\left\langle TrA^{3}TrA^{3}\right\rangle =\frac{1}{4}\frac{1}{4}\sum_{ijk}%
\sum_{lmn}d_{ijk}d_{lmn}A_{i}A_{j}A_{k}A_{l}A_{m}A_{n}%
\]
Now we will use the property of $d$ - it being zero after contraction when two
indices are the same and its symmetry which leaves us only with six equal
terms%
\[
\left\langle TrA^{3}\right\rangle =\frac{1}{4}\frac{1}{4}\left(  \frac{1}%
{4}A_{2}\right)  ^{3}\ast\sum_{ijk}d_{ijk}d_{ijk}=\frac{1}{4}\frac{1}%
{4}\left(  \frac{1}{4}A_{2}\right)  ^{3}6\frac{N^{2}-4}{N}\left(
N^{2}-1\right)  =\frac{5}{64}A_{2}^{3}%
\]%
\[
\left\langle TrA^{3}\right\rangle =\frac{5}{64}A_{2}^{3}%
\]
Or
\[
A_{33}\left(  0\right)  =\frac{5}{64}A_{2}^{3}%
\]

\textbf{Calculation of} $\left\langle A_{2}A_{3}\right\rangle $%
\begin{align*}
\left\langle A_{2}A_{3}\right\rangle  &  =\left\langle \frac{1}{2}\frac{1}%
{4}\sum_{n}\sum_{ijk}A_{n}A_{n}A_{i}A_{j}A_{k}d_{ijk}\right\rangle =\\
&  =\frac{1}{2}\frac{1}{4}3\sum_{n}\sum_{ijk}\left\langle A_{n}A_{n}%
A_{i}\right\rangle \left\langle A_{j}A_{k}\right\rangle d_{ijk}+\\
&  +\frac{1}{2}\frac{1}{4}6\sum_{n}\sum_{ijk}\left\langle A_{n}A_{i}%
A_{j}\right\rangle \left\langle A_{n}A_{k}\right\rangle d_{ijk}+\\
&  +\frac{1}{2}\frac{1}{4}\sum_{n}\sum_{ijk}\left\langle A_{n}A_{n}%
\right\rangle \left\langle A_{i}A_{j}A_{k}\right\rangle d_{ijk}%
\end{align*}%
\begin{align*}
\left\langle A_{2}A_{3}\right\rangle  &  =\\
&  =\frac{1}{2}\frac{1}{4}3\sum_{n}\sum_{ijk}\left\langle A_{n}A_{n}%
A_{i}\right\rangle \frac{1}{4}A_{2}\delta_{jk}d_{ijk}+\\
&  +\frac{1}{2}\frac{1}{4}6\sum_{n}\sum_{ijk}\left\langle A_{n}A_{i}%
A_{j}\right\rangle \frac{1}{4}A_{2}\delta_{nk}d_{ijk}+\\
&  +\frac{1}{2}\frac{1}{4}\sum_{n}\sum_{ijk}\frac{1}{4}A_{2}\left\langle
A_{i}A_{j}A_{k}\right\rangle d_{ijk}%
\end{align*}
The first term falls out completely - contraction of $d$ with two equal indices.%

\begin{align*}
\left\langle A_{2}A_{3}\right\rangle  &  ==\frac{1}{2}\frac{1}{4}6\sum_{n}%
\sum_{ijk}d_{ijk}\delta_{nk}\left\langle A_{n}A_{i}A_{j}\right\rangle \frac
{1}{4}A_{2}+\\
&  +\frac{1}{2}\frac{1}{4}\sum_{n}\sum_{ijk}\frac{1}{4}d_{ijk}A_{2}%
\left\langle A_{i}A_{j}A_{k}\right\rangle
\end{align*}
Summing over $n:$%

\begin{align*}
\left\langle A_{2}A_{3}\right\rangle  &  =\frac{1}{2}\frac{1}{4}6\frac{1}%
{4}A_{2}\sum_{ijk}d_{ijk}\left\langle A_{k}A_{i}A_{j}\right\rangle +\frac
{1}{2}\frac{1}{4}8\frac{1}{4}A_{2}\sum_{ijk}d_{ijk}\left\langle A_{i}%
A_{j}A_{k}\right\rangle =\\
&  =\frac{1}{2}\frac{1}{4}(6+8)\frac{1}{4}A_{2}\sum_{ijk}d_{ijk}\left\langle
A_{k}A_{i}A_{j}\right\rangle =\frac{1}{2}\frac{1}{4}(6+8)A_{2}A_{3}=\frac
{7}{4}A_{2}A_{3}%
\end{align*}
\newpage\textbf{Correlation Functions}

\bigskip Now we turn to the correlation functions. Square brackets denote
contraction and we will silently use the formulae above.%
\[
\left\langle A_{n}\left(  x\right)  A_{m}\left(  0\right)  \right\rangle
=A_{nm}\left(  x\right)  +\left\langle A_{n}\left(  x\right)  \right\rangle
\left\langle A_{m}\left(  x\right)  \right\rangle
\]%
\begin{align*}
A_{24}\left(  R\right)   &  =\left\langle A_{2}\left(  R\right)  ;A_{4}\left(
0\right)  \right\rangle =\frac{1}{2}\left\langle A_{2}\left(  R\right)
;A_{2}^{2}\left(  0\right)  \right\rangle =\\
&  =\frac{1}{2}\left[  \left[  A_{2}\left(  R\right)  ;\frac{N^{2}+1}{N^{2}%
-1}A_{2}^{2}\left(  0\right)  \right]  \right]  =\frac{1}{2}2\frac{N^{2}%
+1}{N^{2}-1}\left[  \left[  A_{2}\left(  R\right)  ;A_{2}\left(  0\right)
\right]  \right]  \left[  \left[  A_{2}\left(  0\right)  \right]  \right]  =\\
&  =\frac{N^{2}+1}{N^{2}-1}A_{22}\left(  R\right)  A_{2}=\frac{5}{4}%
A_{22}\left(  R\right)  A_{2}%
\end{align*}%
\begin{align*}
A_{35}\left(  R\right)   &  =\left\langle A_{3}\left(  R\right)  ;A_{5}\left(
0\right)  \right\rangle =\frac{5}{6}\left\langle A_{3}\left(  R\right)
;A_{3}\left(  0\right)  A_{2}\left(  0\right)  \right\rangle =\\
&  =\frac{5}{6}\left[  \left[  \left\langle \left\langle A_{3}\left(
R\right)  \right\rangle \right\rangle ;\left\langle \left\langle A_{3}\left(
0\right)  A_{2}\left(  0\right)  \right\rangle \right\rangle \right]  \right]
=\\
&  =\frac{5}{6}\frac{N^{2}+5}{N^{2}-1}\left[  \left[  \left\langle
\left\langle A_{3}\left(  R\right)  \right\rangle \right\rangle ;\left\langle
\left\langle A_{3}\left(  0\right)  \right\rangle \right\rangle \left\langle
\left\langle A_{2}\left(  0\right)  \right\rangle \right\rangle \right]
\right]  =\\
&  =\frac{5}{6}\frac{N^{2}+5}{N^{2}-1}\left(  \left[  \left[  A_{3}\left(
R\right)  ;A_{3}\left(  0\right)  \right]  \right]  \left[  \left[
A_{2}\left(  0\right)  \right]  \right]  +\left[  \left[  A_{3}\left(
R\right)  ;A_{2}\left(  0\right)  \right]  \right]  \left[  \left[
A_{3}\left(  0\right)  \right]  \right]  \right)  =\\
&  =\frac{35}{24}A_{3,3}A_{2}%
\end{align*}%
\begin{align*}
A_{44}\left(  R\right)   &  =\left\langle A_{4}\left(  R\right)  ;A_{4}\left(
0\right)  \right\rangle =\left[  \left[  \left\langle \left\langle
A_{4}\left(  R\right)  \right\rangle \right\rangle ;\left\langle \left\langle
A_{2}^{2}\left(  0\right)  \right\rangle \right\rangle \right]  \right]  =\\
&  =\left(  \frac{1}{2}\frac{N^{2}+1}{N^{2}-1}\right)  ^{2}\left[  \left[
A_{2}^{2}\left(  R\right)  ;A_{2}^{2}\left(  0\right)  \right]  \right]  =\\
&  =\left(  \frac{1}{2}\frac{N^{2}+1}{N^{2}-1}\right)  ^{2}\left(
\begin{array}
[c]{c}%
2\left[  \left[  A_{2}\left(  R\right)  ;A_{2}\left(  0\right)  \right]
\right]  \left[  \left[  A_{2}\left(  R\right)  ;A_{2}\left(  0\right)
\right]  \right]  +\\
+4\left[  \left[  A_{2}\left(  R\right)  ;A_{2}\left(  0\right)  \right]
\right]  \left[  \left[  A_{2}\left(  0\right)  \right]  \right]  ^{2}%
\end{array}
\right)  =\\
&  =\left(  \frac{1}{2}\frac{N^{2}+1}{N^{2}-1}\right)  ^{2}\left(
2A_{22}\left(  R\right)  ^{2}+4A_{22}\left(  R\right)  A_{2}^{2}\right)  =\\
&  =\left(  \frac{5}{8}\right)  ^{2}\left(  2A_{22}\left(  R\right)
^{2}+4A_{22}\left(  R\right)  A_{2}^{2}\right)
\end{align*}

\begin{align*}
A_{55}\left(  R\right)   &  =\left\langle A_{5}\left(  R\right)  ;A_{5}\left(
0\right)  \right\rangle =\left[  \left[  \left\langle \left\langle
A_{5}\left(  R\right)  \right\rangle \right\rangle ;\left\langle \left\langle
A_{5}\left(  0\right)  \right\rangle \right\rangle \right]  \right]  =\\
&  =\left(  \frac{5}{6}\right)  ^{2}\left(  \frac{N^{2}+5}{N^{2}-1}\right)
^{2}\left[  \left[  A_{2}\left(  R\right)  A_{3}\left(  R\right)
;A_{2}\left(  0\right)  A_{3}\left(  0\right)  \right]  \right]  =\\
&  =\left(  \frac{5}{6}\right)  ^{2}\left(  \frac{N^{2}+5}{N^{2}-1}\right)
^{2}\left(
\begin{array}
[c]{c}%
\left[  \left[  A_{2}\left(  R\right)  ;A_{2}\left(  0\right)  \right]
\right]  \left[  \left[  A_{3}\left(  R\right)  ;A_{3}\left(  0\right)
\right]  \right]  +\\
+\left[  \left[  A_{2}\left(  R\right)  ;A_{3}\left(  0\right)  \right]
\right]  ^{2}+\\
+\left[  \left[  A_{2}\left(  R\right)  ;A_{2}\left(  0\right)  \right]
\right]  \left[  \left[  A_{3}\left(  R\right)  \right]  \right]  ^{2}+\\
+2\left[  \left[  A_{2}\left(  R\right)  ;A_{3}\left(  0\right)  \right]
\right]  \left[  \left[  A_{3}\left(  R\right)  \right]  \right]  \left[
\left[  A_{2}\left(  R\right)  \right]  \right]  +\\
+\left[  \left[  A_{3}\left(  R\right)  ;A_{3}\left(  0\right)  \right]
\right]  \left[  \left[  A_{2}\left(  R\right)  \right]  \right]  ^{2}%
\end{array}
\right)  =\\
&  =\left(  \frac{5}{6}\right)  ^{2}\left(  \frac{N^{2}+5}{N^{2}-1}\right)
^{2}\left(  A_{22}A_{33}+A_{23}^{2}+A_{22}A_{3}^{2}+2A_{23}A_{3}A_{2}%
+A_{33}A_{2}^{2}\right)  =\\
&  =\left(  \frac{5}{6}\right)  ^{2}\left(  \frac{N^{2}+5}{N^{2}-1}\right)
^{2}\left(  A_{22}A_{33}+A_{33}A_{2}^{2}\right)  =\left(  \frac{35}%
{24}\right)  ^{2}\left(  A_{22}A_{33}+A_{33}A_{2}^{2}\right)
\end{align*}

In the last step we used the symmetry relation (\ref{appsym})


\newpage

\begin{thebibliography}{99}                                                                                               %


\bibitem {I}P. Bialas, A. Morel, B. Petersson, K. Petrov and T. Reisz, ``High
Temperature 3D QCD: Dimensional Reduction at Work'', \textit{Nucl.\ Phys.}
\textbf{B581} (2000) 477.

\bibitem {II}P. Bialas, A. Morel, B. Petersson, K. Petrov and T. Reisz, `` QCD
with Adjoint Scalars in 2D: Properties in the Colourless Scalar Sector '',
\textit{Nucl.\ Phys.} \textbf{B603} (2001) 369.

\bibitem {III}P. Bialas, A. Morel, B. Petersson, K. Petrov and T. Reisz,
``Screening Masses in Dimensionally Reduced (2+1)D Gauge Theory'',
\textit{Nucl.Phys.Proc.Suppl.} \textbf{(2002)} 882-884.

\bibitem {IV}P. Bialas, A. Morel, B. Petersson, K. Petrov and T. Reisz, ``Z(3)
Symmetric Dimensional Reduction of (2+1)D QCD'', submitted to
\textit{Nucl.\ Phys.\ Proc.\ Suppl.}

\bibitem {karschreview}F.Karsch, ``Lattice QCD at High Temperature and
Density'', hep-lat/0106019

\bibitem {eos}G.~Boyd, J.~Engels, F.~Karsch, E.~Laermann, C.~Legeland,
M.~L\"utgemeier, B.~Petersson, Phys.Rev.Lett. 75 (1995) 4169-4172

\bibitem {ginsparg}P.~Ginsparg, \textit{Nucl.\ Phys.} \textbf{B170} (1980) 388.

\bibitem {appelquist}T. Appelquist and R. Pisarski, \textit{Phys.\ Rev.}
\textbf{D23} (1981) 2305.

\bibitem {nadkarni}S. Nadkarni, \textit{Phys. Rev.} $\mathbf{D27}$ (1983) 917;
\textit{Phys. Rev.} $\mathbf{D38}$ (1988) 3287.

\bibitem {landsman}N.~P.~Landsman, \textit{Nucl.~Phys.} \textbf{B322} (1989) 498.

\bibitem {thomas}T. Reisz, \textit{Z. f. Phys.} $\mathbf{C53}$ (1992) 169.

\bibitem {karsch2p1}J. Engels, F. Karsch, E. Laermann, C. Legeland, M.
Lutgemeier, B. Petersson, T. Scheideler, Nucl.Phys.Proc.Suppl.\textbf{53}:420-422,1997

\bibitem {engelsFSS}J.Engels, Nucl.Phys. \textbf{B30 }(1993), 347

\bibitem {su2}P. Lacock, D. E. Miller and T. Reisz, \textit{Nucl. Phys.}
$\mathbf{B369}$ (1992) 501.

\bibitem {su3}L. K\"arkk\"ainen, P. Lacock, D. E. Miller, B. Petersson and T.
Reisz, \textit{Phys. Lett.} $\mathbf{B282}$ (1992) 121.

\bibitem {qcd}L. K\"arkk\"ainen, P. Lacock, B. Petersson and T. Reisz,
\textit{Nucl. Phys.} $\mathbf{B395}$ (1993) 733.

\bibitem {lacock}P. Lacock, T. Reisz, \textit{Nucl. Phys. B (Proc. Suppl.)}
\textbf{30} (1993) 307

\bibitem {kajantie0}K.~Kajantie, M.~Laine, K.~Rummukainen, M.~Shaposhnikov,
Phys. Lett \textbf{77} (1996) 2887, \textit{Nucl. Phys.} \textbf{B493} (1997)
413.\newline M.~Gurtler, E.-M.~Ilgenfritz, A.~Schiller, Phys.
Rev.~\textbf{D56} (1997) 3888.\newline F.~Karsch, T.~Neuhaus, A.~Patkos,
J.~Rank, \textit{Nucl. Phys. Proc. Supl.} \textbf{53} (1997) 623.

\bibitem {kajantie1}K.~Kajantie, M.~Laine, K.~Rummukainen, M.~Shaposhnikov,
\textit{Nucl.~Phys.} \textbf{B503} (1997) 357.

\bibitem {kajantie2}K.~Kajantie, M.~Laine, A.~Rajantie, K.~Rummukainen, and
M.~Tsypin, \textit{JHEP} \textbf{9811} (1998) 11.

\bibitem {kajantie3}K.~Kajantie, M.~Laine, J.~Peisa, A.~Rajantie, K.~
Rummukainen, M.~ Shaposhnikov \textit{Phys.~Rev.~Lett.} \textbf{79} (1997) 3130.

\bibitem {datta}S.~Datta, S.~Gupta, Nucl. Phys. \textbf{B534} (1998) 392,
\textit{Phys.Lett.} \textbf{B471} (2000) 382.

\bibitem {karschpetr}F.~Karsch, M.~Oevers and P.~Petreczky,
\textit{\textquotedblleft Screening masses of hot SU(2) gauge theory from the
3D adjoint Higgs model\textquotedblright}, hep-ph/9902373.

\bibitem {hart}A.~Hart and O.~Philipsen, \textit{\textquotedblleft The
spectrum of the three-dimensional adjoint Higgs model and hot SU(2) gauge
theory,\textquotedblright}, hep-lat/9908041.

\bibitem {space}L.~K\"arkk\"ainen, P.~Lacock, D.E.~Miller, B.~Petersson,
T.~Reisz \textit{Phys.~Lett.} \textbf{B312} (1993) 173.

\bibitem {karsch}G.S.~Bali, J.~Fingberg, U.M.~Heller, F.~Karsch and
K.~Schilling, \textit{Phys.~Rev.~Lett.} \textbf{71} (1993) 3059\newline
F.~Karsch, E.~Laermann, M.~L\"utgemeier \textit{Phys.~Lett.} {B346} (1995) 94.

\bibitem {dhoker}E. D'Hoker, \textit{Nucl. Phys} \textbf{B201} (1982) 401.

\bibitem {lego}C.Legeland, \textit{PhD. Thesis}, \textit{\textquotedblleft
Aspects of (2+1) Dimensional Lattice Gauge Theory\textquotedblright}
(University of Bielefeld, Germany, September 1998).

\bibitem {petereisz}B. Petersson and T. Reisz, \textit{Nucl. Phys.}
$\mathbf{B353}$ (1991) 757.

\bibitem {curci}C. Curci and P. Menotti, \textit{Z. f. Phys.} \textbf{C21}
(1984) 281; C. Curci, P. Menotti and G. Paffuti, \textit{Z. f. Phys.}
\textbf{C26} (1985) 549.

\bibitem {STALG}T.~Reisz, \textit{Jour. Math. Phys.} \textbf{32} (1991) 515

\bibitem {tbook}T.~Reisz, In preparation

\bibitem {frules}A.~Irb\"ack, P.~Lacock, D.~Miller, B.~Petersson, T.~Reisz,
\textit{Nucl.~Phys.} \textbf{B363} (1991) 34.

\bibitem {kark}L. K\"arkk\"ainen, P. Lacock, D.E. Miller, B. Petersson and T.
Reisz, \textit{Nucl. Phys.} \textbf{B418} (1994) 3.

\bibitem {thomas2}T. Reisz, ``Dimensionally Reduced SU(2) Yang-Mills Theory is
Confined'',in \textit{Quantum Field Theoretical Aspects of High Energy
Physics}, 230-235, B. Geyer and E.M. Ilgenfritz Eds., Frankenhausen 1993.
%\cite{Caselle:mt}
\bibitem{Caselle:mt}
M.~Caselle, R.~Fiore, F.~Gliozzi, P.~Guaita and S.~Vinti,
%``On The Behavior Of Spatial Wilson Loops In The High Temperature Phase
%Of Lgt,''
Nucl.\ Phys.\ B {\bf 422}, 397 (1994)
[arXiv:hep-lat/9312056].
%%CITATION = HEP-LAT 9312056;%%

\bibitem {heathbath}N.~Cabibbo and E.~Marinari,
%``A New Method For Updating SU(N) Matrices In Computer Simulations Of Gauge Theories,''
Phys.\ Lett.\ \textbf{B119} (1982) 387.

\bibitem {kp}A.~D.~Kennedy and B.~J.~Pendleton,
%``Improved Heat Bath Method For Monte Carlo Calculations In Lattice Gauge Theories,''
Phys.\ Lett.\ \textbf{B156} (1985) 393.

\bibitem {gross}D. J. Gross and E. Witten, \textit{Phys. Rev.} \textbf{D21}
(1980) 446.

\bibitem {lang}C. B. Lang, P. Salomonson and B. S. Skagerstam, \textit{Phys.
Letters} \textbf{100} (1981) 29.

\bibitem {grad}I.S. Gradshteyn and I.M. Ryzhik, \textit{Table of Integrals,
Series, and Products}, Alan Jeffrey, Ed. (Academic Press).

\bibitem {abram}M. Abramowitz and I.A. Stegun, \textit{Handbook of
Mathematical Functions}, (National Bureau of Standards).

\bibitem {gao}M.~Gao, Phys. Rev. \textbf{D40} (1989) 2708.

\bibitem {kacz}O.~Kaczmarek, F.~Karsch, E.~Laermann and M.~Lutgemeier,
\textquotedblleft\textit{Heavy quark potentials in quenched QCD at high
temperature,}\textquotedblright\ hep-lat/9908010.

\bibitem {reisz}T. Reisz, \textit{Z. f. Phys.} $\mathbf{C53}$ (1992) 169.

\bibitem {rothe}H.~Rothe, Introduction to Lattice Gauge Theories.

\bibitem {swendsen}A. M. Ferrenberg, Robert H.~Swendsen, Phys.Rev.Lett.
\textbf{63 } (1989) 1195-1198.

\bibitem {Creutz}M.Creutz, Quarks, gluons and lattices, Cambrige University
Press (1993)

\bibitem {kark1}L. K\"arkk\"ainen, P. Lacock, D.E. Miller, B. Petersson and T.
Reisz, \textit{Phys. Lett.} \textbf{B282} (1992) 121.

\bibitem {kark2}L. K\"arkk\"ainen, P. Lacock, B. Petersson and T. Reisz,
\textit{Nucl. Phys.} \textbf{B395} (1993) 733.

\bibitem {philipsen}O. Philipsen, ``Static correlation lengths in QCD at high
temperature and finite density'', hep-lat/0011019.

\bibitem {arnold}P. Arnold and L.G. Yaffe, \textit{Phys.~Rev.} \textbf{D52}
(1995) 7208.

\bibitem {hart1}A. Hart, O. Philipsen, J.D. Stack and M. Teper, \textit{Phys.
Lett.} \textbf{B396} (1997) 217.

\bibitem {hart2}A. Hart and O. Philipsen, \textit{Nucl. Phys.} \textbf{B572}
(2000) 243.

\bibitem {parisi}G. Parisi, \textit{Statistical Field Theory} (Addison-Wesley,
New York, 1988)

\bibitem {gleiser}M. Alford and M. Gleiser \textit{Phys. Rev} \textbf{D48}
(1993) 2838; J. Borrill and M. Gleiser, \textit{Nucl. Phys.} \textbf{B483}
(1997) 416.

\bibitem {martin}A.D. Martin and T.D. Spearman. Elementary Particle Theory.
North-Holland Publishing Co., Amsterdam, 1970.

\bibitem {buchmuller}W. Buchmuller and O. Philipsen \textit{Phys. Lett.}
\textbf{B397} (1997) 112.

\bibitem {bauer}We thank M. Bauer for providing us with this simple trick.

\bibitem {pisarski}R. Pisarski, \textquotedblleft\textit{Quark-Gluon Plasma as
a condensate of }$SU(3)$\textit{\ Wilson Lines}\textquotedblright,
hep-ph/0006205. K.~Kajantie, M.~Laine, J.~Peisa, A.~Rajantie, K.~ Rummukainen,
M.~ Shaposhnikov \textit{Phys.~Rev.~Lett.} \textbf{79} (1997) 3130.

\bibitem {etherboot}http://etherboot.sourceforge.net/

\bibitem {lam}http://www.lam-mpi.org
\end{thebibliography}
\end{document}